\documentclass{aa}  

\usepackage{graphicx, array}
\usepackage[varg]{txfonts}
\usepackage{ulem}
\usepackage{natbib,twoopt}
\usepackage[dvipsnames]{xcolor}
\usepackage[breaklinks=true]{hyperref} 
\hypersetup{
    colorlinks=true,
    linkcolor=blue,
    filecolor=magenta,
    urlcolor=cyan,
    citecolor=blue
}

\providecommand{\teff}{\ensuremath{T_{\rm eff}}}
\providecommand{\feh}{[Fe/H]}
\providecommand{\logg}{\ensuremath{\log g}}
\providecommand{\tgm}{(\teff,\logg,\feh)}
\providecommand{\Fbol}{$F_{\rm bol}$}
\providecommand{\diam}{$\theta_{\rm LD}$}

\begin{document}

   \title{Gaia FGK Benchmark Stars: fundamental \teff\ and \logg\ of the  third version
   \thanks{The full catalogue is only available in electronic form at the CDS via anonymous ftp to cdsarc.u-strasbg.fr (130.79.128.5) 
or via http://cdsarc.u-strasbg.fr/viz-bin/qcat?J/A+A/?/?}
}

    \author{
C. Soubiran\inst{\ref{LAB}} \and  
O. Creevey\inst{\ref{OCA}} \and
N. Lagarde\inst{\ref{LAB}} \and  
N. Brouillet\inst{\ref{LAB}} \and  
P. Jofr\'e\inst{\ref{UDP},\ref{eris}} \and
L. Casamiquela\inst{\ref{GEPI},\ref{ICC}} \and
U. Heiter\inst{\ref{UPS}} \and
C. Aguilera--G\'omez\inst{\ref{PUC}} \and
S. Vitali\inst{\ref{UDP}} \and
C. Worley\inst{\ref{NZ},\ref{ioa}} \and
D. de Brito Silva\inst{\ref{UDP}}
}

\institute{
Laboratoire d'Astrophysique de Bordeaux, Univ. Bordeaux, CNRS, B18N, all\'ee Geoffroy Saint-Hilaire, 33615 Pessac, France\label{LAB}
\email{caroline.soubiran@u-bordeaux.fr}
\and
Universit\'e C\^ote d'Azur, Observatoire de la C\^ote d'Azur, CNRS, Laboratoire Lagrange, Bd de l'Observatoire, CS 34229, 06304 Nice cedex 4, France
\label{OCA}
\and
Instituto de Estudios Astrof\'isicos, Facultad de Ingenier\'ia y Ciencias, Universidad Diego Portales, Av. Ej\'ercito Libertador 441, Santiago, Chile 
\label{UDP}
\and
Millenium Nucleous ERIS\label{eris}
\and
GEPI, Observatoire de Paris, PSL Research University, CNRS, Sorbonne Paris Cit\'e, 5 place Jules Janssen, 92190 Meudon, France\label{GEPI}
\and
Institut de Ci\`encies del Cosmos (ICCUB), Departament de F\' isica Qu\`antica i Astrof\'isica, Universitat de Barcelona (UB), Mart\'i i Franqu\`es, 1, 08028 Barcelona, Spain\label{ICC}
\and
Observational Astrophysics, Division of Astronomy and Space Physics, Department of Physics and Astronomy, Uppsala University, Box 516, SE-751 20 Uppsala, Sweden \label{UPS}
\and
Instituto de Astrof\'isica, Pontificia Universidad Cat\'olica de Chile, Av. Vicu\~na Mackenna 4860, 782-0436 Macul, Santiago, Chile \label{PUC}
\and
School of Physical and Chemical Sciences -- Te Kura Mat\=u, University of Canterbury, Private Bag 4800, Christchurch 8140, New
Zealand \label{NZ}
\and
Institute of Astronomy, University of Cambridge, Madingley Road,
Cambridge CB3 0HA, United Kingdom \label{ioa}
}
  
 \date{Received, accepted }

  \abstract
   {Large spectroscopic surveys devoted to the study of the Milky Way, including Gaia, use automated pipelines to massively determine the atmospheric parameters of millions of stars. The Gaia FGK Benchmark Stars are reference stars with \teff\ and \logg\ derived through fundamental relations, independently of spectroscopy, to be used as anchors for the parameter scale. The first and second versions of the sample have been extensively used for that purpose, and more generally to help constrain stellar models.}
   {We provide the third version of the Gaia FGK Benchmark Stars, an extended set intended to improve the calibration of spectroscopic surveys, and their interconnection. }
  {We have compiled about 200 candidates which have precise measurements of angular diameters and parallaxes.  We determined their bolometric fluxes by fitting their spectral energy distribution. Masses were determined using two sets of stellar evolution models. In a companion paper we describe the determination of metallicities and detailed abundances.}
 {We provide a new set of 192 Gaia FGK Benchmark Stars with their fundamental \teff\ and \logg, and with uncertainties lower than 2\% for most stars. Compared to the previous versions, the homogeneity and accuracy of the fundamental parameters are significantly improved thanks to the high quality of the Gaia photometric and astrometric data.}
 {}
 
 \keywords{stars: late-type -- stars: fundamental parameters -- stars: atmospheres -- standards -- surveys}

\maketitle
%

\section{Introduction}
The last decade has been marked by a large observational effort aimed at deciphering the history of our Galaxy based on large samples of stars observed by spectroscopic surveys. This has stimulated the development of efficient methodologies for the massive determination of atmospheric parameters (APs).  In particular the recent Gaia Data Release 3 \citep[][Gaia DR3]{GDR3} just delivered \teff, \logg\ and \feh\ for millions of stars \citep{cre22,fou22}. In particular two datasets were released that mainly include F, G and K-type stars, one for 5.6 million stars with APs based on medium resolution spectra from the Radial Velocity Spectrometer \citep{GSP-Spec}, the other one for 471 million stars with APs based on low resolution spectra from the blue and red prisms, parallax and integrated photometry \citep{GSP-Phot}. The methodologies used for the massive determination of atmospheric parameters rely on stellar models which are not perfect and not able to reproduce exactly real spectra, causing some biases which have to be corrected. 

The Gaia FGK Benchmark Stars (GBS) are reference stars to be used for the calibration and the validation of spectroscopic methods of parametrisation. They are chosen to cover the range of F, G, and K spectral types at different luminosities and metallicities, and to have the necessary observations available to determine their effective temperature and surface gravity independently from spectroscopy, at a precision level of 1-2\%. The determination
of \teff\ and \logg\  is performed through the fundamental relations implying observable quantities (angular diameters directly measured by interferometry, bolometric fluxes and parallaxes) and the mass, the only parameter depending on theoretical assumptions.

The first and second versions of the GBS (hereafter V1 and V2, respectively) were presented in a series of papers. \cite{hei15}, hereafter Paper~I,  describe the initial selection of 34 stars, including the Sun, and the determination of their fundamental effective temperatures and surface gravities, resulting in the GBS V1 sample.   \cite{bla14} (Paper~II) present the library of high-resolution spectra that was assembled and used to determine metallicities \citep[][Paper~III]{jof14} and elemental abundances  of $\alpha-$capture and iron-peak elements \citep[][Paper~IV]{jof15}. One limitation of the V1 sample was the small number of targets, in particular in the metal-poor regime. Metal-poor stars are usually distant and faint, which makes them difficult to observe in interferometry.  In Paper~V, \cite{haw16} proposed a list of ten metal-poor stars to be included in the GBS sample. The GBS V2 sample summarised by \cite{jof18} includes 36 stars, merged from Paper~I and Paper~V. The change in number from 34 to 36 comes from the  addition of five metal-poor stars from Paper V and the removal of some stars from Paper~I because their spectroscopic analysis indicated that they could not be recommended as reference stars. However, V2 was an intermediate version where the fundamental properties of the stars were not redetermined owing to the lack of direct and accurate measurements of angular diameters for some stars.

The material provided in these series of papers consists of accurate APs for stars covering a extensive range of spectral types and metallicities, in addition to a library of high resolution and high signal-to-noise spectra from which line-by-line abundances are also provided. This material can be further exploited in spectroscopic studies. Indeed, Paper VI of the GBS series \citep{jof17}, reports on a collective work using the GBS to investigate the different sources of uncertainties in elemental abundances in order to improve spectroscopic pipelines.

The ultimate goal of the efforts dealing with GBS is to provide to spectroscopic surveys the fundamental \teff\ and \logg\ scales and an external reference for abundances. Despite their limitation in sample size and parallax precision previous to Gaia data, the GBS have been extensively used in the past years. The Gaia astrophysical parameters inference system \citep{bai13, cre22} made use of GBS for the validation of the stellar parameters published in Gaia DR3. The GBS are also a fundamental source of calibration and validation of the Gaia-ESO Survey \citep{gil22, ran22, hou23}, of the RAVE survey \citep{rave, ste20}, and of the GALAH survey \citep{bud21}. The OCCASO project \citep{cas19} has systematically observed two GBS giants, Arcturus and $\mu$ Leo, to validate chemical abundances of open clusters. Upcoming large projects such as WEAVE \citep{jin22} are also making use of the GBS.

Calibrations based on GBS can help to make surveys more homogeneous and mutually compatible so that they can be combined into the most comprehensive database of chemical measurements for the study of the Milky Way stellar populations \citep{jof18}.  The applications of the GBS however can extend far beyond this specific purpose. As for the study presented in Paper VI \citep[see also][]{bla19}, many spectroscopic studies have benefited from the GBS effort. For example, \cite{adi20} used the GBS to assess the performances of the ESPRESSO, PEPSI and HARPS high-resolution spectrographs while \cite{hei21} used some spectra from Paper~II to assess the quality of hundreds of spectral lines and the corresponding atomic and molecular data used for the abundance analyses of FGK-type stars carried out within the Gaia-ESO survey \citep[see also][for lines in the Infrared]{kon19, fuk21}. \cite{ama22} and \cite{lin22} used the GBS to quantify the differences in abundances derived using state-of-the-art 3D non-LTE atmosphere models and the standard 1D LTE models.

In addition to spectroscopy, the GBS help to constrain better stellar evolution models. For example, \cite{sah19} determined ages of the GBS as a way to test the reliability of the determination of stellar ages for various stellar populations; \cite{ser17} used GBS to validate their asteroseismic analysis performed on dwarfs and subgiants.  The GBS have also been used as validation for the PLATO stellar analysis pipeline \citep{gen22}. Many of the lessons learnt from the GBS are further discussed in \cite{jof19}.

However, we are aware that the current sample of GBS is still imperfect and too small to make a satisfactory interconnection between surveys. This is why an extension of the sample is required. The V1 and V2 GBS samples were also limited by the parallax accuracy needed for a fundamental \logg\ determination. This is not anymore an issue thanks to the exquisite astrometric quality of the Gaia data \citep{gaia}.

In this Paper VII of the series, we present the extended sample and third version of the GBS (GBS V3) that includes about 200 stars. We took advantage of recent interferometric studies that provided new measurements of angular diameters for large samples of stars \citep[e.g.][]{lig16, bai18,bai21,van21} and for metal-poor stars \citep[e.g.][]{cre15, kar18, kar20}. As explained in Sect. \ref{s:selection} we selected new GBS candidates based on quality criteria applied on interferometric measurements. Sect. \ref{s:teff} describes the compilation of angular diameters and fluxes that are needed to compute the fundamental \teff. Bolometric fluxes (\Fbol) were homogeneously computed by the method of spectral energy distribution (SED) fitting  based on a large collection of (spectro)photometric data. Sect. \ref{s:logg} deals with the determination of \logg\ with parallaxes from Gaia DR3 \citep{GDR3}, or Hipparcos \citep{hip-2} for the brightest stars, and with masses inferred from a state-of-the-art methodology and stellar tracks. At each of these different steps we assess the uncertainties of the stellar parameters. Sect. \ref{s:sample} provides an overview of the sample properties and shows some comparisons to \teff\ and \logg\ from different catalogues, before our concluding remarks in Sect. \ref{s:conclusion}. All the compiled and computed parameters of this work are given in the form of a catalogue distributed by the CDS. We note that these parameters still require a last iteration considering \feh\ values, needed for the estimation of \Fbol\ and masses and here adopted from the literature, that are consistent with our fundamental parameters. This is a necessary step to recommend our parameters for reference \citep{hei15}.  
The accompanying Paper VIII (Casamiquela et al. in preparation)  presents homogeneous determinations of \feh\ and of  detailed abundances of the GBS V3 derived from a spectroscopic analysis. For this purpose a large dataset of high-resolution, high signal-to-noise spectra was collected from public archives and through our own observing programs. The recommended parameters and abundances of the GBS are appropriately updated at the CDS.

\section{Star selection}
\label{s:selection}
In order to determine \teff\ and \logg\ through the fundamental relations with a minimum of assumptions and theoretical input, our principal criterion was to choose F, G and K stars with a high quality measurement of angular diameter. Ideally we want our GBS sample to homogeneously cover the \tgm\ space, which implies a special effort to add metal-poor stars. 
We have therefore searched the literature for GBS candidates fulfilling these criteria.

First, we considered the GBS from V1 and V2 \citep{hei15, jof18}. 
The GBS V1 sample has 29 FGK-type stars (including the Sun), four giants with \teff\ around 4000~K, corresponding to late K and early M spectral types, and one cooler M giant. 
The V2 sample was built from the V1 one, with the addition of metal-poor stars. For the V3 list we considered 
all the 39 V1 and V2 stars including several stars with indirect determinations of angular diameters. For all we searched for new direct determinations of angular diameters as well as other data needed to update their fundamental \teff\ and \logg. We added to this list eight metal-poor stars from \cite{kar20,kar22a}, not part of V1 and V2, and two targets recently observed with the CHARA interferometer (Creevey et al. in preparation). This sample of 49 stars was our initial set.

To further extend the GBS sample, we searched for new candidates observed in interferometry. We used the compilation from the Jean-Marie Mariotti Center (JMMC), the JMMC Measured Stellar Diameters Catalogue \citep[JMDC, ][]{duv16}. This catalogue, regularly updated, intends to be as exhaustive as possible in listing all the measurements of stellar apparent diameters made with direct techniques. It is therefore a very appropriate resource to extend the GBS sample. The JMDC is a bibliographical catalogue which implies that some stars have multiple entries, resulting from studies with different instruments, in different bands and with different precisions. Deciding which value of angular diameter is the most appropriate for a given star can be challenging, in particular owing to non-homogeneous uncertainties listed in the JMDC. In addition, there are many stars in the JMDC which are not appropriate for our purpose, such as some classes of variable stars, hot stars, spectroscopic binaries, and fast rotators. In addition, some very uncertain measurements of angular diameters could propagate large uncertainties to \teff\ and should be discarded. Therefore we made a first selection to reject stars and measurements not relevant for our purpose.

To do so, we followed \cite{sal20}  who established accurate surface brightness-colour relations for different spectral types and luminosity classes. They applied three types of rejection criteria on the JMDC data. First they examined the stellar characteristics to reject variable and semi-regular pulsating stars, spectroscopic binaries and other multiple stars, fast rotators and stars with a doubtful luminosity class. Second, they used criteria on the quality of the interferometric measurements that we apply similarly (see Sect. \ref{s:diam}). The third type of criterion is based on the uncertainty of the K magnitude. We considered their list of 106 carefully selected F5 to K7 dwarfs and giants that we added to the initial set (five stars were already there). However, the study of \cite{sal20} does not take into account the metallicity of the stars since their objective is to infer radii of stars and planets in the context of the PLATO mission which mainly focuses on solar-like stars. For us the metallicity is essential since the GBS should be representative of all the Milky Way stellar populations. We aim to improve the sampling of the GBS in \teff\ and \logg\ but also in \feh\
with as many GBS candidates as possible on the metal-poor side. We noticed that the criteria used by \cite{sal20}, in particular the photometric one, tend to reject metal-poor stars. The only star with \feh$<-$1.0 in \cite{sal20}'s sample is the well-known benchmark star HIP76976 (HD 140283), part of GBS V1, which has \feh=$-$2.36$\pm$0.10 in Paper~IV. 

We then searched for additional stars in the September 2021 version of JMDC available at the CDS which includes 2013 measurements of 1062 stars, a significant increase compared to the February 2020 version used by \cite{sal20}. In order to find stars in the appropriate range of atmospheric parameters, we used the PASTEL catalogue \citep{sou16}  and its recent version which provides mean atmospheric parameters for 14\,181 FGK stars \citep{sou22}. We expect PASTEL to be complete for metal-poor stars brighter than V$\sim$8.25, which is the limiting magnitude of FGK-type stars with an interferometric measurement in JMDC. Among the $\sim$500 stars in common between PASTEL and JMDC, we considered 63 additional stars to include in our sample, because they fill gaps in the AP space, and their interferometric angular diameters fulfill the criteria of \cite{sal20}.

The resulting list of selected candidates for GBS V3 includes 201 stars (the Sun is not considered here) 
They are all members of the Hipparcos catalogue (ESA 1997) and only the ten brightest ones are missing in Gaia DR3 \citep{gaia}. We keep the Sun in the GBS V3 since it is an obvious benchmark star, although it is not observable in the same conditions as other stars. We do not discuss the Sun in the present paper, keeping its fundamental \teff\ and \logg\ determined in Paper~I (we also note that a nominal value for the effective temperature of the Sun was adopted at the XXIXth IAU General Assembly, see \citealt{2015arXiv151007674M,2016AJ....152...41P}).

In the following, metallicities \feh\ are needed for the determination of \Fbol\ from SEDs (to initialize the minimization process, see Sect. \ref{s:fbol}), and for the determination of masses from stellar evolutionary tracks (see Sect. \ref{s:mass}).  We have adopted \feh\ from the literature for the 201 stars, mainly from the PASTEL catalogue. For a sake of homogeneity, we have not adopted \feh\ from Papers~III and V for stars in V1 and V2 because they are corrected from non-local thermodynamic equilibrium (NLTE) effects, while for all the other stars the literature values are assuming LTE. It is the purpose of the forthcoming Paper VIII to provide precise and homogeneous abundances of Fe and other elements.  This will imply some iterations to get the recommended \teff\ and \logg\ of our targets.

\section{Fundamental effective temperature}
\label{s:teff}
The luminosity $L$, the radius $R$, and the effective temperature \teff\ of a given star are linked through the fundamental relation $L = 4\pi R^2 \sigma T_{\rm eff}^4$, where  $\sigma$ is the Stefan-Boltzmann constant. The fundamental relation can be expressed in a way that gives \teff\ as a function of the limb-darkened angular diameter \diam\ and the bolometric flux \Fbol\ which are measurable quantities:

\begin{equation}
\centering
 T_{\rm eff} = \left(\frac{F_{\rm bol}}{\sigma}\right)^{0.25} (0.5\,\theta_{\rm LD})^{-0.5} = 2341\left(\frac{F_{\rm bol}}{\theta_{\rm LD}^2}\right)^{0.25}
 \label{e:teff}
\end{equation}

where \diam\ is in milliarcseconds (mas) and \Fbol\  in $10^{-8}$~erg~s$^{-1}$~cm$^{-2}$ or $10^{-11}$~W~m$^{-2}$.

In the following subsections we describe our compilation of measured angular diameters and fluxes. The latter were used to compute \Fbol\ by means of SED fitting. Subsequently, Eq.~(\ref{e:teff}) was used to obtain \teff\ for the selected stars.

\subsection{Compilation of angular diameters}
\label{s:diam}
As explained in Sect. \ref{s:selection}, the selection of GBS V3 stars was mainly based on the JMDC which provides one or several values of \diam\ for each star. In particular, we considered the 106 targets that \cite{sal20} used as calibration stars to define precise surface brightness–colour relations. \cite{sal20} applied interferometric criteria to remove non reliable values of \diam\ in the JMDC. They rejected measurements with a relative uncertainty on the angular diameter larger than 8\%, and those based on observations in the 8-13 micron band or having a bad observation quality and/or a poor spatial frequency coverage in the visibility curve. They also rejected stars with inconsistent redundancies.  We adopted their selected values of \diam\ for the 106 stars. For HIP112748 and HIP54539, provided with two values of \diam\ differing by less than 1\%, we adopted the one with the lowest uncertainty.

For the remaining stars, we queried JMDC and the recent literature in order to retrieve the latest values of \diam\ fulfilling the interferometric criteria applied by \cite{sal20}. When provided we inspected the visibility curves to evaluate the reliability of the measurement.

We found recent and precise \diam\ for ten of the GBS V1 and V2. In particular, for three of the metal-poor benchmark stars new measurements are available, for HD103095 (HIP57939) and HD122563 (HIP68594) by \cite{kar20}, and for HD140283 (HIP76976) by \cite{kar18}. The two components of the binary $\alpha$ Cen were remeasured by \cite{ker17}, while four other targets were found in \cite{bai18,bai21}. Among the six stars which had no direct \diam\ in Paper~I, only one (HIP48455, $\mu$ Leo) was observed in interferometry by \cite{bai18}. Among the five metal-poor stars from Paper~V with indirect values of \diam, one (HIP92167) was observed by \cite{kar20}. Thus, we are left with nine stars from V1 and V2 that are still without any direct measurement of \diam. We keep them in a separate table for continuity of the samples, but we do not consider them as GBS anymore.

The final version of the GBS V3 includes 192 stars with a direct measurement of \diam. For each we provide the limb darkened angular diameter with its uncertainty and the corresponding reference in Table \ref{t:fund_teff_logg} of Appendix~\ref{s:appendix} and in the catalogue available at the CDS. The sample includes stars with small angular diameters such as HIP97527 (\diam=0.231$\pm$0.006 mas) and HIP93427 (\diam=0.289$\pm$0.006 mas), both of which are asteroseismic targets observed with the CHARA/PAVO instrument by \cite{hub12}. The sample also includes Aldebaran (HIP21421) and Arcturus (HIP69673) which have angular diameters as large as $\sim$20 mas. The median angular diameter of the sample is 1.12 mas.

\begin{figure}[ht]
\centering
 \includegraphics[width=0.48\textwidth]{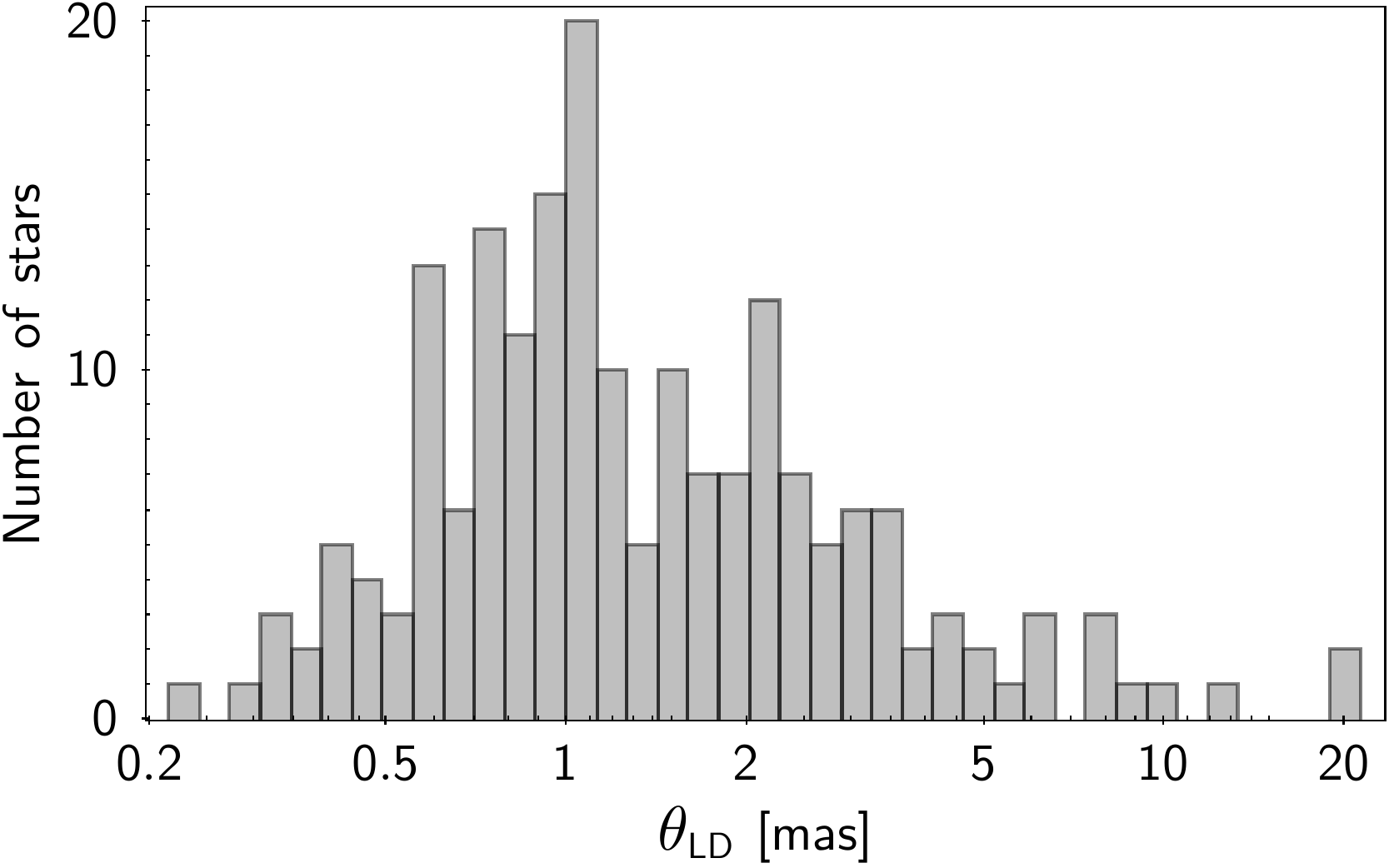} 
 \includegraphics[width=0.48\textwidth]{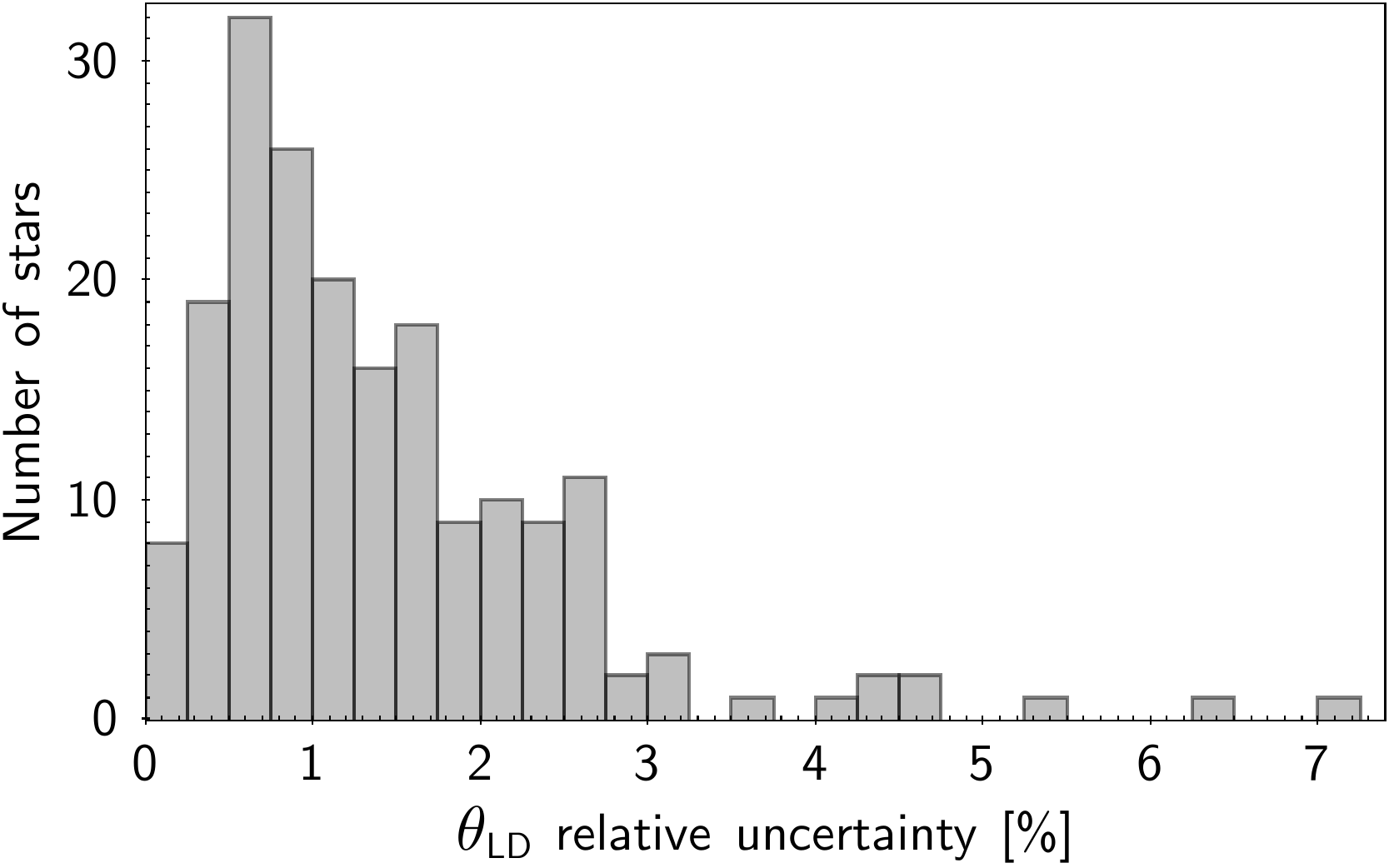} 
 \caption{Histogram of \diam\ (top panel) and its relative uncertainty (bottom panel) for the 192 GBS V3 stars having interferometric measurements.}
 \label{f:e_ThetaLD_histo}
\end{figure}

The relative \diam\ uncertainties range from 0.1\% (HIP87808) to 7\% (HIP25993) with a median value of 1.1\% (see histogram in Fig. \ref{f:e_ThetaLD_histo}). Two other stars have relative uncertainties larger than 5\%, HIP7294 and HIP14838. 
In absolute values, the largest uncertainties occur for the two giants $\psi$ Phe (HIP8837) and Arcturus (HIP69673), with  \diam=8.0$\pm$0.2 mas and \diam=21.0$\pm$0.21 mas, respectively. The two stars do not seem to have been re-observed recently, so that their \diam\ is still that of Paper~I.

\begin{figure}[ht]
\centering
 \includegraphics[scale=0.3]{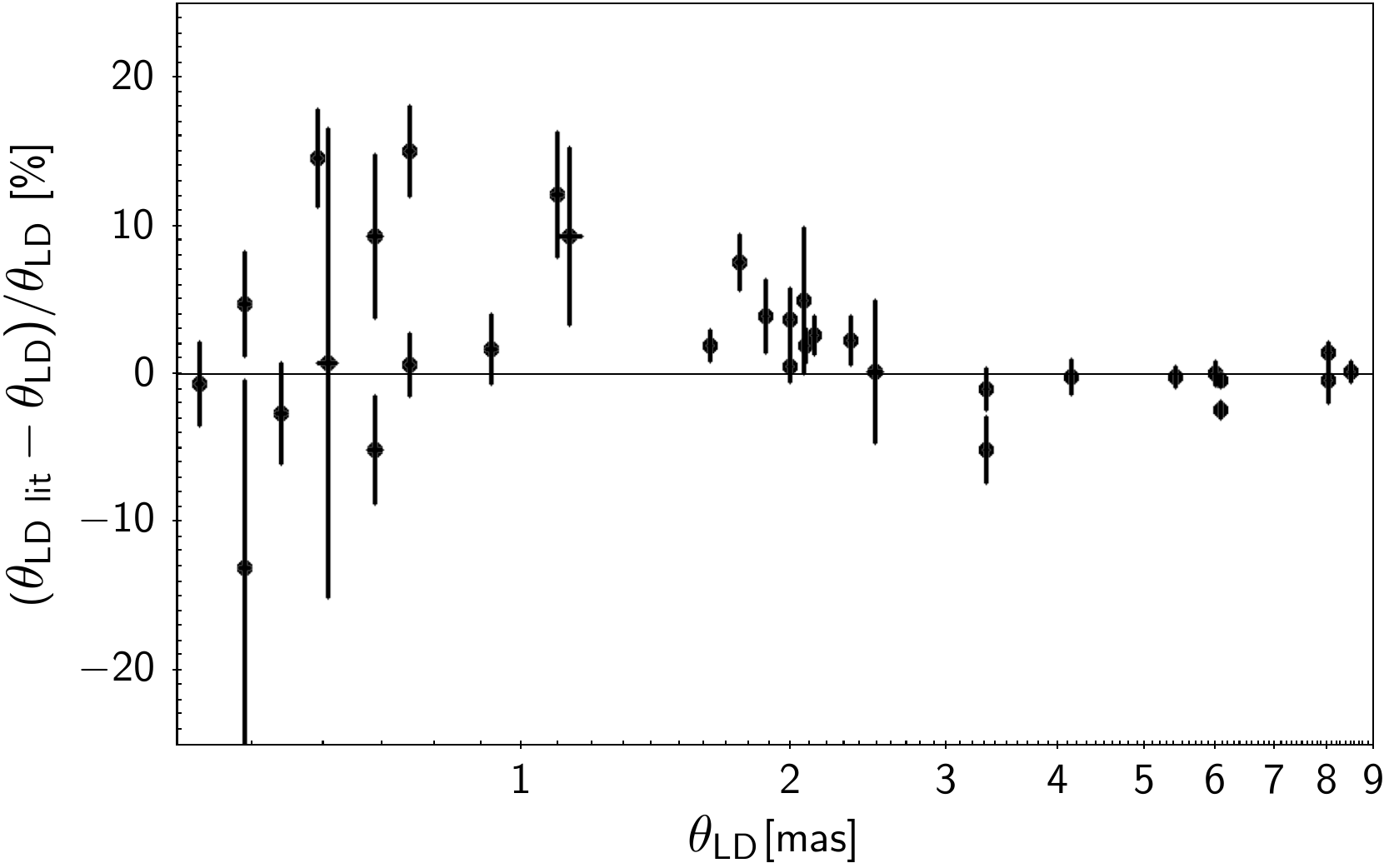} 
 \caption{Difference between \diam\ adopted for this work and other values in JMDC fulfilling the selection criteria by \cite{sal20}.}
 \label{f:comparison-angular-diameters}
\end{figure}

We note that, for a small fraction of stars, we had to make a choice between the two or more values of \diam\ fulfilling the adopted quality criteria. As shown in Fig. \ref{f:comparison-angular-diameters}, several small diameters (typically \diam$<$1.5 mas) disagree by more than 10\%, but in general the agreement is at the 2$\sigma$ level. We note three stars with estimations of their angular diameters differing by more than 3$\sigma$: HIP96441, HIP57939, HIP108870. 

HIP96441 has three values of \diam\ reported in the JMDC, that fulfill the interferometric criteria of \cite{sal20}: 0.861$\pm$0.015 mas in the K band \citep{boy12a}, 0.753$\pm$0.009 mas in the R band \citep{whi13} and 0.749$\pm$0.007 mas in the H band \citep{lig16}. The first determination is not compatible with the two others, but \cite{boy13} mention a calibration problem and discarded this star. Between the two other values we adopted the most recent one by \cite{lig16}. 

HIP57939 (HD103095) is a well-known metal-poor dwarf studied by several authors. We adopted the latest determination, \diam=0.593$\pm$0.004 mas, by \cite{kar20} who used the combination of two instruments, VEGA and PAVO on the CHARA interferometer giving a high confidence to their result. 

For HIP108870 we adopted the value \diam=1.758$\pm$0.012 mas by \cite{rai20} which significantly differs from that previously reported by \cite{ker04}, \diam=1.89$\pm$0.02 mas. \cite{rai20} have analysed this discrepancy, considering that they obtained tighter constraints on the angular diameter by better resolving the star, thanks to the configuration now available at the VLTI. 

These cases of disagreement also illustrate the inhomogeneity of uncertainties listed in JMDC, which sometimes only reflect the precision of a fit, or also include systematic effects identified at the calibration level. The dispersion among measurements available for a given star is critical for small angular diameters, typically below $\sim$1.5 mas, because it corresponds to  discrepancies that can reach 10 to 15\%. This illustrates the limitations of measuring interferometric diameters in the sub-mas regime. Some inhomogeneity can also arise from different recipes applied for the limb darkening correction. According to Eq.~(\ref{e:teff}), a variation of 10\% in \diam\ translates into a variation of 5\% in \teff. Inversely, a 1\% precision on \teff\ implies angular diameters obtained at the 2\% level.

\subsection{Compilation of magnitudes and fluxes}

In order to build a SED for each star and measure the corresponding \Fbol\ we compiled fluxes using the VOSA tool\footnote{\url{http://svo2.cab.inta-csic.es/theory/vosa/}} \citep{vosa}. VOSA allowed us to collect all the photometry available in the Virtual Observatory (VO) for our list of 201 stars (including the nine stars from V1 and V2 with an indirect \diam) and to convert magnitudes into fluxes thanks to an exhaustive description of all the existing filters. 
We only kept the photometry from the VO catalogues that contain at least fifty of our targets, namely 2MASS \citep{2mass}, AKARI \citep{akari}, Gaia DR3 \citep{GDR3}, GALEX \citep{GALEX}, Str\"omgren photometric catalogues \citep{hau98,pau15}, Johnson UBV  \citep{mer87}, IRAS \citep{1984ApJ...278L...1N}, Hipparcos \citep{hip97}, Tycho-2 \citep{hog00} and WISE \citep{WISE}. 

 We note that the components of the bright binary star $\alpha$~Cen A and B are not resolved in the 2MASS catalog\footnote{\url{https://www.ipac.caltech.edu/2mass/releases/allsky/doc/sec4_4a.html}}, and the magnitudes given for $\alpha$~Cen~A contain actually the combined flux of both components. We therefore used J, H, and K magnitudes from \citet{1981A&AS...45....5E}, 
which are given for each component separately, and converted them to flux values using the VOSA tool.

An interesting new feature of the latest VOSA version (July 2022 update) is to provide synthetic photometry based on Gaia DR3 BP/RP spectra analysed with the GaiaXPy tool \citep{dea22,mon22}. We therefore collected through VOSA the synthetic photometry from Gaia which is provided in 13 passbands corresponding to the filters of the Hubble Space Telescope, Sloan Digital Sky Survey, PanSTARRS1 and Johnson UBVRI systems. Also from Gaia BP/RP spectra and GaiaXPy, VOSA computes fluxes in the 65 bands of the OAJ/J-PAS and OAJ/J-PLUS surveys. However we noticed that a small fraction of the Gaia synthetic photometry was affected by saturation, causing the corresponding SED to be deformed. We had to remove the Gaia spectrophometry, totally or partially, for about thirty bright stars with G$\simeq$4. Finally we added to the compilation the fluxes in the range 320--1080~nm from the Pulkovo spectrophotometric catalog \citep{ale96}, adopting a homogeneous uncertainty of 1\% for each value of flux (this value allowed us to give these data an appropriate weighting in our analysis). The Pulkovo catalogue provides 167 or 305 flux values, depending on the star. The details of the number of stars retrieved in each passband or catalogue are provided in Table \ref{t:summary_fluxes}.
The number of flux values per star ranges from only 15 for HIP14135 ($\alpha$ Cet) to 404 for HIP7294 ($\chi$ Cas). The median number of fluxes per star is 101. Fluxes used for the determination of \Fbol\  are available at the CDS.

\begin{table}[htbp]
\centering
\small{
\caption{Passbands (VOSA designation) or catalogues with the corresponding number of stars having a valid value of flux (N). 
 }
\begin{tabular}{lr | lr  }
\hline
Catalogue  &  N  & Catalogue  &  N    \\
\hline\hline
2MASS/2MASS.H           	 & 197  & Generic/Stromgren.b     	 & 125   \\
2MASS/2MASS.J           	 & 198  & Generic/Stromgren.u     	 & 125   \\
2MASS/2MASS.Ks          	 & 197  & Generic/Stromgren.v     	 & 125   \\
AKARI/IRC.L18W          	 & 181  & Generic/Stromgren.y     	 & 125   \\
AKARI/IRC.S9W           	 & 192  & Hipparcos/Hipparcos.Hp  	 & 201   \\
GAIA/GAIA3.G            	 & 192  & IRAS/IRAS.100mu         	 & 86    \\
GAIA/GAIA3.Gbp          	 & 192  & IRAS/IRAS.12mu          	 & 188   \\
GAIA/GAIA3.Grp          	 & 192  & IRAS/IRAS.25mu          	 & 165   \\
GAIA/GAIA3.Grvs              & 151  & IRAS/IRAS.60mu          	 & 116   \\
GALEX/GALEX.FUV         	 & 62   & TYCHO/TYCHO.B           	 & 198   \\
GALEX/GALEX.NUV         	 & 63   & TYCHO/TYCHO.V           	 & 198   \\
GCPD/Stromgren.b        	 & 140  & WISE/WISE.W1            	 & 53    \\
GCPD/Stromgren.u        	 & 140  & WISE/WISE.W2            	 & 59    \\
GCPD/Stromgren.v        	 & 140  & WISE/WISE.W3            	 & 115   \\
GCPD/Stromgren.y        	 & 140  & WISE/WISE.W4            	 & 116   \\
Generic/Johnson.B       	 & 198  & Gaia DR3 J-PAS Synt.Phot.   & 116  \\
Generic/Johnson.U       	 & 196  & Synt.Phot. from Gaia DR3    & 117  \\
Generic/Johnson.V       	 & 199  & Pulkovo                     & 52   \\
\hline
\end{tabular}
\label{t:summary_fluxes}
}
\end{table}

An illustration of the obtained SEDs is given for two stars in Fig. \ref{f:fbol_fit} in Sect. \ref{s:fbol}.  HIP103598 is a K4 giant with a metallicity of $-0.36$ according to PASTEL, which has a well constrained SED thanks to Gaia and Pulkovo spectrophotometry. HIP50564 is a F6 turn-off star with a metallicity of +0.10 according to PASTEL, having only broad-band photometric observations. We chose these stars to illustrate both the SED shape variations due to different temperatures, and the more or less good coverage of the SED depending on the availability of spectrophotometric data.

\subsection{Extinction}\label{ssec:extinction}
The extinction towards each of the 201 targets was estimated thanks to the recent 3D maps provided by \cite{ver22}, based on the inversion of large spectroscopic and photometric catalogues including Gaia DR3. We chose the closest map, covering a volume of 3 kpc x 3 kpc x 800 pc at a resolution of 10~pc, which is particularly well adapted for our sample of nearby stars. 

\begin{figure}[h]
\centering
 \includegraphics[width=0.48\textwidth]{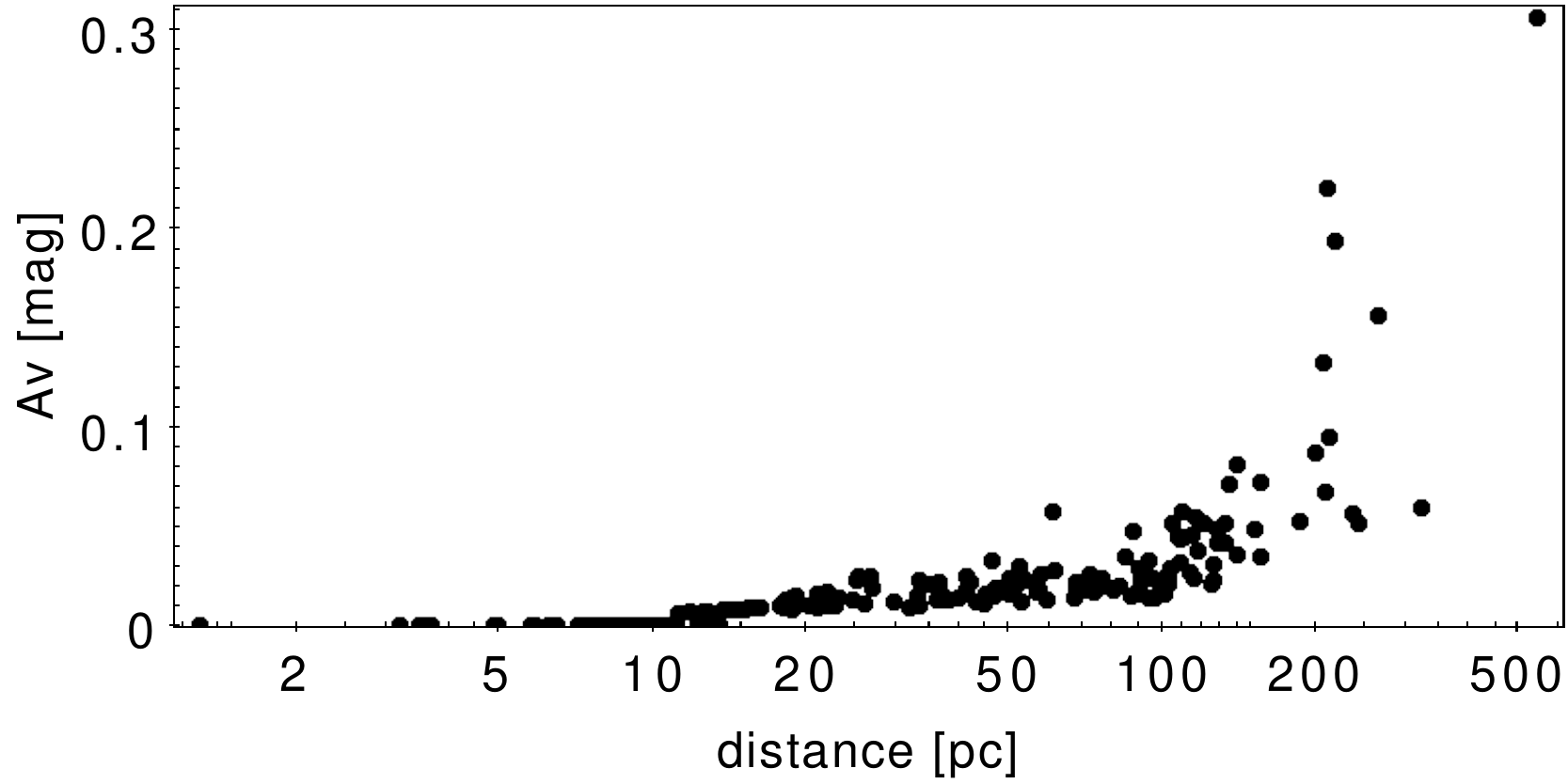} 
 \caption{Extinction $A_{\rm V}$ deduced from 3D maps of \cite{ver22} as a function of distance, for the 201 targets.}
 \label{f:Av}
\end{figure}

The extinction is low for most of the stars (90\% of them have $A_{\rm V}<0.05$) which is not surprising owing to the small distances of the GBS V3 from the Sun. Our GBS span distances from 3 pc to 550 pc (deduced from parallaxes, see Sect. \ref{s:par}).   Five giants have the highest extinction values, between 0.1 and 0.31 mag. As expected, $A_{\rm V}$  is well correlated to the distance, as shown in Fig. \ref{f:Av}.

\subsection{SED fitting and bolometric fluxes}
\label{s:fbol}

\begin{figure*}
    \centering
    \includegraphics[width=0.48\textwidth]{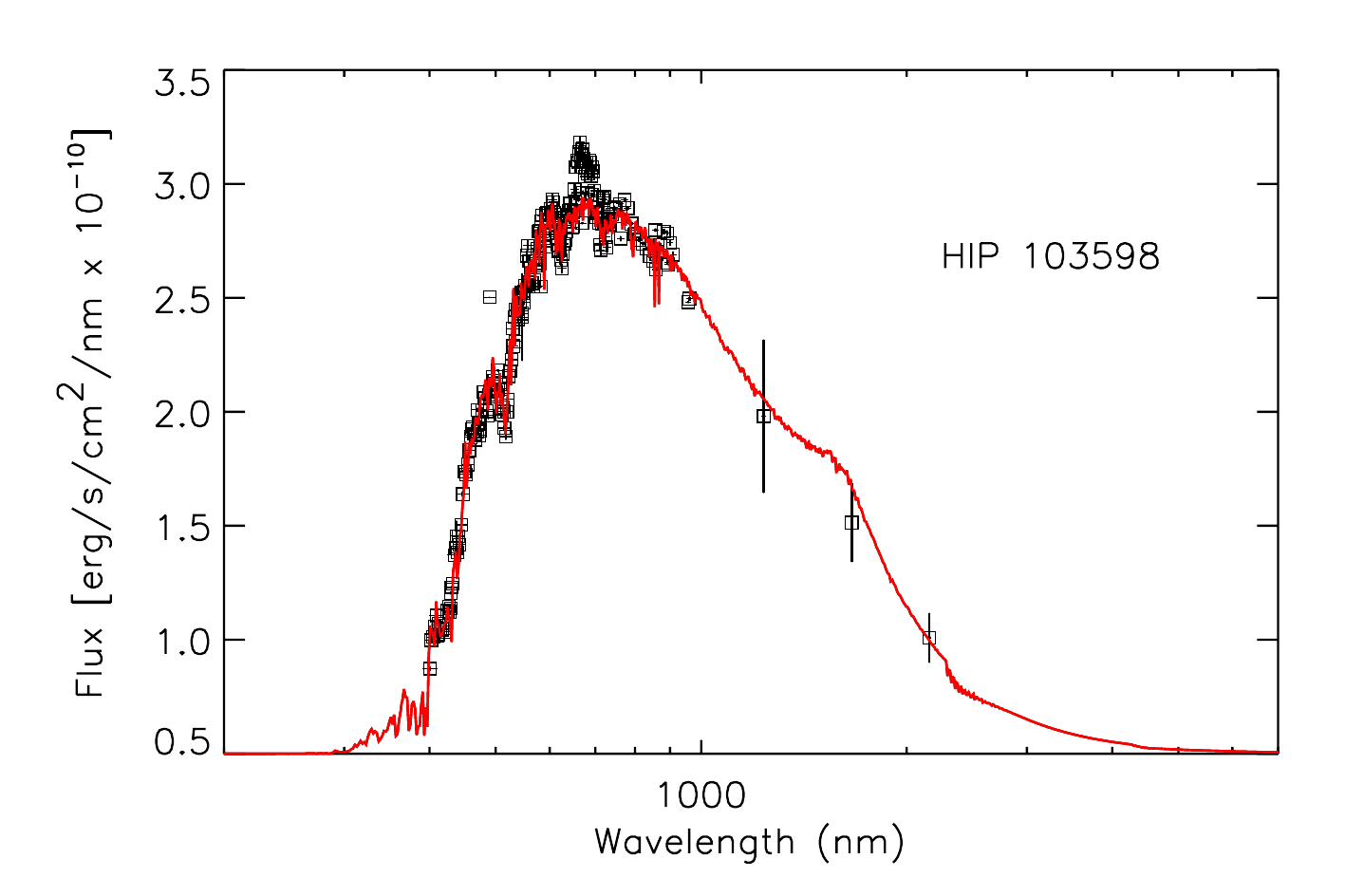}
    \includegraphics[width=0.48\textwidth]{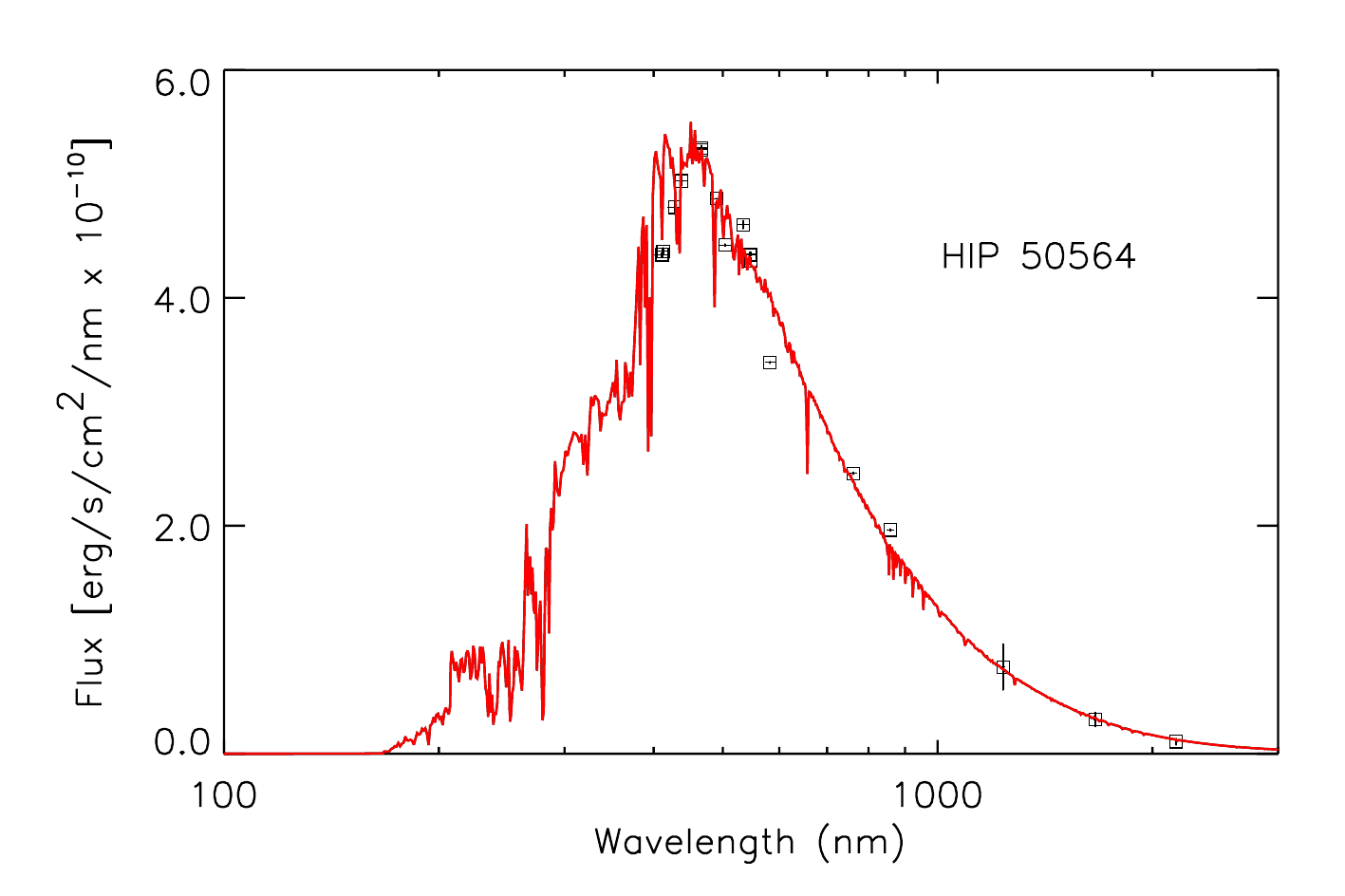}
    \includegraphics[width=0.48\textwidth]{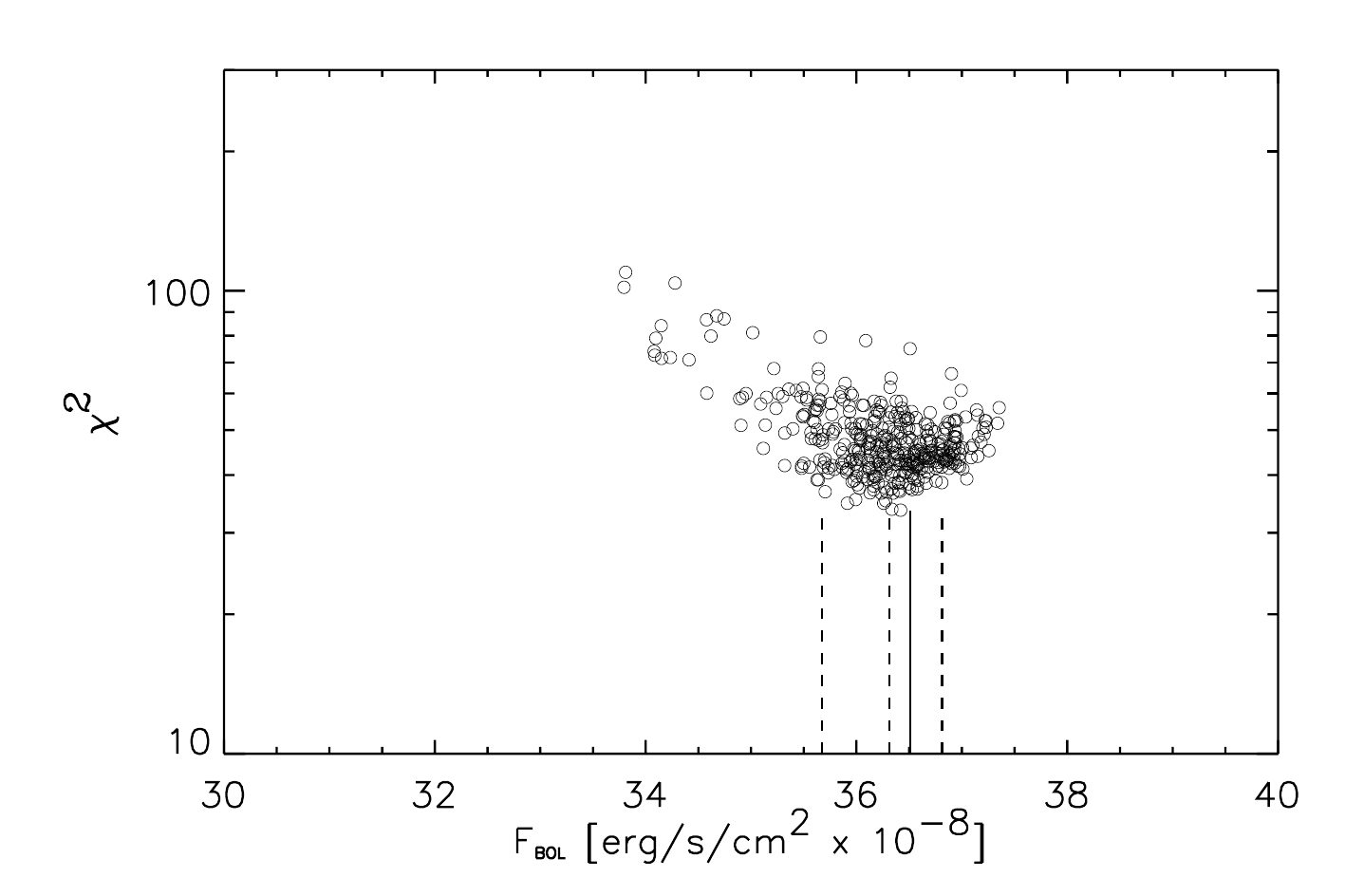}  
    \includegraphics[width=0.48\textwidth]{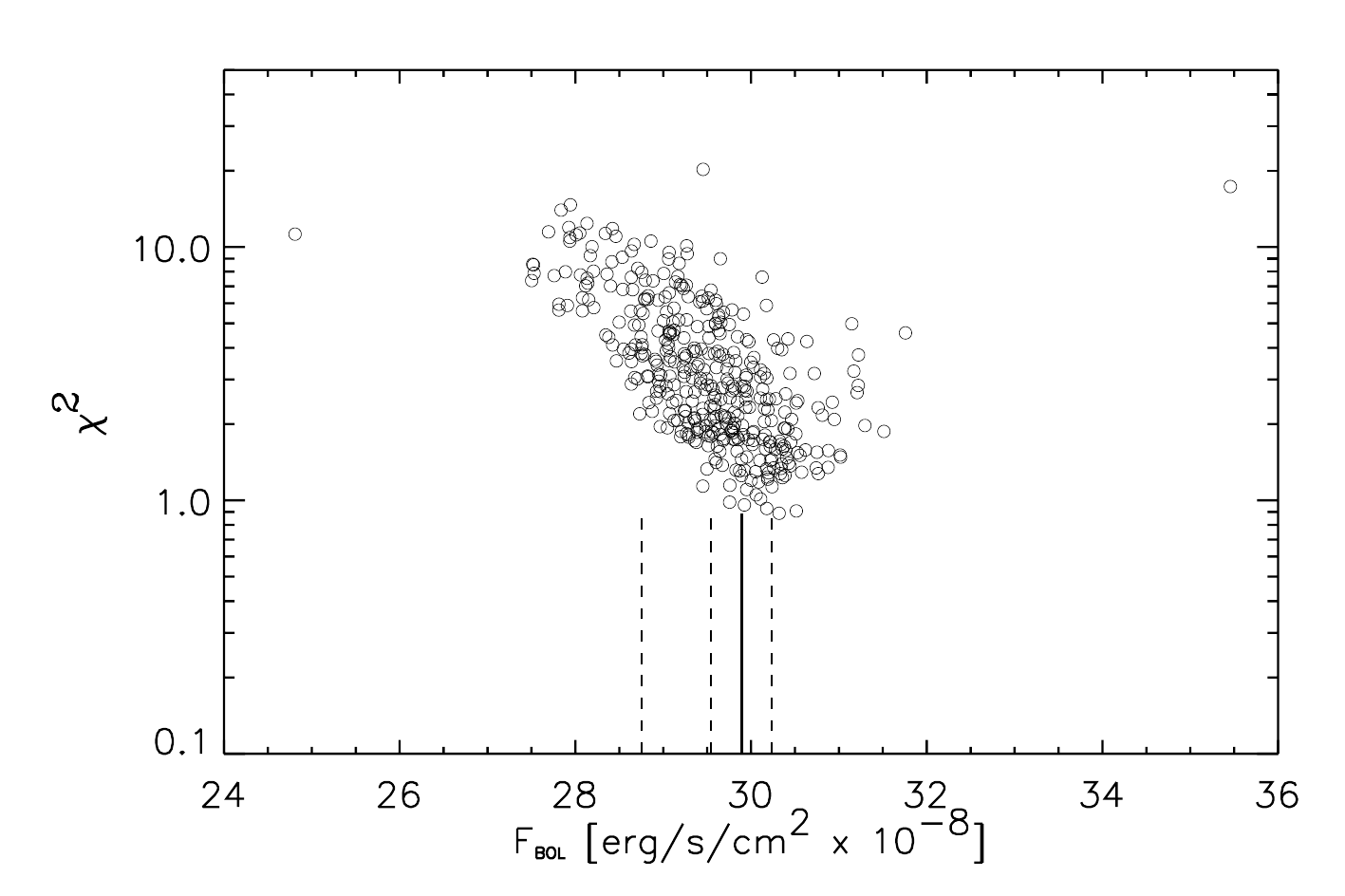}  
    \caption{
    Example of fits of the observed (reddened) data to the (reddened) semi-empirical spectra for \object{HIP\,103598} (left) and  \object{HIP\,50564} (right).  The bolometric flux is calculated by integrating the un-reddened spectrum. The bottom panels illustrate the distribution of $\chi^2_R$ versus \Fbol\ for the  400 simulations for the two stars with the 16$^{\rm th}$, 50$^{\rm th}$ (median), and 84$^{\rm th}$ percentiles indicated by the dashed lines.  
     }
    \label{f:fbol_fit}
\end{figure*}

Although the observed fluxes compiled for the GBS cover a wide range of wavelength, some extrapolation of the SED is needed to integrate the full distribution and measure the total flux from the star received at the Earth, \Fbol. To do so we followed the SED fitting method previously used by \cite{cre15} and \cite{lig16}, based on the BASEL empirical library of spectra \citep{lej97},   a highly cited library in the astrophysical community.
 Our choice for these models is based upon the work in \cite{cre15} where one star was analysed in detail using different approaches and models.   A 1\% flux difference was found using the BASEL and PHOENIX libraries, with the former being in best agreement with other literature results using different methodologies.

The BASEL library covers the following parameter ranges: 
$3\,500 < \teff < 50\,000$ K, $ 0.00 < \logg < 5.00$, and --5.0 $<$ [M/H] $<$ +1.0.  It extends to 2\,000 K for a subset of the \logg\ and [M/H].  The wavelength range spans 9.1 to 160\,000 nm on a non-evenly sampled grid of 1221 points, with a mean
resolution of 100 nm in the UV and 200 nm in the visible.  
Beyond 10\,000 nm the resolution is 20\,000 nm and to avoid issues with numerical integration we interpolate on a log scale before performing the integration.  
A Levenberg-Marquardt minimisation algorithm finds the optimal template that fits the observed flux points. \Fbol\ is then calculated by integrating the optimal fitted spectrum. 
Recent improvements of the method include the weighting of the fluxes 
and the determination of \Fbol\ uncertainties through Monte-Carlo simulations. 

The parameters of the model are the atmospheric parameters: \teff, \logg, \feh, the extinction, and the scaling factor  (stellar radius scaled according to the distance).  We used the atmospheric parameters from PASTEL to initialise the minimization and the extinction from Sect.~\ref{ssec:extinction}.   
 To account for extinction in our method we implemented the {\sc IDL} routine {\tt ccm\_unred}\footnote{This routine is distributed as part of the IDL Astronomy User's Library at \url{https://github.com/wlandsman/IDLAstro}.} which dereddens theoretical fluxes, and  requires colour-excess on input.  To convert extinction to colour-excess we adopted $R_0 = 3.1$.  Most of these stars are nearby and as such have little or no extinction.  In order to be complete in our analysis, in the catalogue available at the CDS we also provide \Fbol\ for the full sample of stars by assuming zero extinction.

All of the above parameters can be fitted, but in practice due to degeneracies between the parameters, the \teff\ and the scaling factor are the only free parameters, while \logg, \feh, and the extinction are fixed each time a minimization is performed.       
In order to include the impact of the uncertainties of the parameters \logg\ and \feh, and of the fluxes, we performed a bootstrapped-based method where we (a) perturbed these parameters by their uncertainty multiplied by a random number drawn from a Gaussian distribution, and (b) perturbed the fluxes by their uncertainties using the same approach.
These simulations were done 400 times where  400 was a balance between computing time and having a significant sample size (the results with 200 simulations were equivalent within the uncertainties and the standard deviation of the  400 simulations reproduced the uncertainties of the atmospheric parameters).
For the fitted parameters, the result is a distribution of stellar parameters that fit the observational data, and from these fitted parameters we integrated the corresponding semi-empirical flux distribution. We therefore obtained a distribution of \Fbol\ for each star, and from these distributions we calculated the medians and the symmetric 68\% confidence intervals, and report half of the latter as the uncertainty.

Two examples of the data and the best-fitted model SED 
are shown in Fig.~\ref{f:fbol_fit}, left and right panels.  The left is an example of a star with many observational points (in this case HIP103598), while the right panel shows an example where relatively few data points are available; in this case HIP50564.
The lower panels show the distribution of the fitted \Fbol\ and the individual $\chi^2$ values from the  400 Monte Carlo simulations, along with the value of the adopted median and 16 and 84 percentile confidence levels (dashed lines).  We defined the uncertainty as the half of the distance between the upper and lower confidence levels.

\begin{figure}[h!]
\centering
 \includegraphics[width=0.48\textwidth]{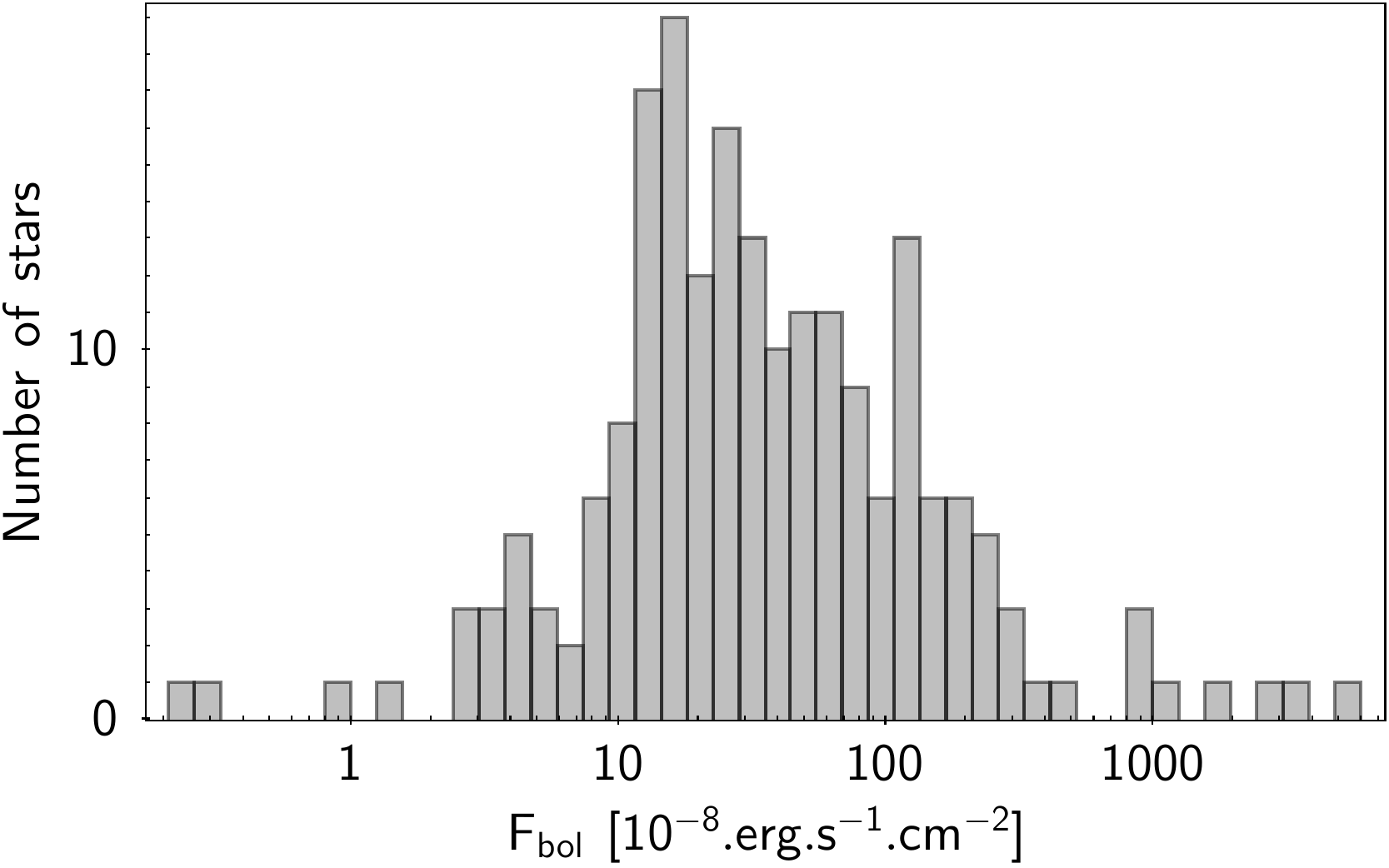} 
 \includegraphics[width=0.48\textwidth]{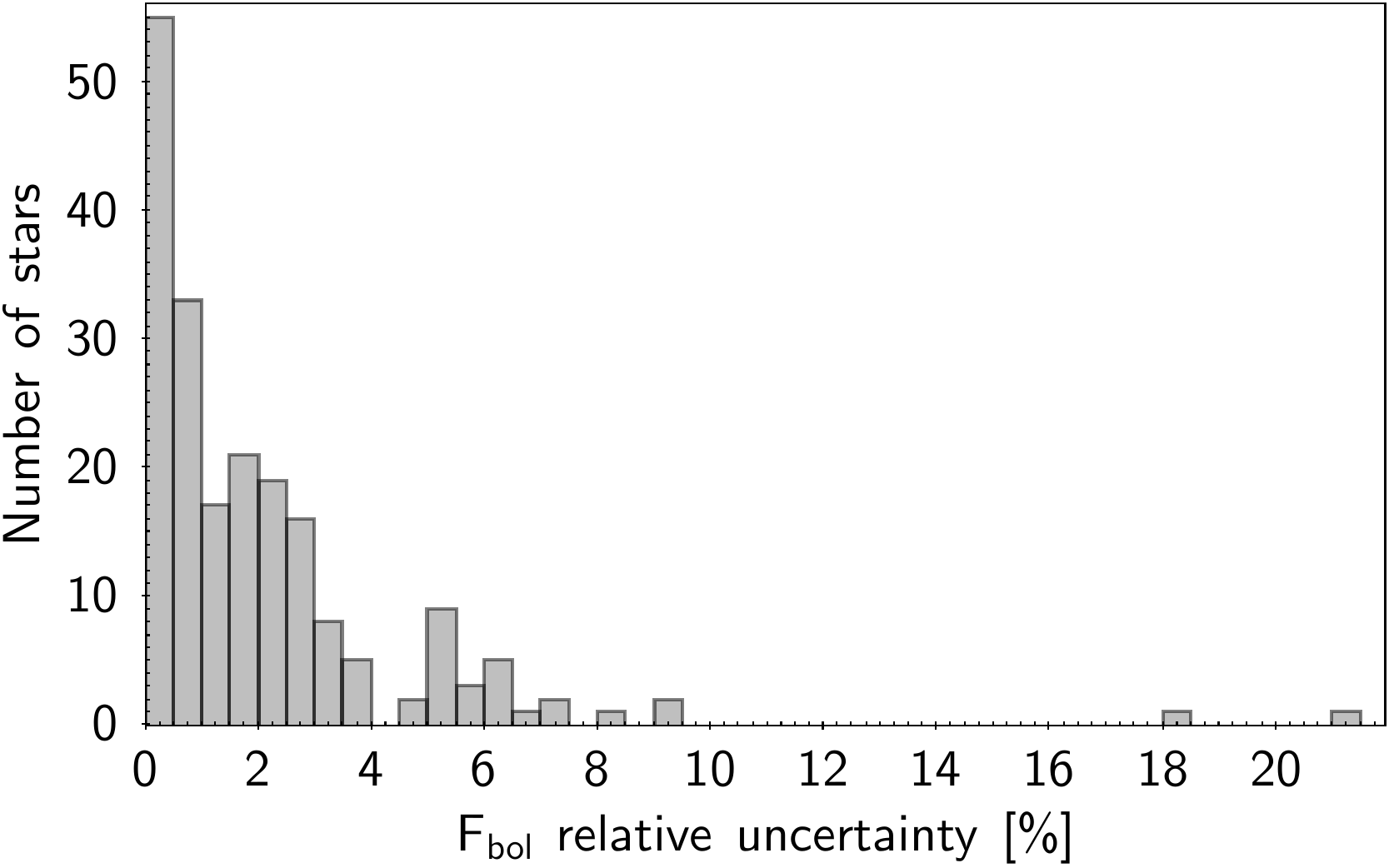} 
 \caption{Histogram of \Fbol\ (top panel) and its relative uncertainties (bottom panel).}
 \label{f:fbol_histo}
\end{figure}

The distribution of \Fbol\ and relative uncertainties is shown in Fig. \ref{f:fbol_histo}. The histogram of uncertainties shows a clear peak in the first bin corresponding to uncertainties lower than 0.5\%.  The relative uncertainties have a median value of   1.4\%, and they are lower than 10\% except for two stars, namely HIP8837 ($\psi$ Phe) and HIP14135 ($\alpha$ Cet). These two M giants combine a low \teff\ and a lack of spectrophotometric data which make their SED poorly constrained, resulting in a relative uncertainty of about  21\% and  18\% respectively. They had uncertain \Fbol\ in Paper~I as well. We note that the stars with uncertainties larger than  4\% have their SED made of broad-band photometry only, while the majority of stars have spectrophotometry from Gaia and/or Pulkovo, resulting in a very precise \Fbol\ determination.

 In this procedure to determine \Fbol, we use \logg\ and metallicity from the PASTEL catalog, a compilation of literature work. We have evaluated the impact of not knowing precisely these parameters. To do so, we made two tests. One test is to adopt a large uncertainty of 0.15 on both \logg\ and \feh\, inducing a different distribution of the fitted \Fbol\ from the 400 Monte Carlo simulations. The other test is to change the literature values of \logg\ and \feh\ by an amount of 0.15 dex, in the eight possible configurations, for seven stars selected to cover the parameter space. In this test, the largest effect ($<$1\%) is reached when adding 0.15 dex to \feh\ for the hottest stars. Varying \logg\ has more effect on the coolest stars.
When combining the variations of \logg\ and \feh\ the effect remains at the level of 1\% for the coolest and the hottest stars. Interestingly the most metal-poor star chosen for that test, HIP48152, is less affected by a change of \logg\ and \feh. In the other test, enlarging the \logg\ and \feh\ uncertainties in the Monte-Carlo simulations also has a low impact on the derived value of \Fbol. We note four stars with \Fbol\ changed by 1-2\%, while 90\% of the sample changes by less than 0.5\%. We conclude that our procedure weakly depends on the input values of \logg\ and \feh. A change of 1\% in \Fbol\ induces a change of 0.2\% on \teff. However a more rigorous treatment will be performed through iterations once the spectroscopic analysis of the targets will be performed to derive \feh\ homogeneously (Paper VIII in preparation). This will lead to self-consistent parameters.

In Paper~I, the \Fbol\ values of the V1 stars were compiled from the literature and therefore not as homogeneous as here. This is another important improvement of the GBS V3, in addition to the larger number of stars.  We still have a good agreement between V1 and V3, with a slight offset of   1.4\%, and a typical dispersion of 2.1\% (median absolute deviation, MAD).

\begin{figure*}[h!]
\centering
 \includegraphics[width=0.48\textwidth,clip=true,trim= 0cm 0cm -1cm 0cm]{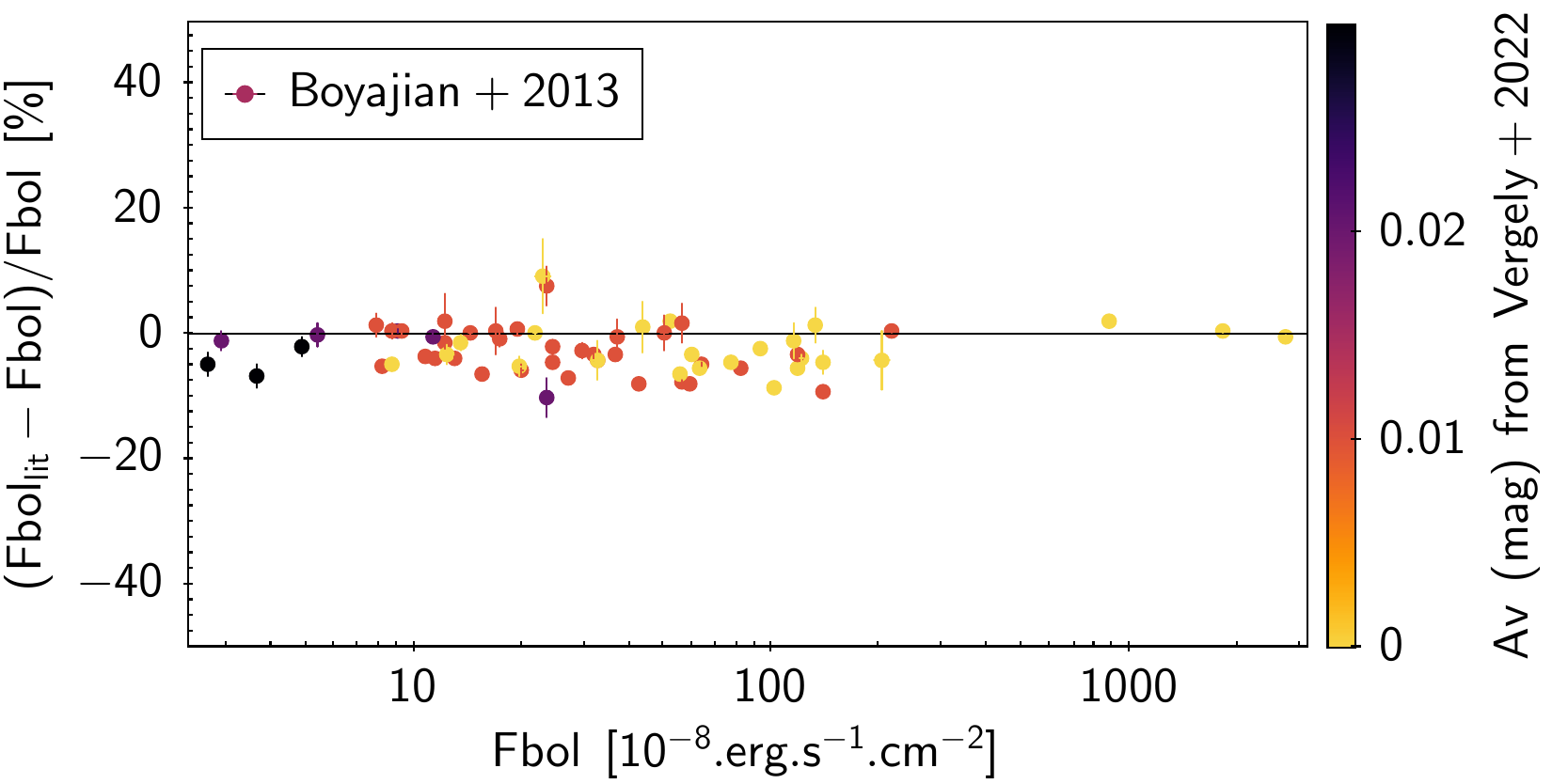} 
 \includegraphics[width=0.48\textwidth,clip=true,trim= 0cm 0cm -1cm 0cm]{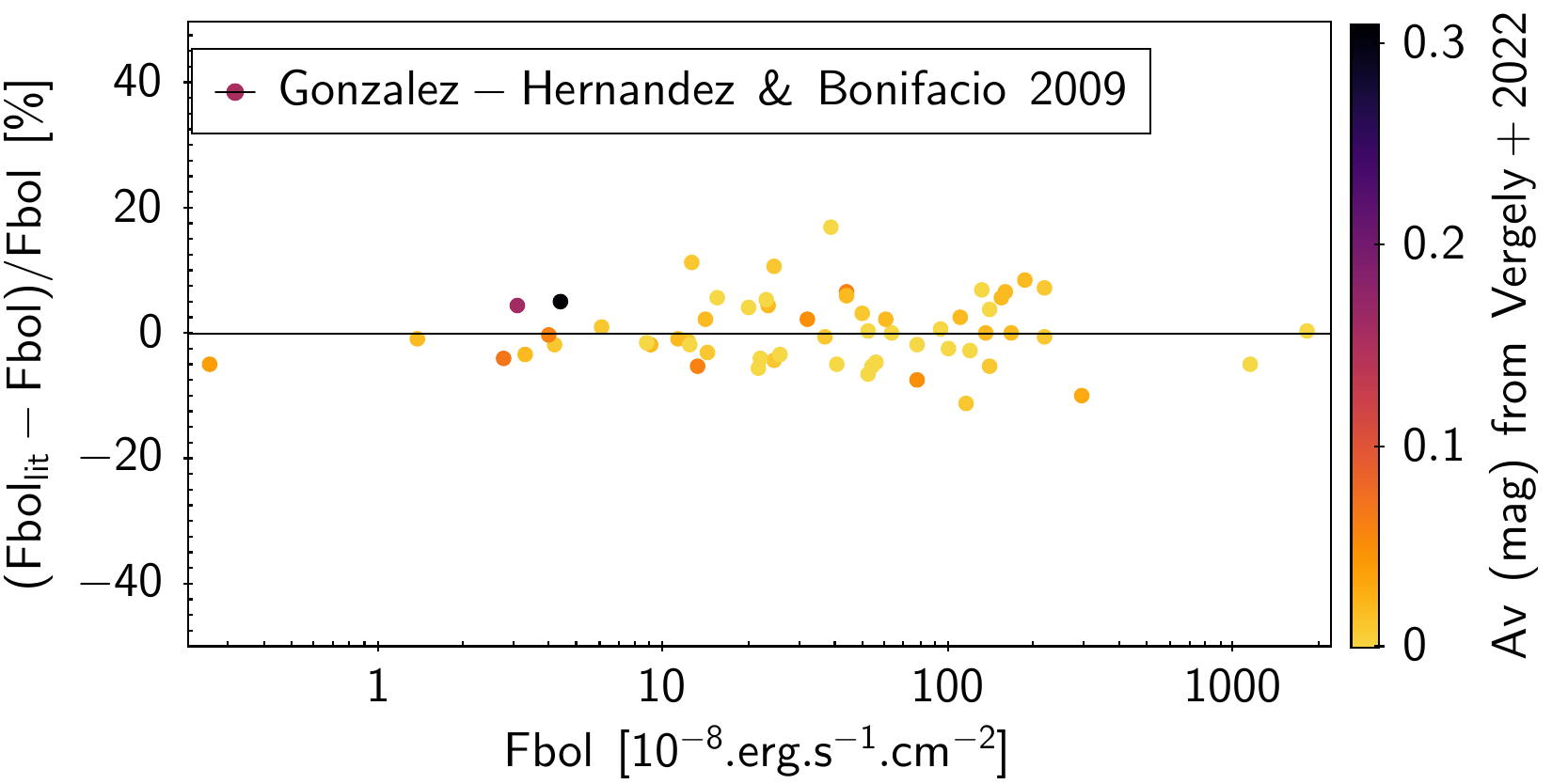} 
 \includegraphics[width=0.48\textwidth,clip=true,trim= 0cm 0cm -1cm 0cm]{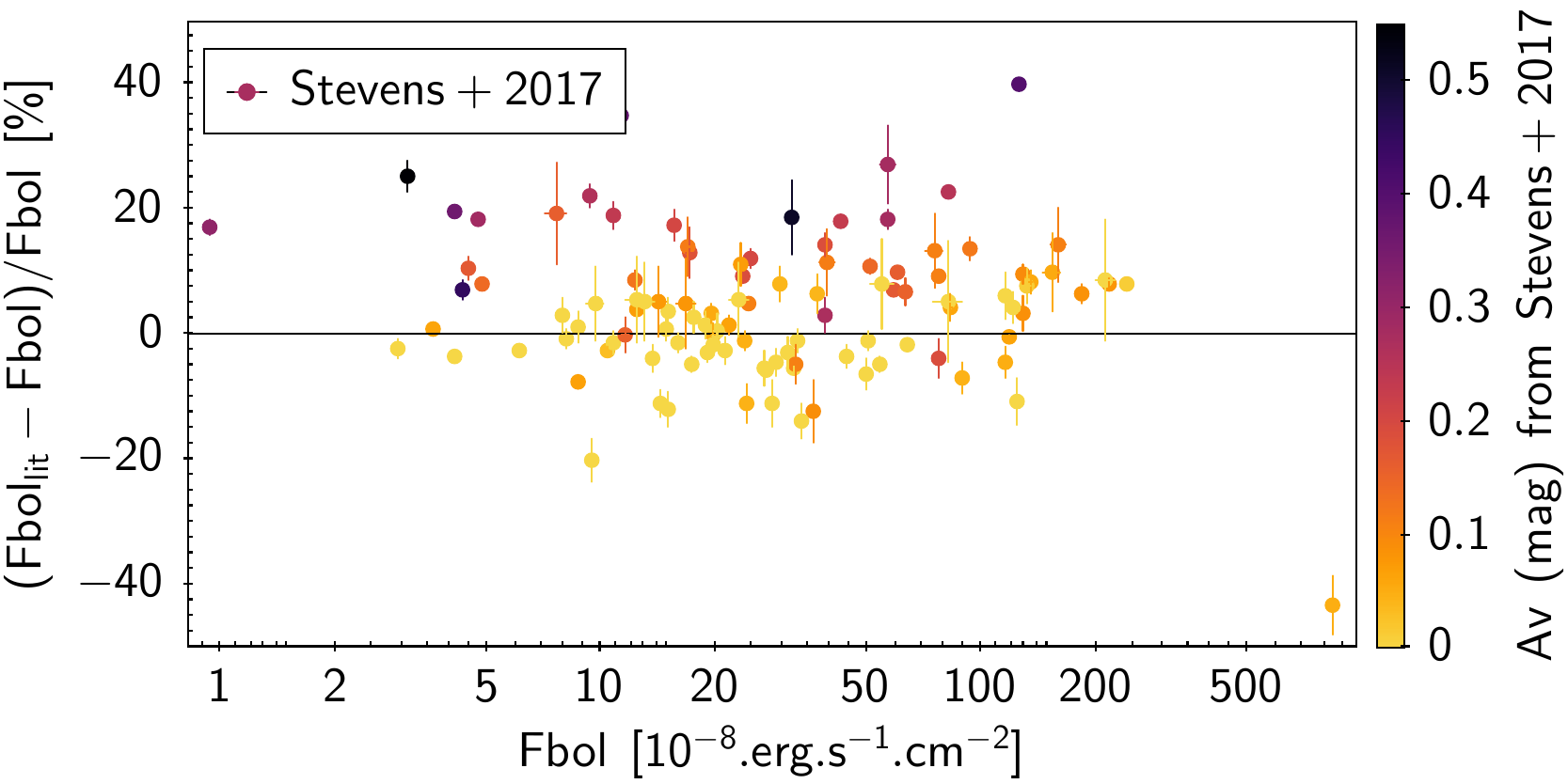} 
\includegraphics[width=0.48\textwidth,clip=true,trim= 0cm 0cm -1cm 0cm]{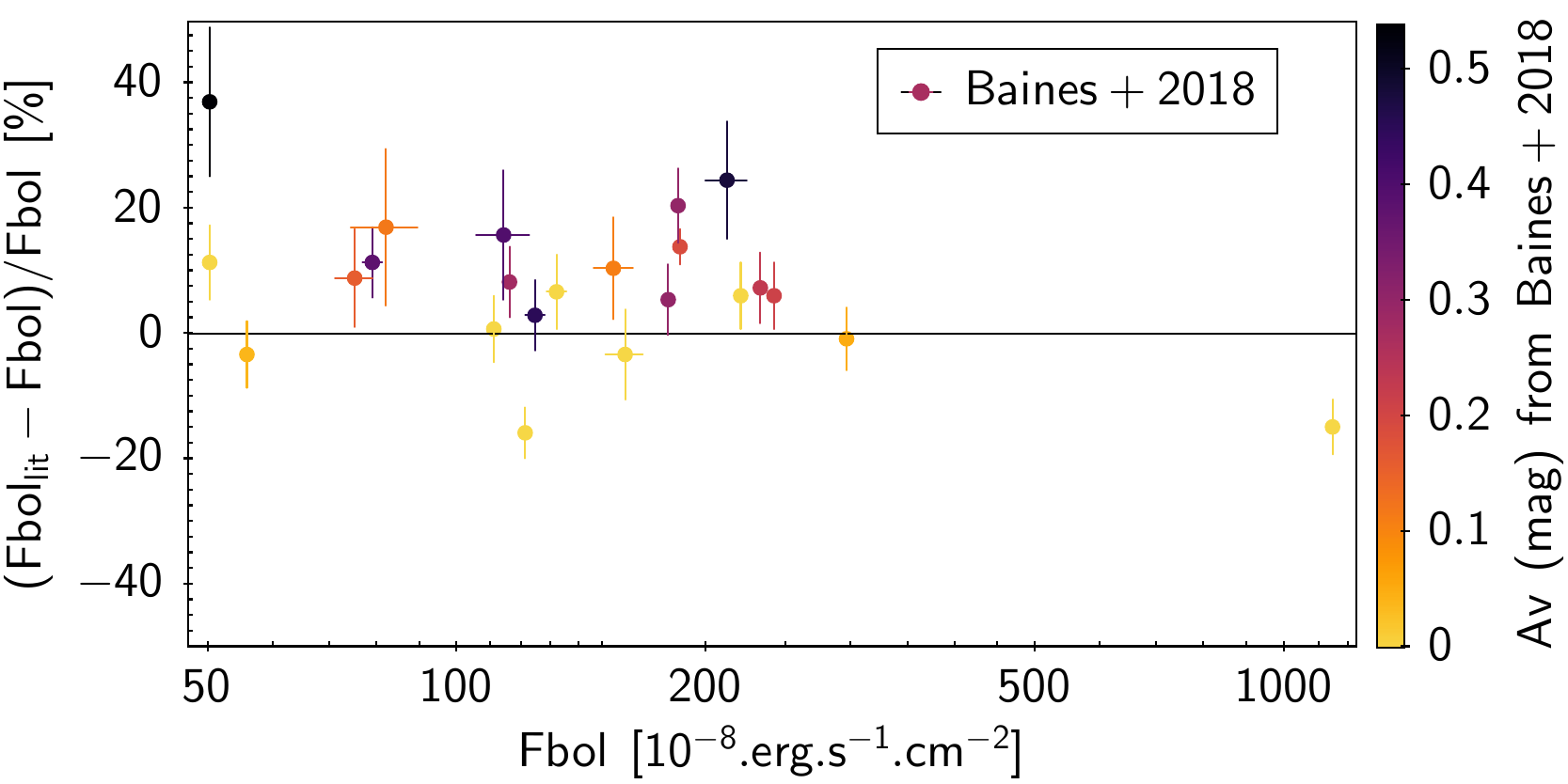}  
\caption{Comparison of \Fbol\ obtained in this work with literature. The colour code relates to the extinction. Several extreme outliers are out of the figure boundaries but they are discussed in the text.}
 \label{f:comp_fbol}
\end{figure*}

Our approach is similar to that of \cite{boy13} and \cite{bai18} who collected broadband photometric measurements available in the literature, extended by spectrophotometry when available. They also applied the SED fitting method with reference templates taken from the library of  \cite{pic98} which is made of observed spectra, whereas we used a hybrid library of synthetic stellar spectra calibrated from observations \citep{lej97}. Another difference comes from the Gaia spectrophotometry recently made available, which constrains very well the SED shape in the optical range. We have 66 stars in common with \cite{boy13} and 24 with \cite{bai18}. The \Fbol\ comparison is shown in Fig. \ref{f:comp_fbol}. The agreement with \cite{boy13} is very good, with differences within 10\%. On average our \Fbol\ values are higher than their values by  3.3\%, with a typical dispersion of 2.5\% (MAD). The offset does not seem correlated with extinction which is lower than 0.03 mag for the stars in common according to our estimations, and that they have not considered given the close distance of the stars.  It is likely that the small offset observed between our \Fbol\ determinations and those of \cite{boy13} is related to their use of magnitudes from photometric catalogues, with a maximum of 17 values per star and less than 12 values in most cases, while we have typically ten times more flux values, mostly from Gaia spectrophotometry, providing SEDs of better quality. In addition they did not take photometric uncertainties into account for the fit, while we do. \cite{bai18} determine a high extinction for some stars which seems correlated with a larger positive offset. HIP47431 and HIP90344 are the most extreme cases with $A_{\rm V}$=0.7 mag and $A_{\rm V}$=0.54 mag respectively in \cite{bai18} while we get $A_{\rm V}$=0.02 mag and $A_{\rm V}$=0.03 mag from the 3D maps of \cite{ver22}, leading to a difference of 50\% and 37\% on \Fbol\ (HIP47431 is not shown in Fig. \ref{f:comp_fbol}). Considering the 24 stars in common,  \cite{bai18}  find \Fbol\ higher than us by 7.6\% (median) with a dispersion of 6.6\% (MAD).

We also compared our \Fbol\ determinations to those of \cite{gon09} who implemented the infrared flux method (IRFM) based on 2MASS magnitudes (see Fig. \ref{f:comp_fbol}). The 61 stars in common generally agree well with an offset less than 1\% and a dispersion of 3.9\% (MAD). The extinction is low for the majority of these nearby stars.

Finally, we also made a comparison with the catalog of empirical bolometric fluxes and angular diameters of 1.6 million Tycho-2 stars built by \cite{ste17} which has 119 stars in common with us. This work is based on the flux-colour relations of \cite{cas10} with \teff\ and $A_{\rm V}$ being determined separately in an iterative way. Their \Fbol\ are globally larger than ours by 4.9\%, with a dispersion of  7.5\% (MAD).  Similarly to the tendency observed in the comparison with \cite{bai18}, the larger differences correspond to stars with the largest values of $A_{\rm V}$ in \cite{ste17} which significantly differ from our lower extinctions. Eight stars do not appear in Fig. \ref{f:comp_fbol}, given their difference larger than 50\%, up to 360\% for HIP112731 and HIP96837. They are found highly reddened by \cite{ste17} with $A_{\rm V}$$\geq$0.6 mag, up to more than 2 mag for the two most extreme stars HIP112731 and HIP96837. We therefore suspect that some extinctions are overestimated by \cite{ste17} and \cite{bai18}, leading to overestimated bolometric fluxes.

\subsection{Assessment of \teff}
\begin{figure}[h]
\centering
 \includegraphics[width=0.48\textwidth]{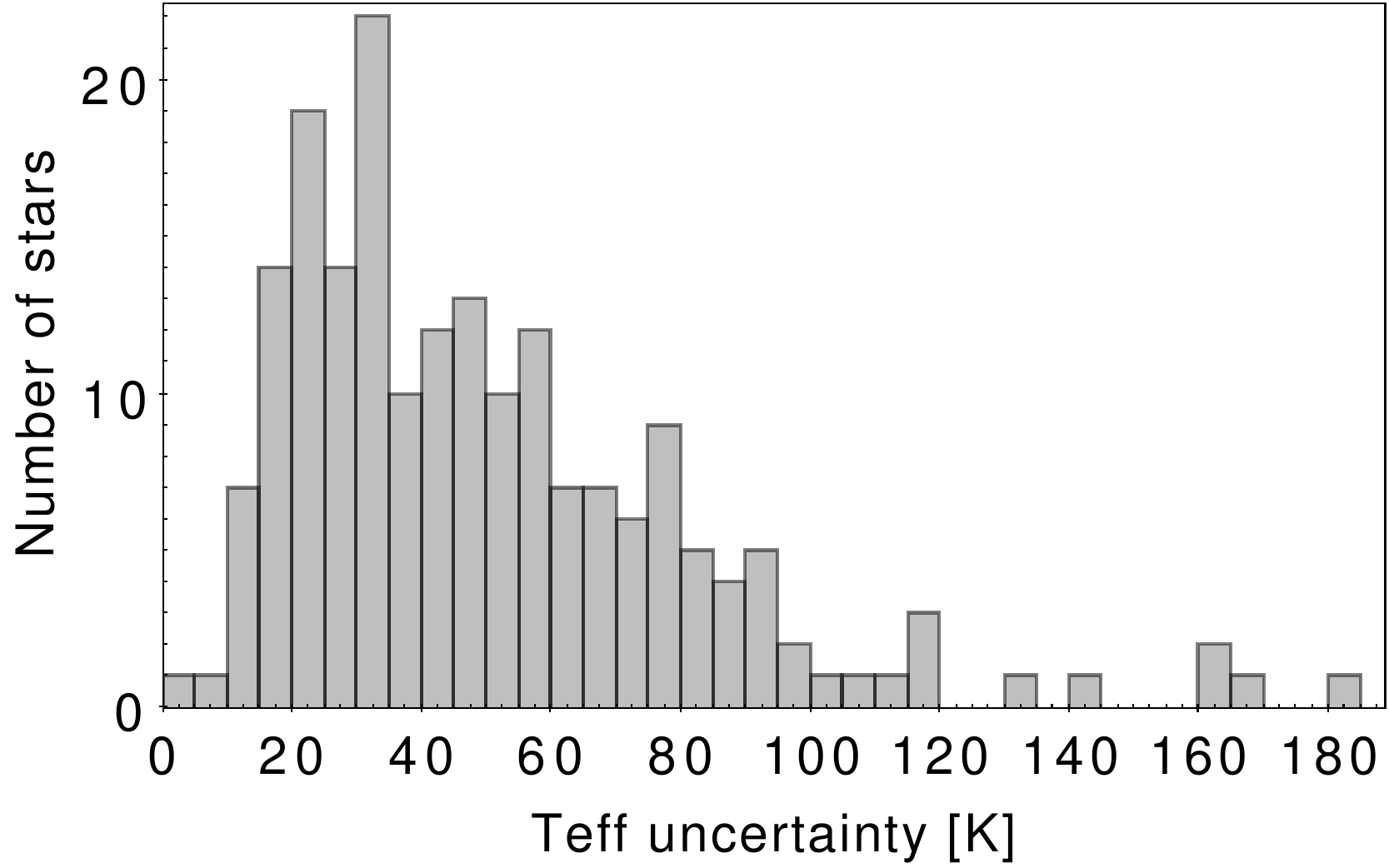} 
  \includegraphics[width=0.48\textwidth]{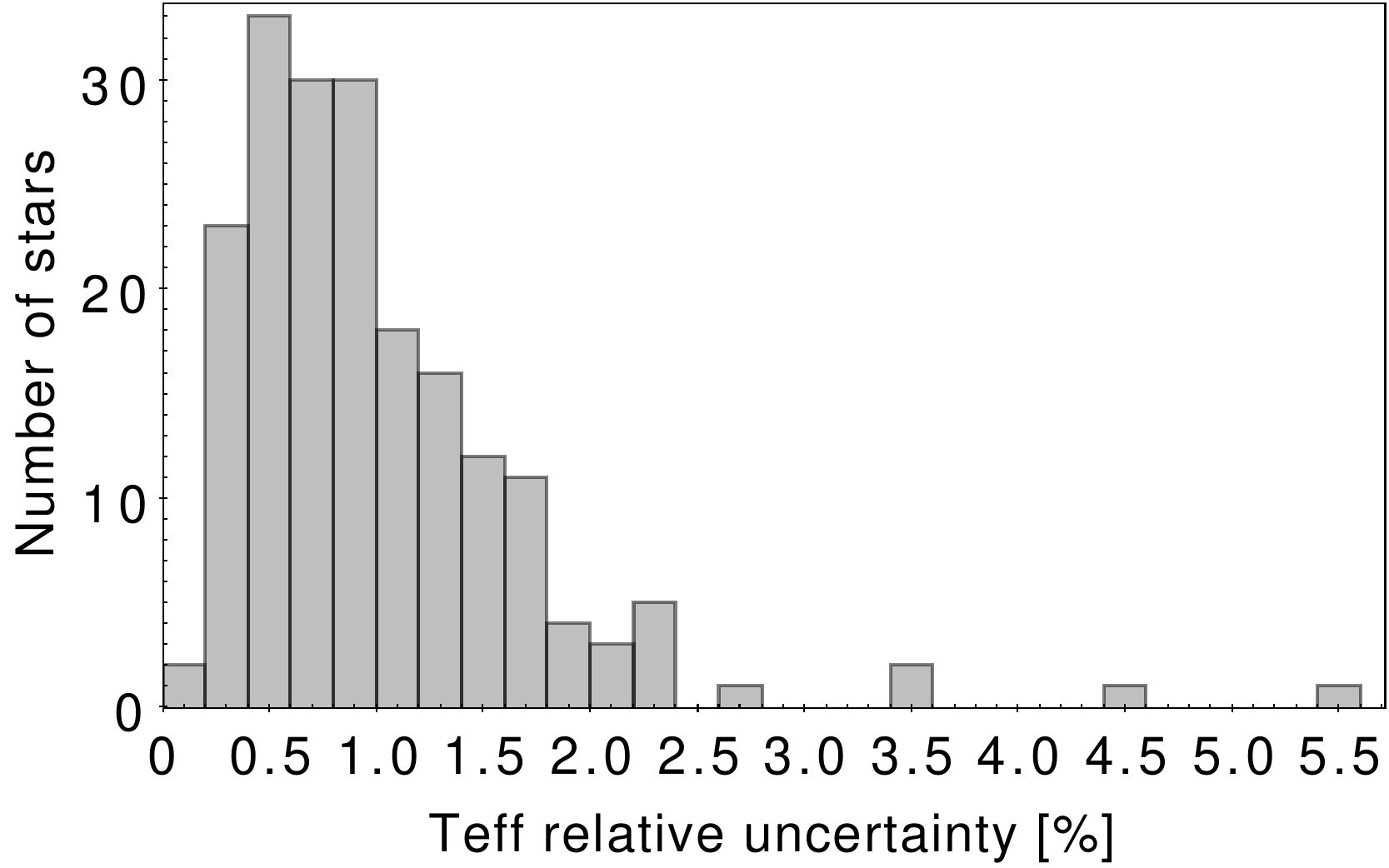} 
 \caption{Histogram of \teff\ absolute (top panel) and relative (bottom panel) uncertainty. The four outliers with \teff\ uncertainty larger than 150 K (or 3\%) are discussed in the text.}
 \label{f:e_teff_histo}
\end{figure}

We computed the fundamental \teff\ of each star by applying Eq.~(\ref{e:teff}) with the values of \diam\ and \Fbol\ obtained as described above. \teff\ uncertainties were deduced by propagating the \diam\ and \Fbol\ uncertainties in Eq.~(\ref{e:teff}). We consider here the 192 stars with a direct value of \diam. The resulting uncertainties on \teff\ span 5 K to 183 K, with a median value of 43 K (see histogram in Fig. \ref{f:e_teff_histo}). Only four giants present a relative uncertainty larger than 3\% (absolute uncertainty larger than 150~K): the two M giants HIP8837 ($\psi$ Phe) and HIP14135 ($\alpha$ Cet) previously mentioned for their large \Fbol\ uncertainty resulting in \teff\ uncertainties of $\sim$5\%, and the K giants HIP25993 and HIP14838 previously mentioned for their large  uncertainty on \diam\ resulting in \teff\ uncertainties of $\sim$3.5\%. These four stars clearly stand as outliers in

\begin{figure}[h]
\centering
 \includegraphics[width=0.48\textwidth]{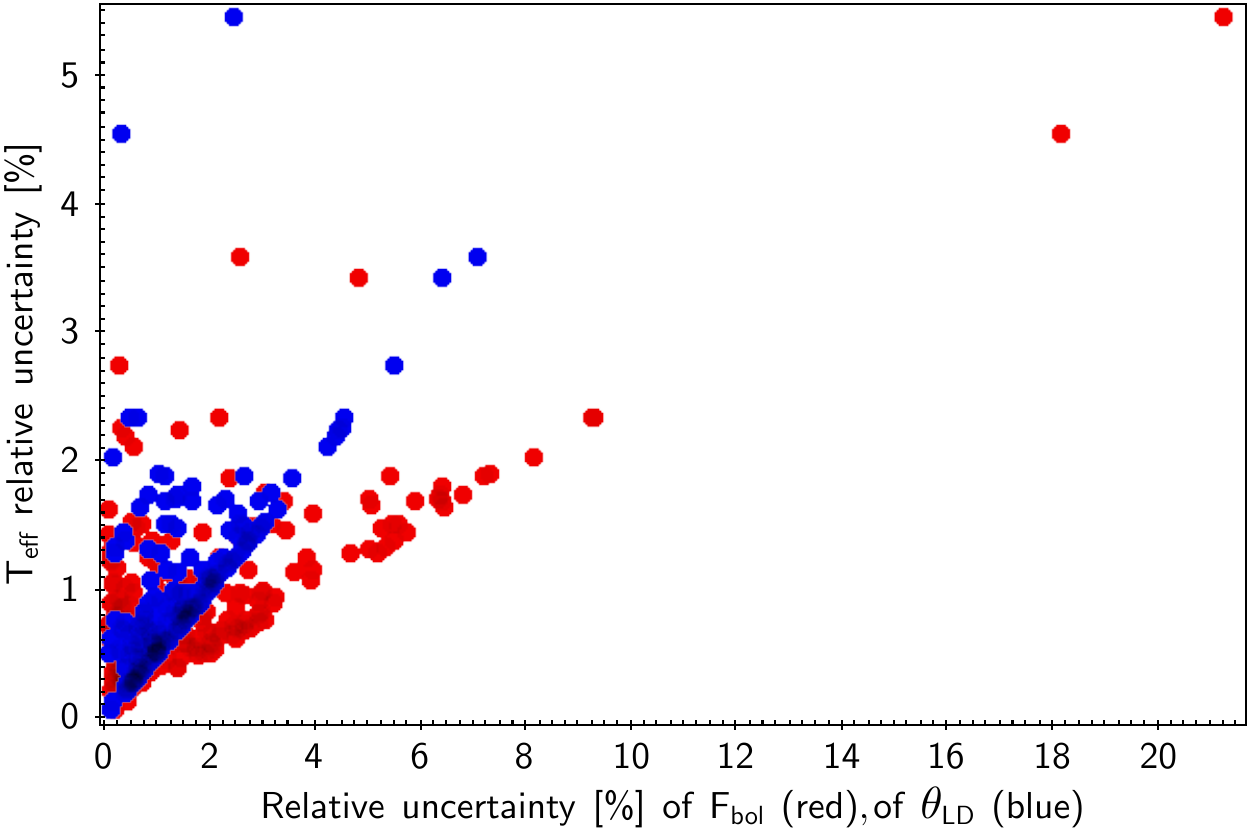} 
 \caption{Propagation of \diam\ and \Fbol\ relative uncertainties on \teff.}
 \label{f:uncertainties_teff}
\end{figure}

Fig. \ref{f:uncertainties_teff} which  shows how the relative uncertainties on \diam\ and \Fbol\ propagate on \teff. In order to reach a 1\% accuracy on \teff\ one should restrict the sample to stars with measurements better than $\sim$2\% in \diam\ and $\sim$4\% in \Fbol. We have 127 stars fulfilling this condition, while 179 of the 192 stars have \teff\ uncertainties better than 2\%.

\begin{figure*}[h]
\centering
 \includegraphics[width=0.48\textwidth,clip=true,trim= 0cm 0cm -1cm 0cm]{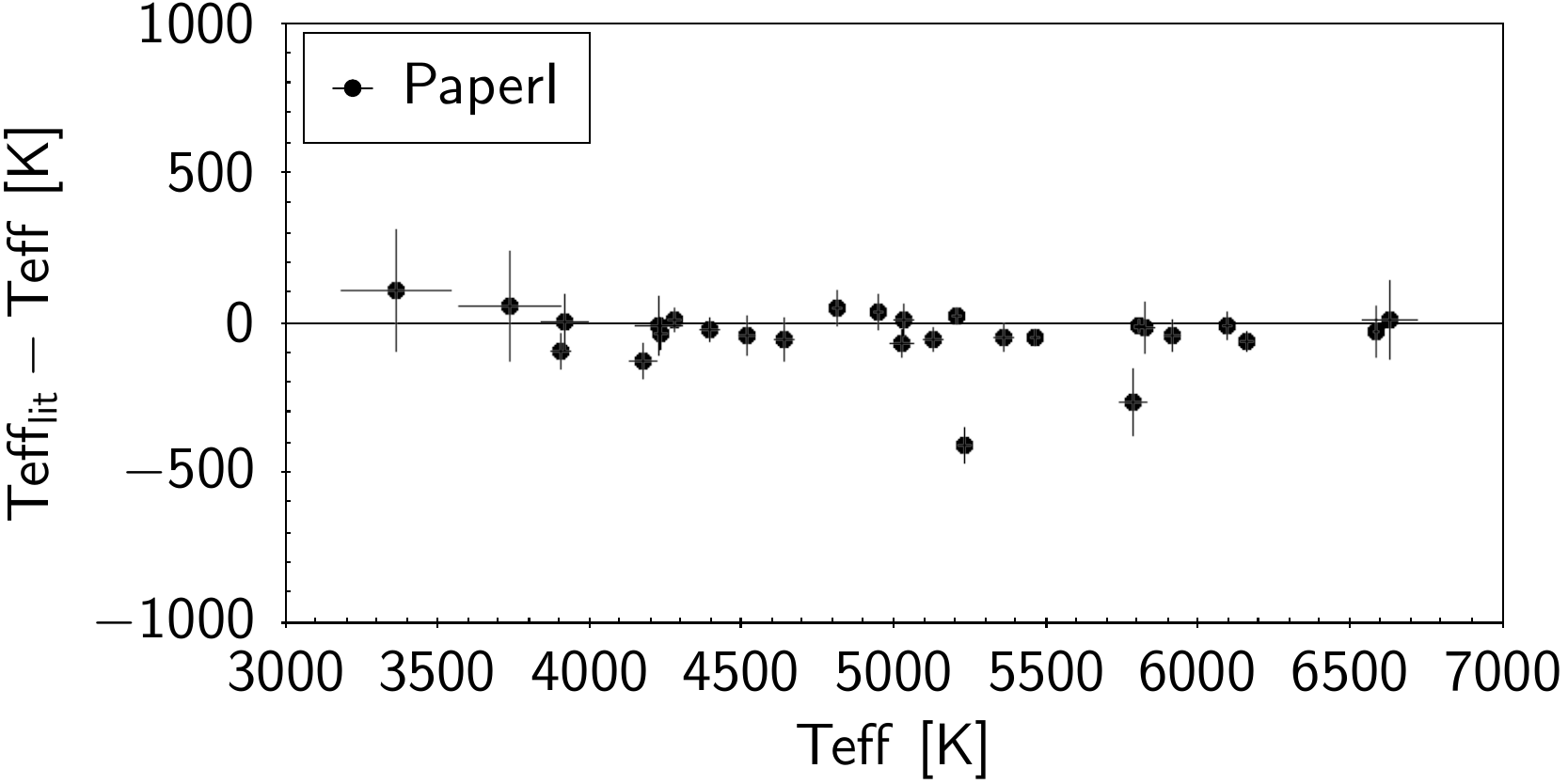} 
 \includegraphics[width=0.48\textwidth,clip=true,trim= 0cm 0cm -1cm 0cm]{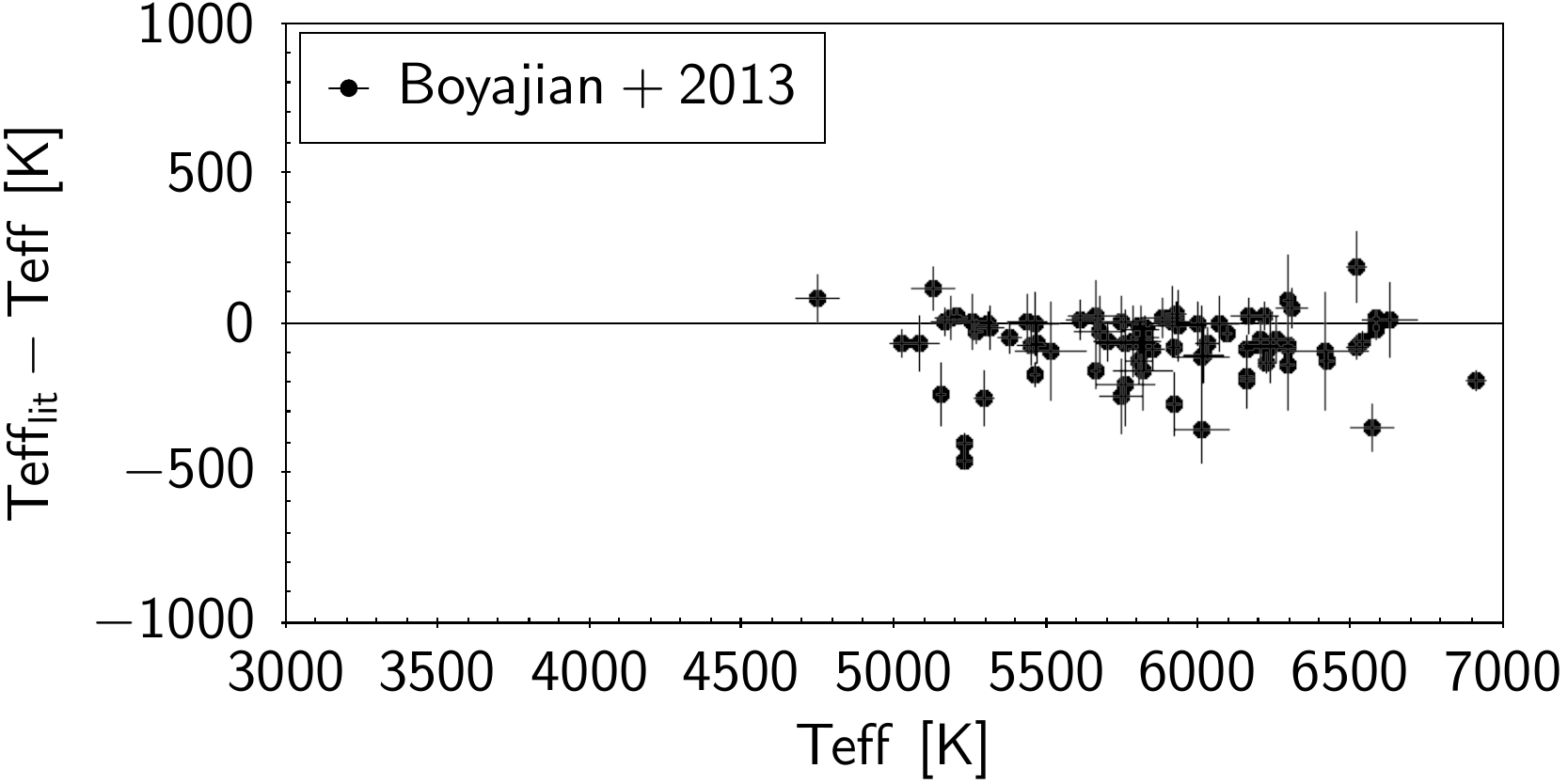} 
 \includegraphics[width=0.48\textwidth,clip=true,trim= 0cm 0cm -1cm 0cm]{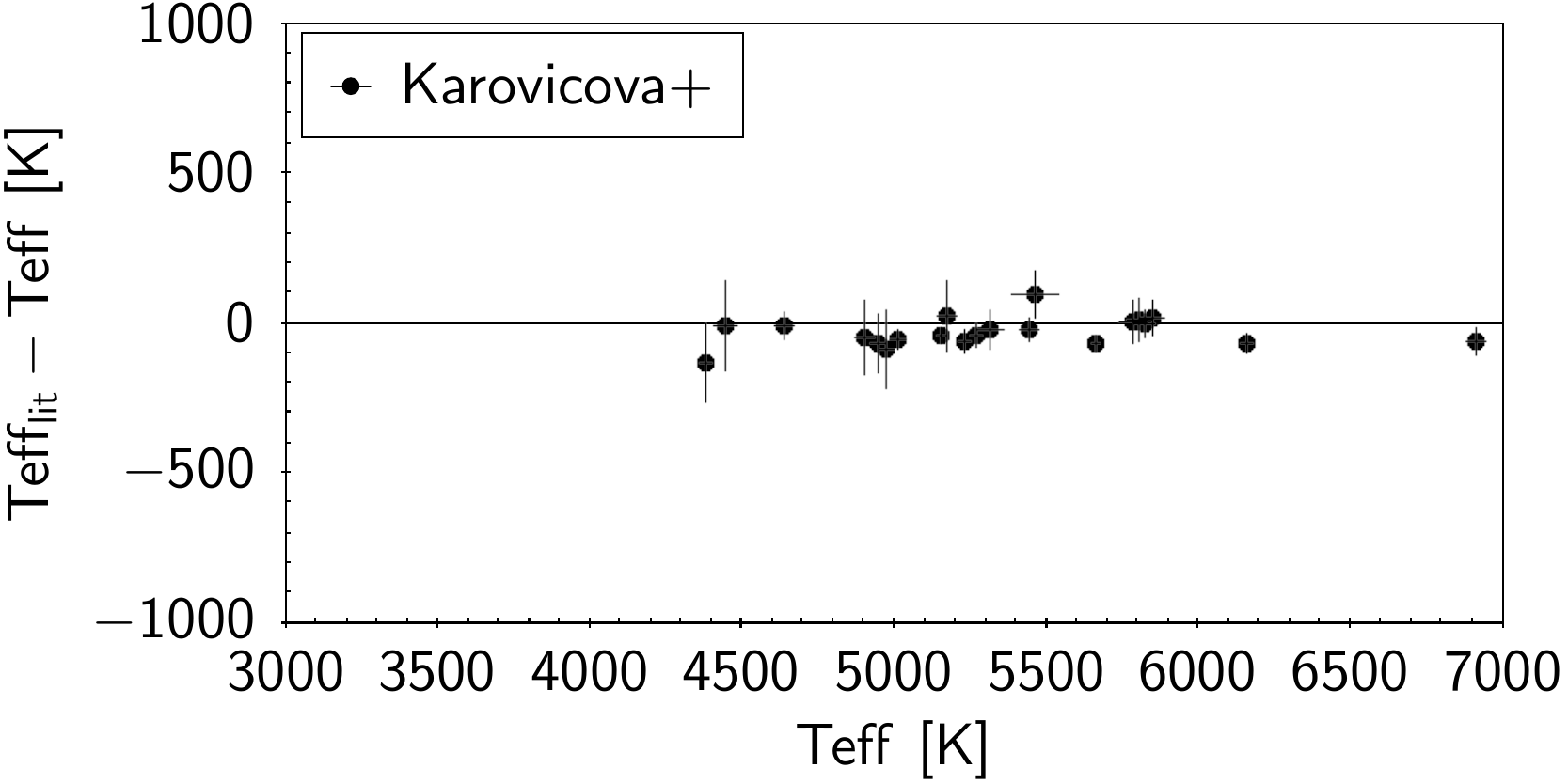} 
 \includegraphics[width=0.48\textwidth,clip=true,trim= 0cm 0cm -1cm 0cm]{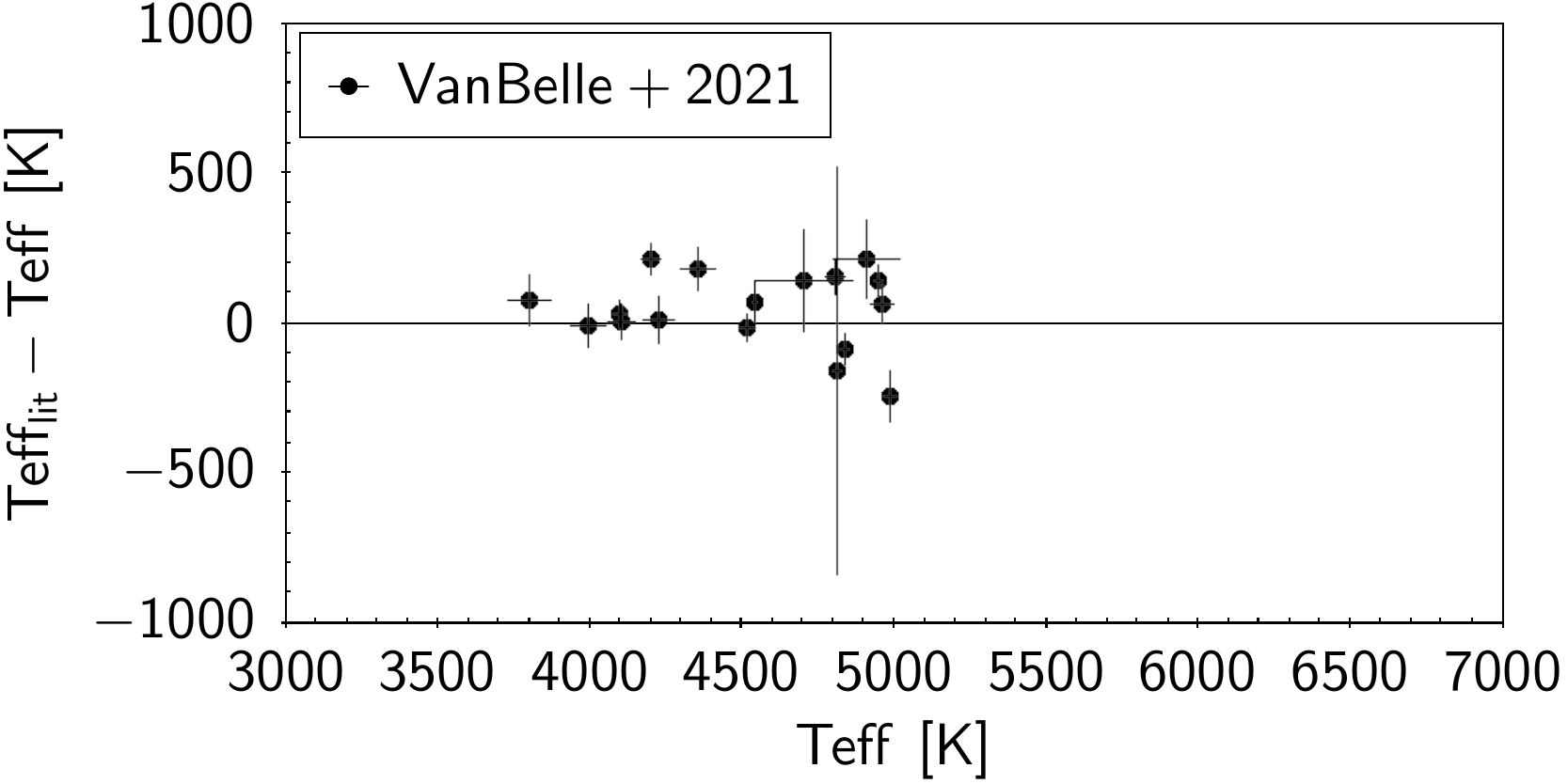} 
\caption{Comparison of our fundamental determinations of \teff\ with other fundamental determinations from the literature in Paper~I and \cite{boy13, kar20, kar22a, kar22b, van21}. }
 \label{f:comp_teff}
\end{figure*}

\begin{table}
\caption[]{Median difference (MED) and median absolute deviation (MAD) between direct determinations of \teff\ from the literature and our values (literature minus this work), for N stars in common.}
\label{t:comp_teff}
\begin{tabular}{l  r r r}
\hline
\noalign{\smallskip}
Reference & N & MED & MAD \\ 
          &   & (K) & (K) \\
\noalign{\smallskip}
\hline
\noalign{\smallskip}
\cite{hei15} - Paper~I & 28 & -26 & 32 \\
\cite{boy13} & 82 & -59 & 58 \\
\cite{kar20, kar22a, kar22b} & 21 & -39 & 30 \\
\cite{van21} & 17 & 61 & 77 \\
\noalign{\smallskip}
\hline
\end{tabular}
\end{table}

In Fig.~\ref{f:comp_teff} 
we compare our values of \teff\ to other direct determinations from the literature, including those in Paper~I. The values of offset (median difference) and dispersion (MAD) are given in Table~\ref{t:comp_teff}. The dispersion is remarkably low  (MAD$\simeq$30~K) for the comparison to Paper~I and \cite{kar20, kar22a, kar22b}, our determinations being larger by 26~K and 39~K respectively. The agreement is therefore at the 1\% level in general.
The two outliers in the comparison to Paper~I (upper left panel of Fig.~\ref{f:comp_teff}) are the metal-poor benchmark stars HIP57939 and HIP76976 (HD103095 and HD140283). Our new \teff\ values are about 400~K and 250~K higher than in Paper~I, where their sub-mas angular diameters were quoted as very uncertain. Both stars have been remeasured by \cite{kar18,kar20} leading to more precise \diam\ and higher \teff. Our determination for HD103095 (\teff=5235$\pm$18~K) is larger by  61~K than that of \cite{kar20}. Since we use their determination of \diam, the difference is only due to \Fbol. As noted in Sect. \ref{s:diam}, the angular diameter of HD103095 measured by \cite{kar20} from the combination of two instruments is very reliable. For HD140283 we find \teff=5788$\pm$45~K, lower by 4~K than their value. Three other stars differ by 2-3\% from Paper~I: $\psi$ Phe, 61 Cyg B and $\gamma$ Sge. Only $\gamma$ Sge has a new angular diameter measured by \cite{bai21}, while for the other ones we used the same \diam\ as in Paper~I, indicating that the difference comes from the new determination of \Fbol, which we expect to be more accurate than the previous determination.

Table~\ref{t:comp_teff} and Fig.  \ref{f:comp_teff} exhibit larger discrepancies in the comparison to \cite{boy13} with an offset of 59~K and a scatter of 58~K. 
We note that we have 66 stars in common but 82 measurements since \cite{boy13} provide a compilation of their own \diam\ together with other values from the literature (we removed discrepant values quoted by them for HD146233 and HD185395). Among the stars that differ by more than 300~K, we have again HD103095 which is the largest outlier. As explained above, the recent \diam\ determination by \cite{kar20} gives a higher \teff\ which is in better agreement with our value for that star. For HIP61317 \cite{boy13} give two values of \teff, only one being in significant disagreement with ours. For HIP89348 our values of \Fbol\ and \diam\ (the latter adopted from \citealt{lig16}) are larger and smaller, respectively, by $\sim$10\% than those of \cite{boy13}, resulting in a significantly different \teff. Our value of \teff=6569$\pm$69~K seems however more consistent with spectroscopic values listed in the PASTEL catalogue than their lower value of \teff=6221$\pm$39~K.

The comparison to \cite{van21} gives an offset of 61~K, this time our values being lower, with a dispersion of 77~K. This relies on 17 giants in common. These large differences could partly be due to disagreement in extinction values for some stars. We note four stars (HIP7607, HIP111944, HIP74666, HIP3031) that \cite{van21} found significantly reddened ($A_{\rm V}$ from about 0.15 to 0.30~mag) while our $A_{\rm V}$ determinations are below 0.05~mag. This possibly explains the \teff\ differences from 150~K to 220~K. On the other hand, HIP22453 has $A_{\rm V}$=0.36~mag in \cite{van21} and $A_{\rm V}$=0.08~mag in our work, but the \teff\ difference is only 31~K.

We note that we use the same determination of angular diameter as in the literature for some of the stars. Hence, the comparison data sets are not completely independent from ours.

\section{Surface gravity}
\label{s:logg}
We determined the surface gravity \logg\ with the fundamental relation expressed as: 

\begin{equation}
\centering
\log g = \log\left(\left(\frac{M}{M_\odot}\right)\left(\frac{R}{R_\odot}\right)^{-2}\right) + \log g_\odot
\label{e:logg}
\end{equation}

\noindent where $M/M_\odot$ and $R/R_\odot$ are the
mass and radius of the star in solar units. For the Sun, we adopt for the surface gravity $\log g_\odot=4.4380\pm0.0002 $ dex\footnote{The units of surface gravity $g$ are $\mathrm{cm\,s^{-2}}$. However, throughout the article, we omit the unit or use the unit dex when specifying values of \logg.} determined in Paper~I. The linear radius of each star is deduced from its angular diameter (see Sect. \ref{s:diam}) and its distance is inferred from its parallax (see below). Masses, which cannot be directly measured, are estimated from evolutionary tracks, using our fundamental \teff, luminosities (from \Fbol\ and parallaxes), radii (from \diam\ and parallaxes) and metallicities from the literature as input. We consider in this section the full sample of 201 GBS V3, including the nine stars with an indirect \diam.

\subsection{Parallaxes, linear radii and luminosities}
\label{s:par}

\begin{figure}[h]
\centering
\includegraphics[scale=0.3]{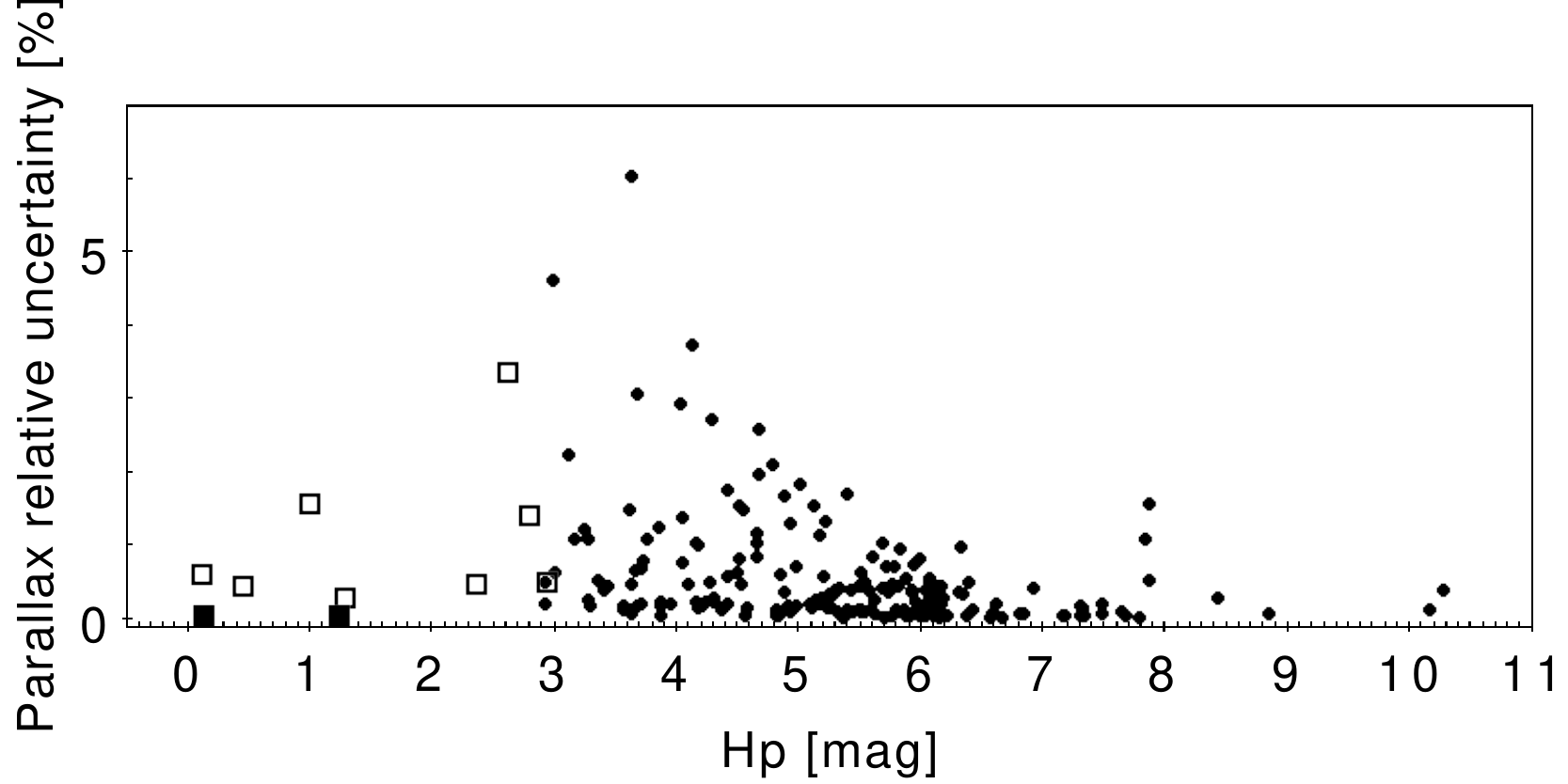} 
\caption{Distribution of parallax relative uncertainties versus Hipparcos magnitudes for the GBS V3 sample. Parallaxes are mainly from Gaia DR3, but from Hipparcos for 8 stars (open squares), and from \cite{ake21} for $\alpha$ Cen A \& B (filled squares).}
 \label{f:parallax}
\end{figure}

The  parallax of the stars is needed to convert their angular diameter into linear radius, and their bolometric flux into luminosity. All the targets have a Hipparcos parallax, and the majority of them have also an even more precise and accurate Gaia DR3 parallax. Only four stars have a Gaia parallax with an uncertainty larger than 3\%, the largest value being 6\% for HIP55219.  
The ten brightest stars not in Gaia DR3 have a precision of their Hipparcos parallax better than 3.5\%. For $\alpha$~Cen~A \& B (HIP71683 \& HIP71681) we adopt the high precision determination by \cite{ake21} instead of the Hipparcos one. Figure~\ref{f:parallax} shows the distribution of the parallax relative uncertainties as a function of the Hipparcos magnitude Hp. The four faintest stars with Hp$>$8~mag have an indirect \diam. We applied the zero-point correction derived by \cite{lin21} to the Gaia parallaxes.

\begin{figure}[h]
\centering
 \includegraphics[width=0.48\textwidth]{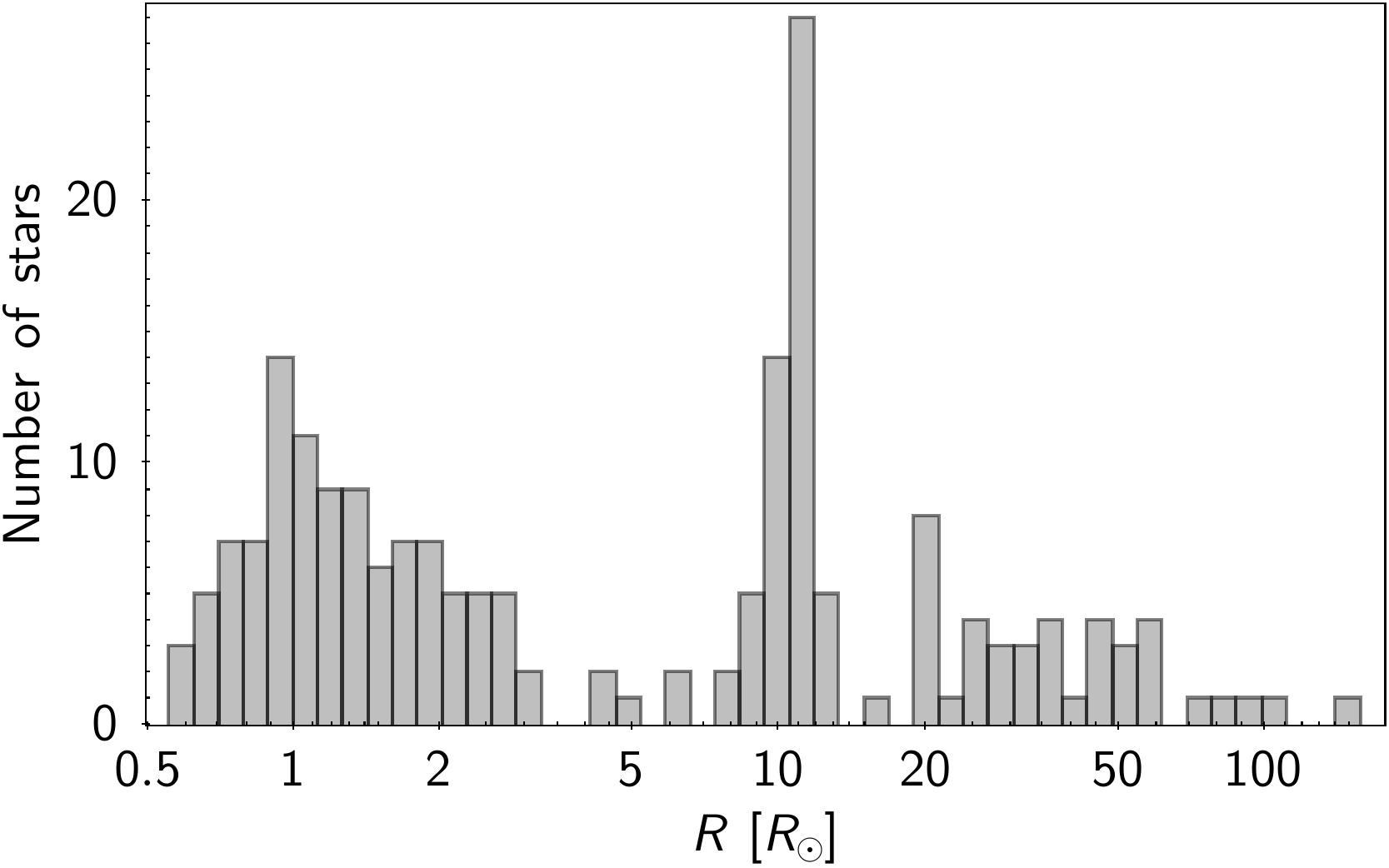} 
  \includegraphics[width=0.48\textwidth]{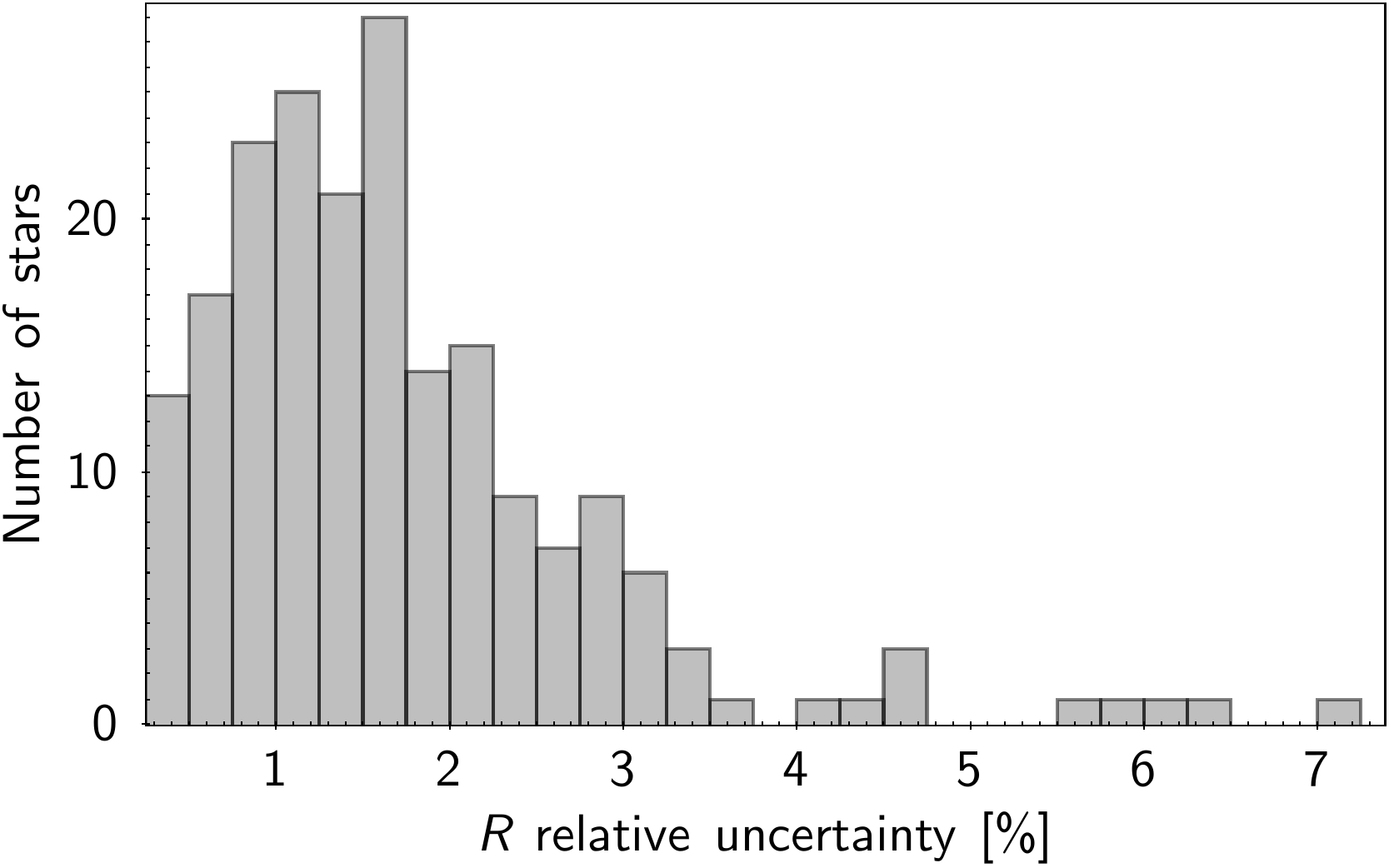} 
 \caption{Histogram of linear radii $\mathit{R}$  (top panel) and their relative uncertainty  (bottom panel). }
 \label{f:radius_histo}
\end{figure}

\begin{figure}[h]
\centering
\includegraphics[width=0.48\textwidth]{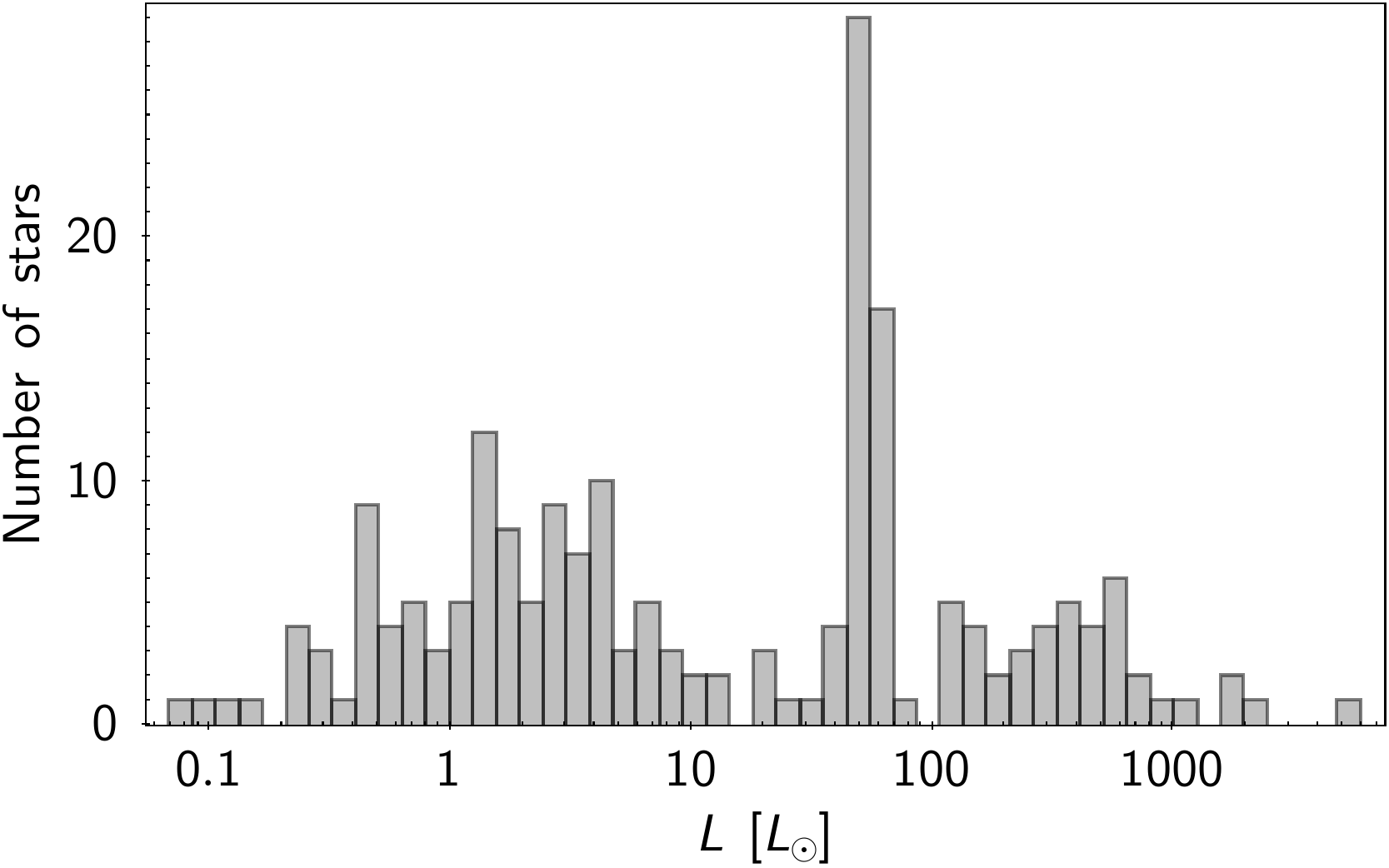} 
\includegraphics[width=0.48\textwidth]{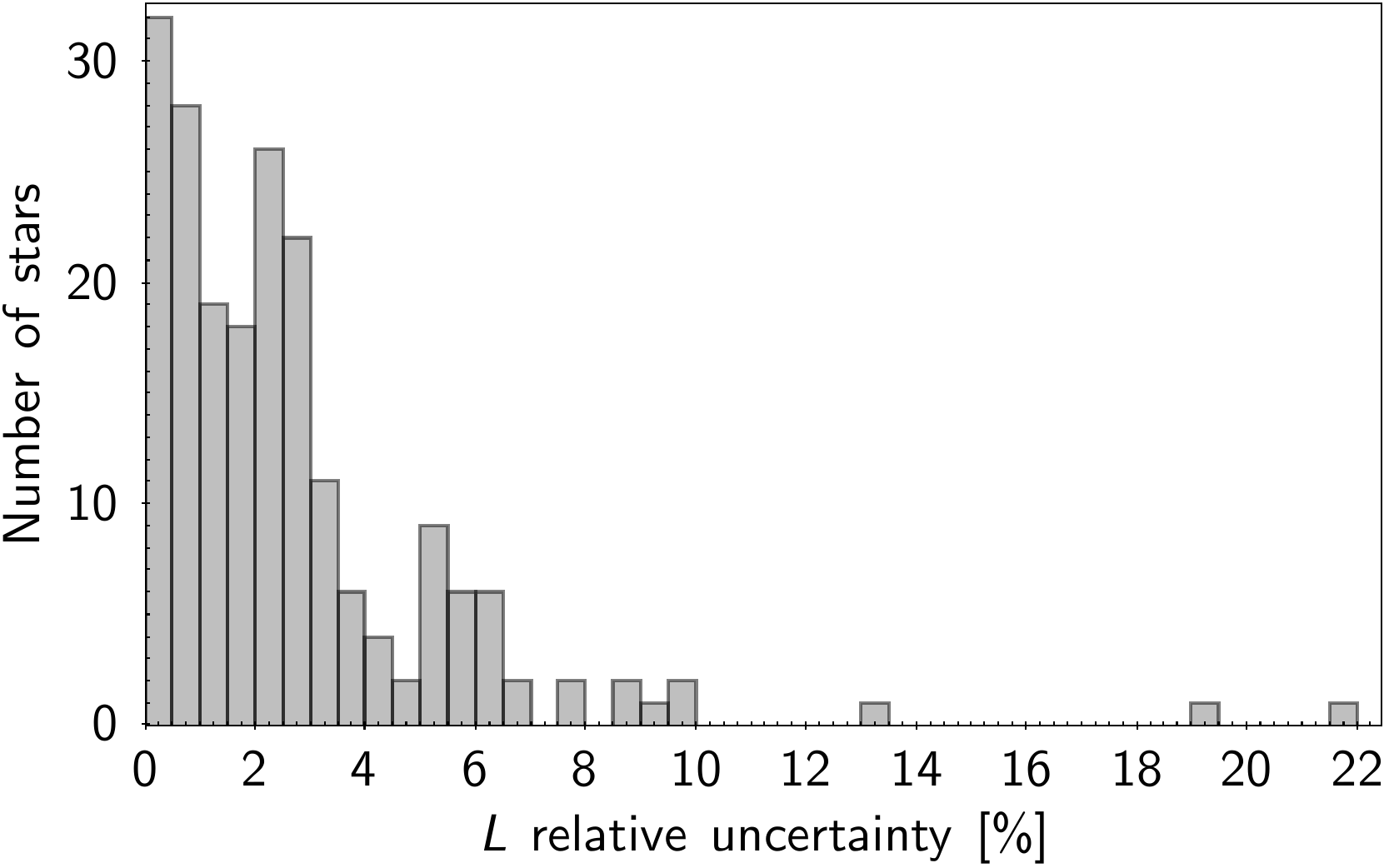} 
 \caption{Histogram of luminosities $\mathit{L}$ (top panel) and their relative uncertainty  (bottom panel). }
 \label{f:lum_histo}
\end{figure}

With parallaxes $\pi$ and \diam\ we computed linear radii $\mathit{R}$ and their uncertainties, while we used parallaxes and \Fbol\ to compute luminosities $\mathit{L}$ and their uncertainties. Adopting the solar radius and luminosity from the 2015 B3 IAU resolution\footnote{\url{https://www.iau.org/static/resolutions/IAU2015_English.pdf}} the equations are:

\begin{equation}
\label{e:radius}
\centering
    \frac{\mathit{R}}{\mathit{R}_\odot}=\frac{1}{0.00930093} \times \frac{\theta_{\rm LD}}{\pi}
\end{equation}

\begin{equation}
\label{e:lum}
\centering
\frac{\mathit{L}}{\mathit{L}_\odot}=312.564 \times \frac{F_{\rm bol}}{\pi^2}
\end{equation}

with \diam\ and $\pi$ expressed in mas, and \Fbol\ in $10^{-8}$~erg~s$^{-1}$~cm$^{-2}$.

 The radii of the GBS V3  span 0.6 to $\sim$140 $R_\odot$ (see Fig. \ref{f:radius_histo}).  The  luminosities span 0.08 to  nearly 6000 $L_\odot$ (see Fig. \ref{f:lum_histo}). 

\begin{figure}[h]
\centering
 \includegraphics[scale=0.3]{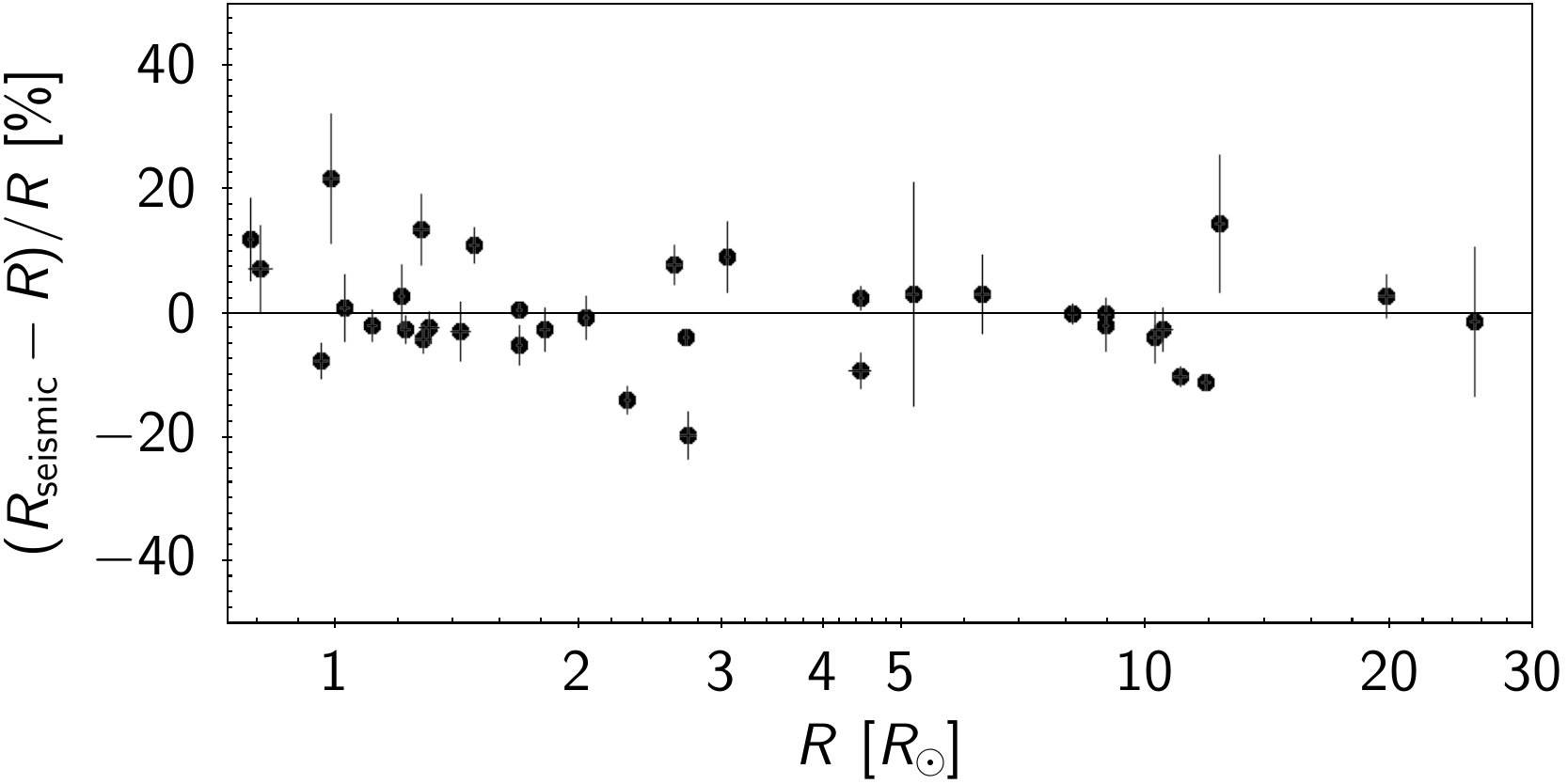} 
\caption{Linear radius difference between our determination of $\mathit R$ from Eq. (\ref{e:radius}), and seismic estimations from Eq. (\ref{equ:seismic_radius}). }
 \label{f:comp_radius_seismic}
\end{figure}

Solar-like oscillations provide robust constraints to the radius of G and K dwarfs and giants \citep{chap13}, giving us an opportunity to compare our determinations with others obtained in a different way.  We estimated seismic radii using the following scaling relation \citep[e.g.][]{mig12} when the asteroseismic parameters, the so-called large frequency separation $\Delta\nu$ and the frequency of maximum oscillation power $\nu_{\rm max}$, were available:

\begin{equation}
   \label{equ:seismic_radius}
\frac{R}{R_\odot} \approx \left(\frac{\nu_{\rm max}}{\nu_{{\rm max}\odot}}\right)\left(\frac{\Delta\nu}{\Delta\nu_\odot}\right)^{-2}\left(\frac{T_{\rm eff}}{T_{{\rm eff},\odot}}\right)^{1/2},
\end{equation}

where we adopt the fundamental \teff\ determined in Sect. \ref{s:teff} and the solar parameters as in Paper~I: $\Delta\nu_\odot = 135.229\pm0.003~\mu$Hz, $\nu_{\rm max, \odot} = 3160\pm40~\mu$Hz, $T_{\rm eff, \odot}=5771\pm1$K. We have compiled $\Delta\nu$ and $\nu_{\rm max}$ from the literature and found determinations of both parameters for 37 stars. The comparison is shown in Fig. \ref{f:comp_radius_seismic}. There is a small systematic offset, the fundamental radii being larger than the seismic ones by 0.7\%, with a typical dispersion (MAD) of 3.3\%. Several stars show discrepancies larger than 10\%, up to 22\% for HIP92984. There is however an ambiguity about the seismic parameters of HIP92984, measured by \cite{mos09} from CoRoT observations, because \cite{hub12} did not detect solar-like oscillations. The other discrepant stars have error bars that still give an agreement at the 3$\sigma$ level. We also note that \cite{sha16} and \cite{hon22} proposed some corrections to the scaling relations to obtain a better agreement for giants. It is however out of the scope of this paper to apply such corrections. We retain from this comparison the general good agreement, with no systematics, between our values and seismic ones, at the level of $\sim$4\%.

\begin{figure}[h]
\centering
  \includegraphics[scale=0.3]{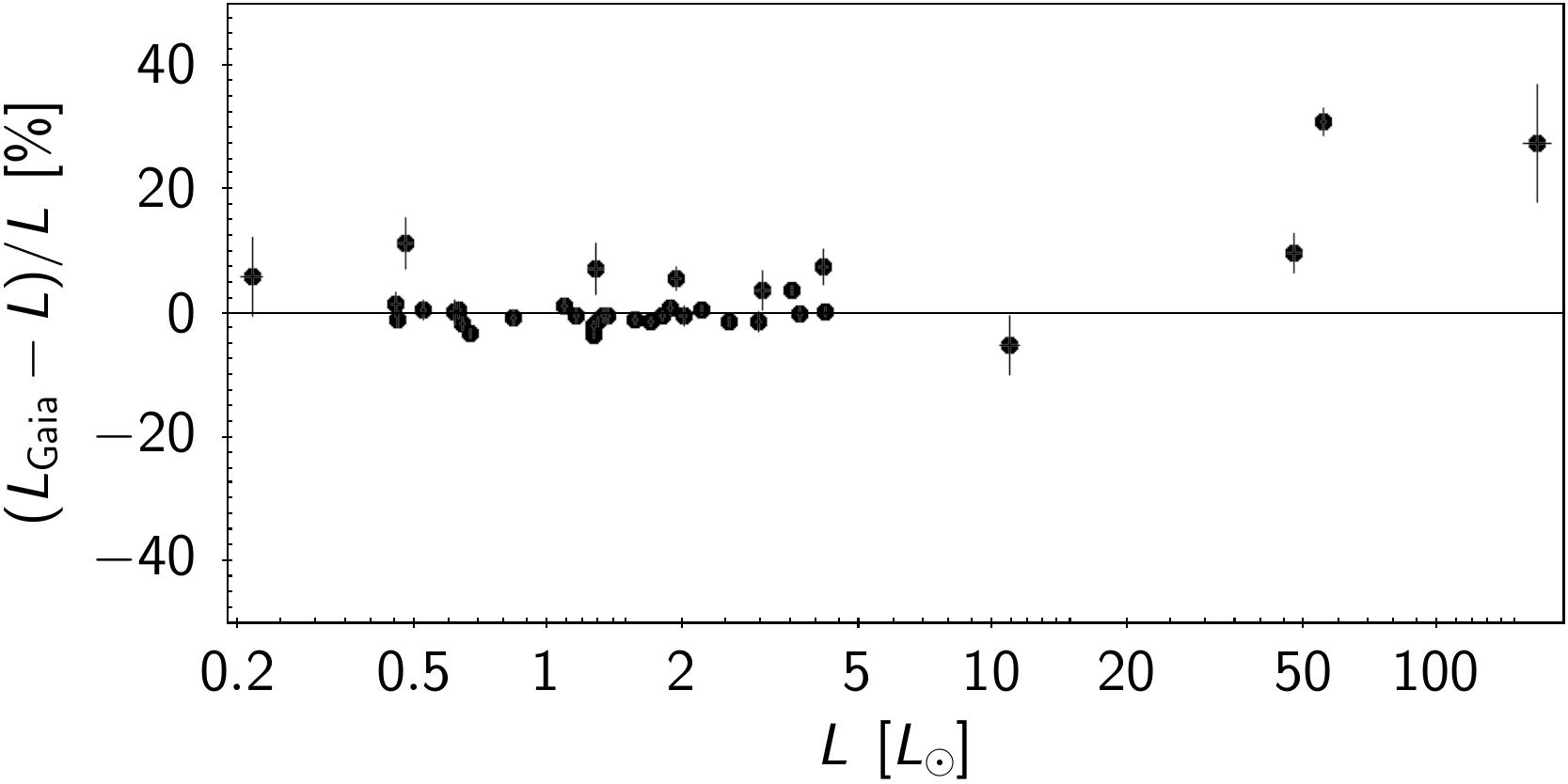} 
\caption{Luminosity difference between our determinations $\mathit{L}$ from \Fbol\ and distance, and Gaia DR3 estimations based on G magnitudes and bolometric corrections for  36 stars in common in the Golden Sample of Astrophysical Parameters \citep{gold}.}
 \label{f:comp_lum_gold}
\end{figure}

Figure \ref{f:comp_lum_gold} shows our derived luminosities compared to those available for 36 stars in the Gaia DR3 Golden Sample of Astrophysical parameters for FGKM stars \citep{gold}. Gaia luminosities were computed from the parallax, the G magnitude and a bolometric correction \citep{cre22} and are therefore different from our determinations, although not completely independent.  The three most luminous stars in common are found brighter by Gaia by more than 10\%, up to 30\% for HIP70791, known as a horizontal branch star. Only that star shows a discrepancy significantly larger than 3$\sigma$. We note six other stars with Gaia luminosities significantly larger than our values, with differences ranging from 5\% to 10\%. These discrepancies cannot be explained by the extinction that we find negligible for these nine stars. 
For the other stars, we find luminosities slightly larger than those from Gaia, by 0.35\% (median), with a typical dispersion of 0.9\% (MAD). 

\subsection{Masses}
\label{s:mass}
Masses were computed with the SPInS code \citep{leb20} implemented with the stellar evolutionary tracks from BaSTI \citep{BASTI_I,BASTI_II}, and 
from STAREVOL \citep{lag12,lag17}. 
We implemented the two grids in order to make comparisons owing to the different behaviour of the tracks in some parts of the HR-diagram (HRD), like the clump. The determinantion of the logg is less accurate at clump luminosity (logg$\simeq$2.2) because this is a point in the HR diagram where the evolutionary tracks of different masses and \feh\ overlap. In the following, we detail the main differences between these two grids that may have an impact on the position on the HRD and thus on the mass determination with SPInS. 
\begin{itemize}
\item \textit{BaSTI} - We use stellar tracks coming from the non-canonical grid covering a mass range between 0.5 M$_\odot$ and 10.0 M$_\odot$ and a metallicity range [Fe/H] $\in$ [$-2.27$, +0.40] without $\alpha$-enhancement.  This grid takes into account core convective overshooting during the H-burning phase. The overshoot parameter is set to 0.2 
for a stellar mass higher than 1.7 M$_\odot$, no overshooting is considered for a mass lower than 1.1 M$_\odot$, and a linear variation is assumed in-between. The solar mixture comes from \citet{GreSau93}.  We tried the $\alpha$-enhanced tracks ([$\alpha$/Fe]=+0.4) for metal-poor stars (\feh<-0.70 dex) leading to masses higher by 30\% on average. However, as explained later, we got a wrong mass for $\mu$ Cas, the only metal-poor binary with a reliable dynamical mass. This convinced us to adopt the tracks without $\alpha$-enhancement for the whole sample.  
\item \textit{STAREVOL} - This stellar grid covers a mass range between 0.6 and 6.0 M$_\odot$ and a metallicity range [Fe/H] $\in$ [$-2.14$, +0.51] without $\alpha$-enhancement, with the exception of [Fe/H]=-2.14 and -1.2 where [$\alpha$/Fe]=+0.3. Except for convection, additional mixing effects such as rotation-induced mixing are not taken into account. The overshoot parameter  
is set to 0.05 or 0.10 for stars with masses below or above 2.0~M$_\odot$, respectively; no overshooting is considered for masses lower than 1.1~M$_\odot$. The stellar grid is constructed using the solar mixture coming from \citet{Asplund09}. 
\end{itemize}

\begin{figure}[h]
\centering
\includegraphics[width=0.48\textwidth]{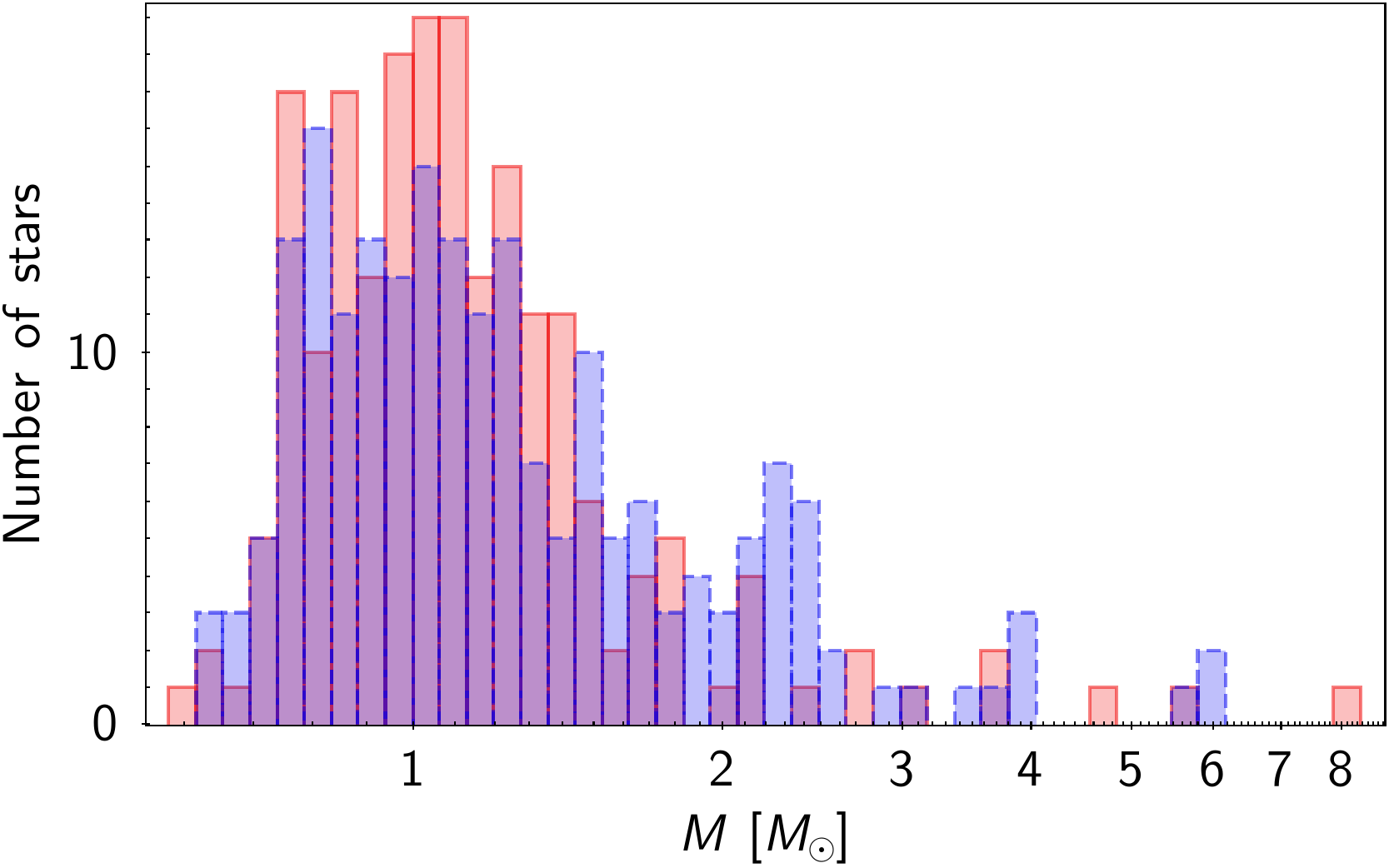} 
\includegraphics[width=0.48\textwidth]{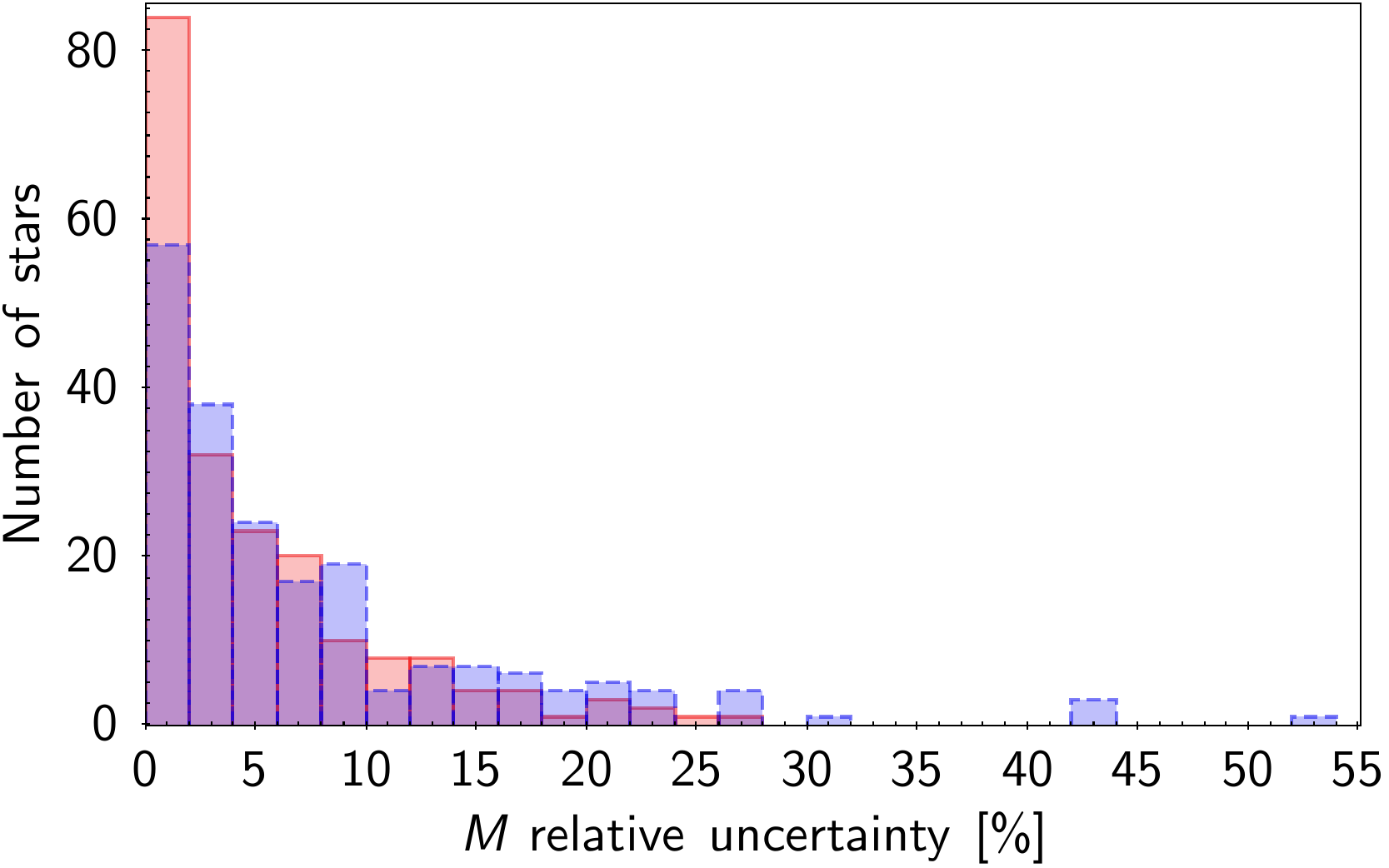} 
 \caption{Histogram of masses (top panel) and their relative uncertainty  (bottom panel), red for BaSTI, blue for STAREVOL.}
 \label{f:mass_histo}
\end{figure}

The Kroupa initial mass function \citep[][IMF]{kro01,kro13} was used as a prior, as well as a truncated uniform star formation rate between 0 and 13.8 Gyr, that is, roughly the age of the Universe. The stellar properties used as an input to SPInS are: (1) our fundamental \teff\ determinations (Sect. \ref{s:teff}); (2) luminosities deduced from \Fbol\ and parallaxes; (3) metallicities from the literature, and (4) radii deduced from \diam\ and parallaxes. Radii are not independent of \teff\ and luminosities, but still add a useful constraint to the mass since the correlations are lost in the way we determined the three parameters. The resulting masses and their uncertainties are shown in Fig.~\ref{f:mass_histo}. Most of the stars have masses $<2$~${\rm M}_\odot$ but STAREVOL finds more stars in the range 2--2.5~${\rm M}_\odot$ than BaSTI. STAREVOL gives a more extended distribution of relative uncertainties with fewer very low values, and five stars within 30-55\%.

For the validation of the mass determination, we compared the SPInS results to other determinations. This includes dynamical masses of binary stars, seismic masses, as well as mass determinations based on evolutionary tracks and methods different from those we used. 

\begin{figure}[h]
\centering
 \includegraphics[width=0.35\textwidth]{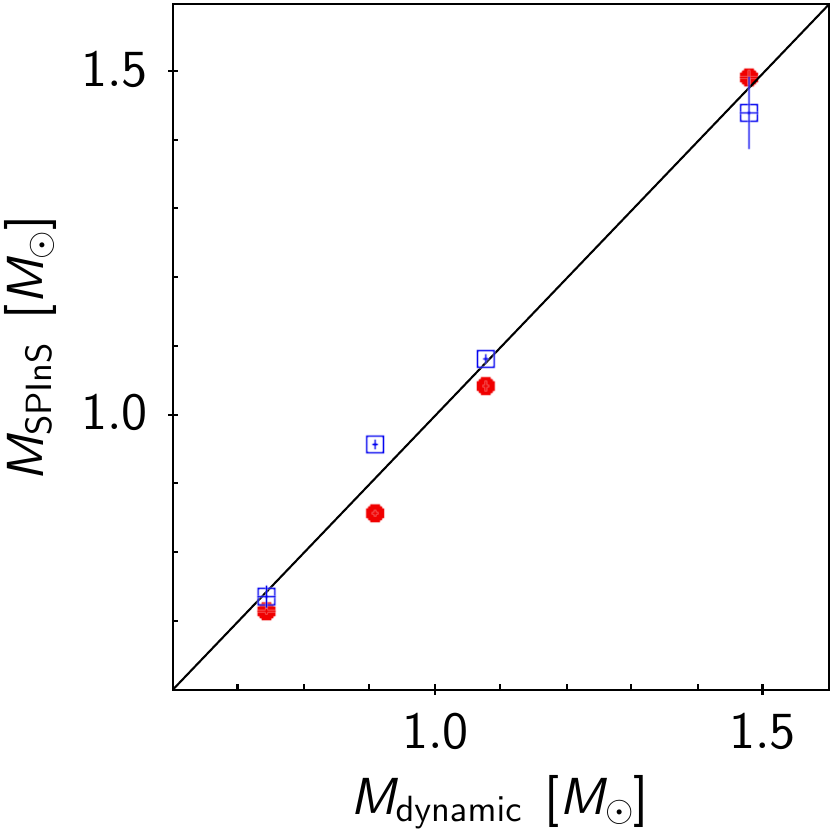} 
\caption{Comparison of masses determined with SPInS (red dots for BaSTI, blue open squares for STAREVOL) to dynamical masses for $\mu$ Cas, $\alpha$ Cen B, $\alpha$ Cen A and Procyon, ordered by increasing mass. }
\label{f:mass_dyn}
\end{figure}

Dynamical masses are available for four stars, $\mu$ Cas, $\alpha$ Cen A \& B, and Procyon. We did not consider 61 Cyg A \& B since their masses are not well established \citep{ker22}. The comparison to SPInS masses is shown in Fig.~\ref{f:mass_dyn}. The orbit of the binary $\alpha$ Cen has been studied by \cite{ake21} who determined masses of 1.0788$\pm$0.0029~$M_\odot$ and 0.9092$\pm$0.0025~$M_\odot$ for the A (HIP71683) and B (HIP71681) components, respectively. The agreement is at the 0.4\% level for the STAREVOL mass and 3.4\% for the BaSTI mass, for the component A. Both sets of evolutionary tracks give masses that differ by 5\% for the B component, in opposite directions.
The dynamical mass of the metal-poor (\feh=-0.83 dex) visual binary $\mu$ Cas (HIP5336) results from an astrometric study with the Hubble Space Telescope by \cite{bon20} who determined a value of 0.7440$\pm$0.0122~$M_\odot$. BaSTI and STAREVOL underestimate it by 4\% and  1.3\% respectively. Running SPInS with the $\alpha$-enhanced BaSTI tracks ([$\alpha$/Fe]=+0.4) for that star led to an overestimation of its mass by 32\%. This convinced us not to adopt the $\alpha$-enhanced tracks for metal-poor stars. Hence, we have opted to exclusively rely on the BaSTI tracks that do not incorporate any alpha-enrichment. This underscores the importance of presenting mass values obtained from both BaSTI and STAREVOL tracks, since it offers an understanding of the inherent errors linked to relying solely on a single stellar evolution model.
The orbit of Procyon (HIP37279) based on Hubble Space Telescope astrometry \citep{bon15,bon18}, yields a dynamical mass of 1.478$\pm$0.012~$M_\odot$. The BaSTI mass differs by 0.9\% while the STAREVOL mass is lower by 2.6\%. There is therefore a satisfactory agreement between the SPInS masses and the dynamical masses for these four stars, whatever the set of evolutionary tracks, considering the few constraints we use with the models, and given the inherent  model assumptions of e.g. the chemical enrichment law which constrain the position of the tracks in the HRD.

\begin{figure}[h]
\centering
 \includegraphics[width=0.48\textwidth]{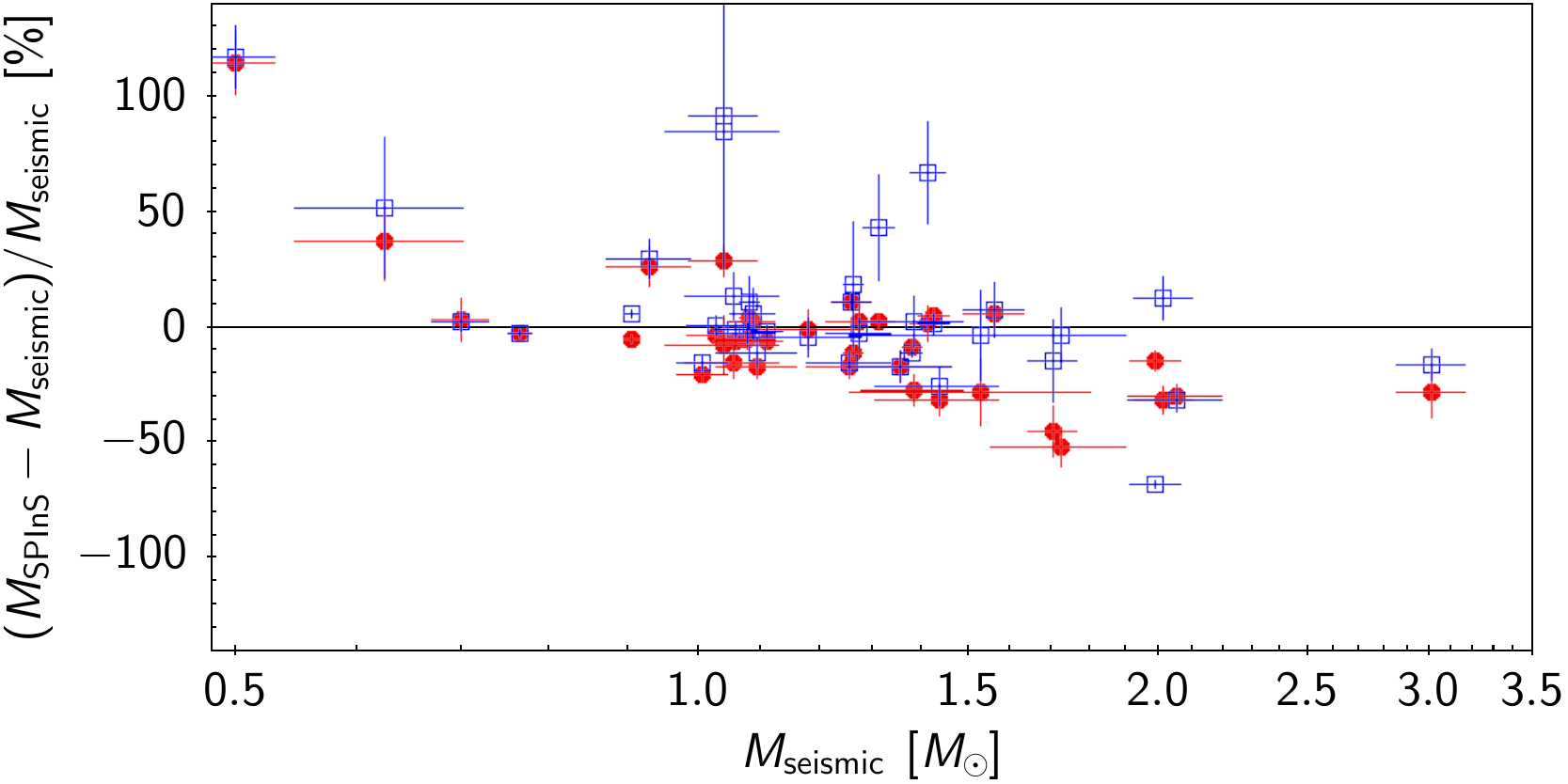} 
\caption{Comparison between masses from SPInS (red dots for BaSTI, blue open squares for STAREVOL) and seismic masses from Eq. (\ref{e:massA}). }
\label{f:comp_mass_seismic}
\end{figure}

We estimated seismic masses using the following scaling relation \citep[e.g.][]{mig12} for the 37 stars having a determination of the asteroseismic parameter $\Delta\nu$ available in the literature (see Sect.~\ref{s:par} for the solar values)

\begin{equation}
   \label{e:massA}
\frac{M}{M_\odot} \approx \left(\frac{\Delta\nu}{\Delta\nu_\odot}\right)^2 \left(\frac{R}{R_\odot}\right)^3,
\end{equation}

where R is the linear radius computed in Sect. \ref{s:par}. The resulting comparison is shown in Fig. \ref{f:comp_mass_seismic}. Although the agreement between SPInS and seismic masses is good in general in the range 1--1.5~M$_\odot$, there is a trend in the sense that SPInS tends to overestimate masses smaller than 1~M$_\odot$ and to underestimate those larger than 1.5~M$_\odot$. This is true for both sets of evolutionary tracks, with more outliers with STAREVOL. However, seismic masses may not necessarily be more accurate than those deduced from evolutionary tracks, given that the range of validity of the scaling relation is not yet clear \citep[e.g.][]{sha16}.

\begin{figure}[h]
\centering
 \includegraphics[width=0.48\textwidth]{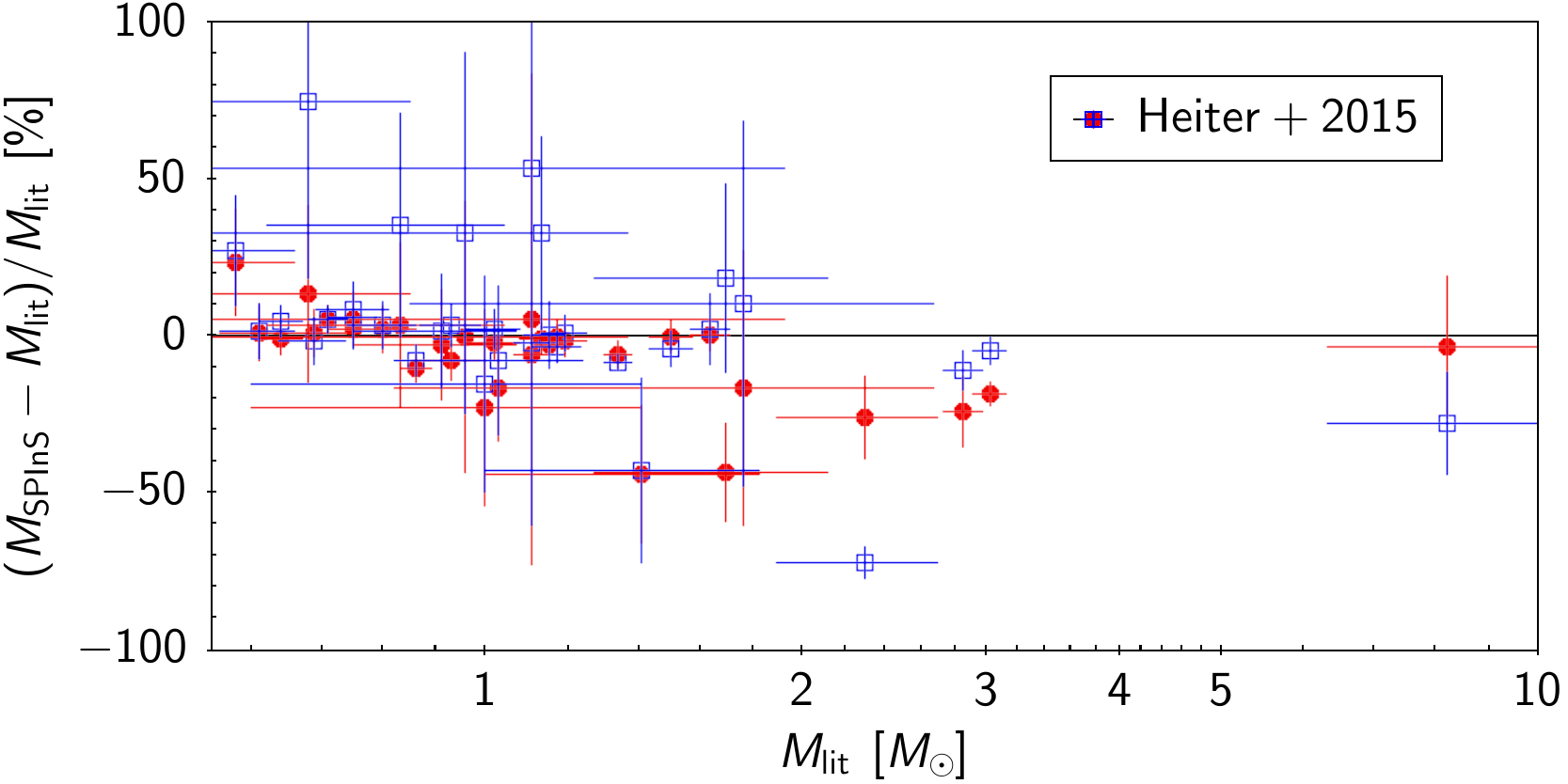} 
 \includegraphics[width=0.48\textwidth]{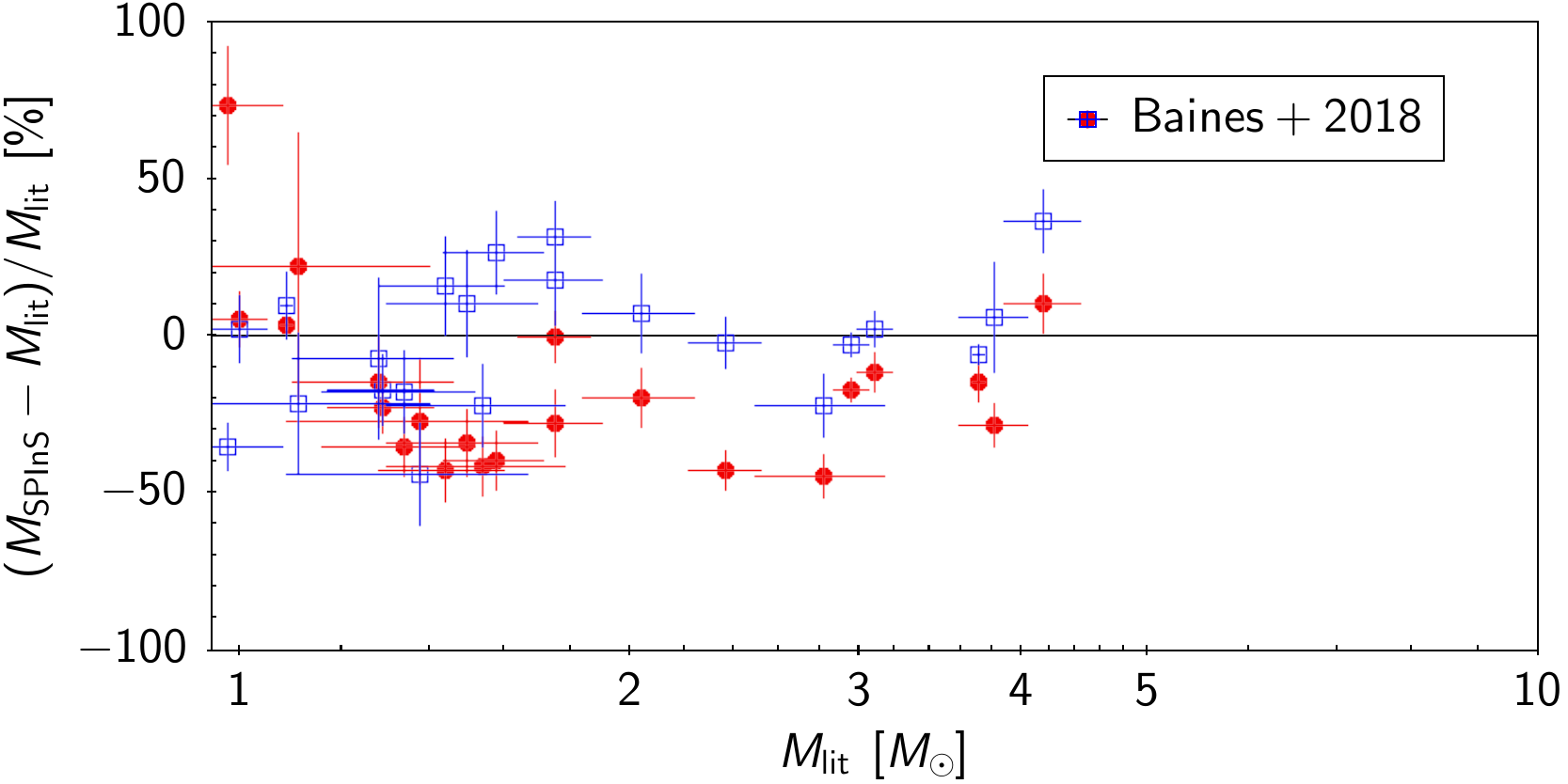} 
 \includegraphics[width=0.48\textwidth]{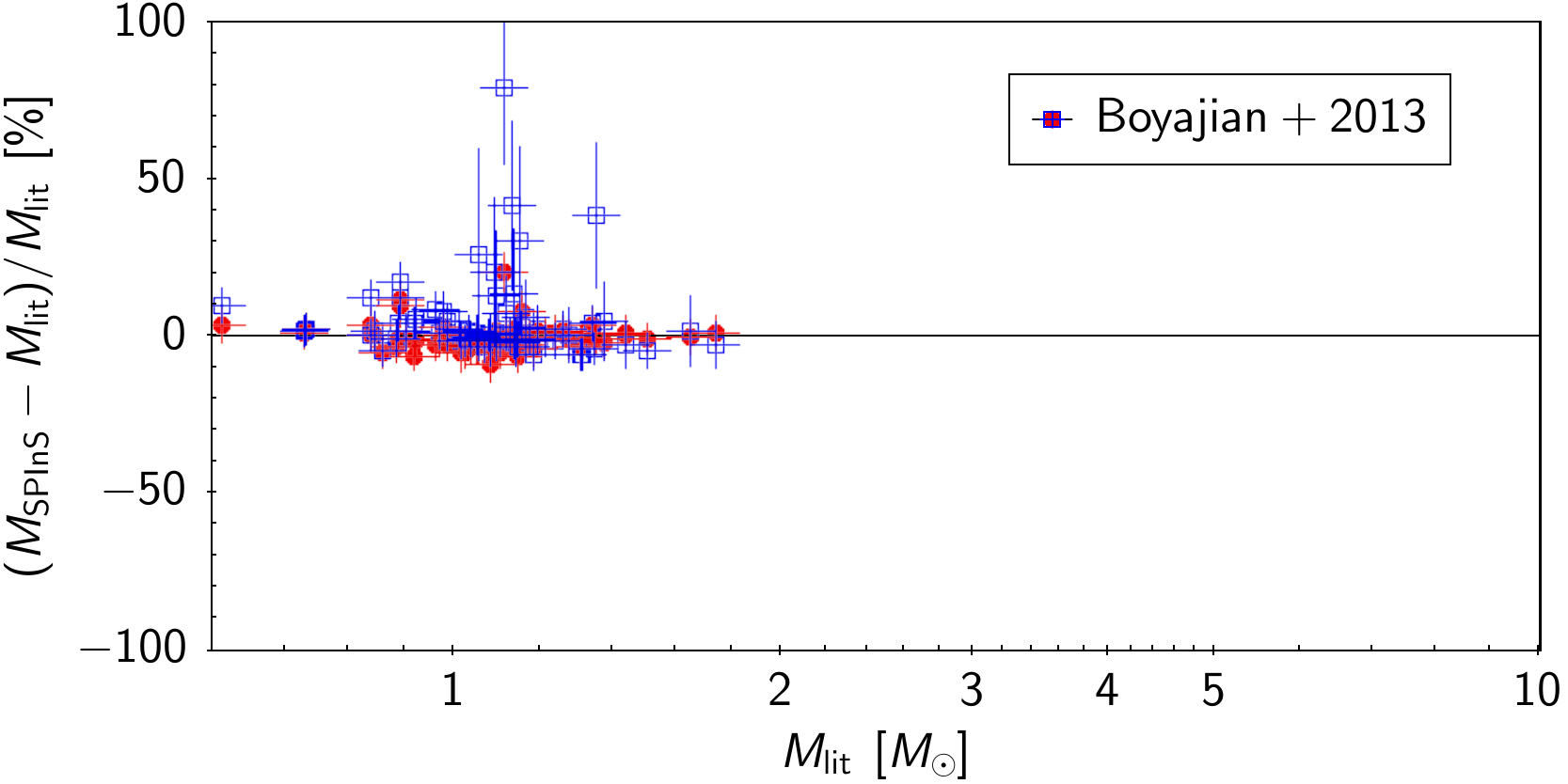} 
\caption{Comparison of masses determined with SPInS (red dots for BaSTI, blue open squares for STAREVOL) to those available in the literature, also based on evolutionary tracks.}
\label{f:comp_mass_lit}
\end{figure}

We also compared the two sets of SPInS masses to masses from the literature, based on different evolutionary tracks and methods. Figure \ref{f:comp_mass_lit} shows comparisons to masses from Paper~I, \cite{bai18} and \cite{boy13}. In Paper~I masses were determined by visual interpolation in two grids, the Padova grid \citep{ber08, ber09} and the Yonsei-Yale grid \citep{yi03,dem04}, the adopted value being the average of the two. \cite{bai18} used a Bayesian method with the PARSEC isochrones developed by \cite{das06,bre12}. \cite{boy13} used the Yonsei-Yale isochrones. The resulting comparisons reflect the nature of the stars in common, with a good agreement for dwarfs and a large dispersion for giants. A larger dispersion is expected for giants, in particular at the clump, because of the overlap of the evolutionary tracks of different masses and \feh. Degeneracies in evolutionary tracks of evolved stars can also lead to different masses for a giant depending on whether it is on the red-giant branch or on the horizontal branch. The sample of \cite{boy13} is made of dwarfs and exhibits a small dispersion, despite a few outliers with STAREVOL. The sample of \cite{bai18} is mostly made of giants and exhibits a large dispersion, while GBS V1 are a mixture of dwarfs and giants. 

\begin{figure}[h]
\centering
 \includegraphics[width=0.48\textwidth]{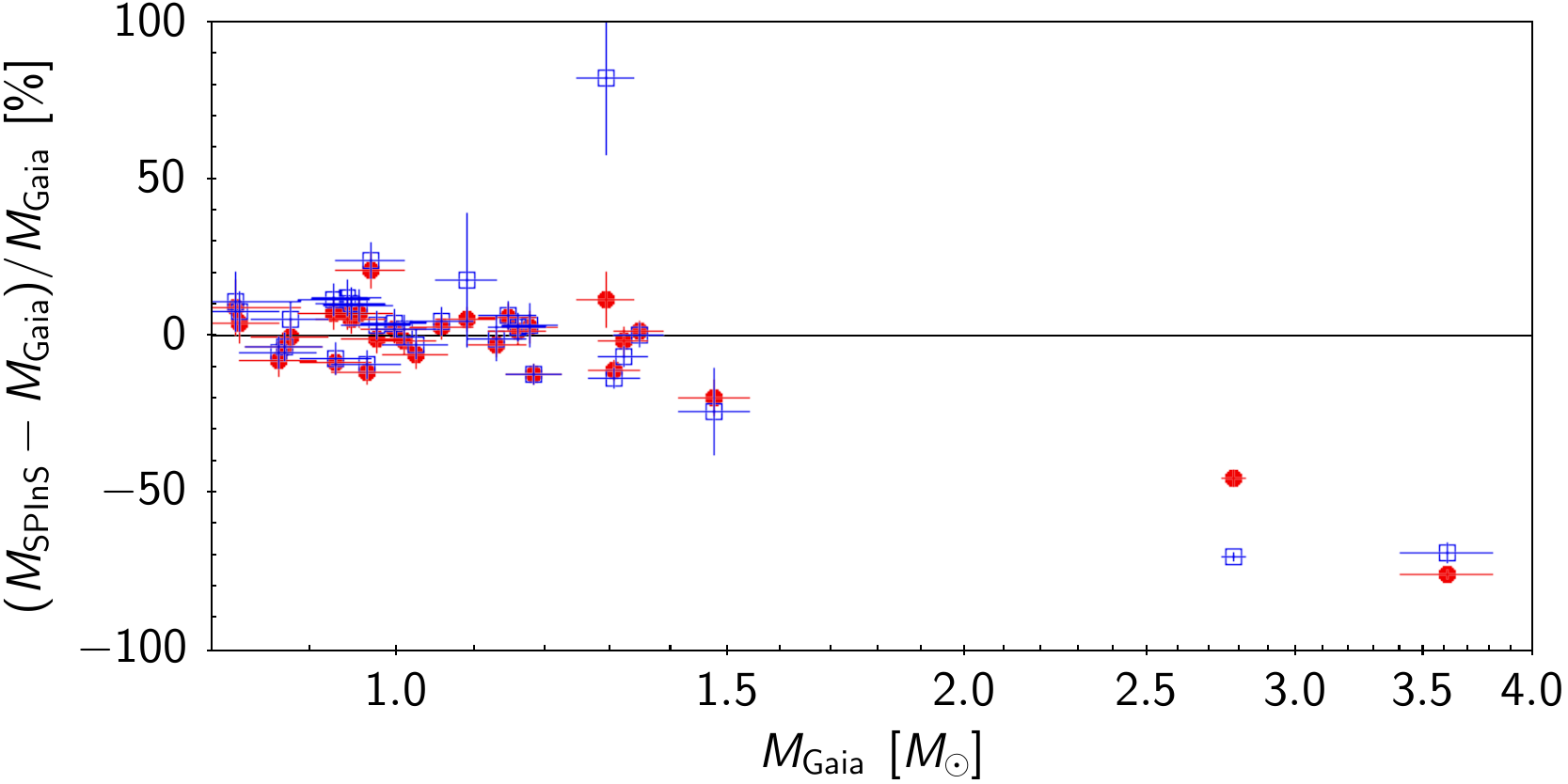} 
\caption{Comparison between masses from SPInS (red dots for BaSTI, blue open squares for STAREVOL) and masses from the Gaia Golden Sample of Astrophysical Parameters \citep{gold}. }
\label{f:comp_mass_gold}
\end{figure}

Figure \ref{f:comp_mass_gold} shows our derived masses compared to those available for 30 stars in the Gaia DR3 Golden Sample of Astrophysical parameters for FGKM stars \citep{gold}. Gaia masses were derived by comparing Gaia photometric effective temperatures and Gaia luminosities to BaSTI solar metallicity stellar evolution models \citep{2018ApJ...856..125H,cre22}, and are therefore similar to our determinations. We find an excellent agreement between our two sets of masses and the Gaia ones, except for the three stars in common with the highest masses ($\gtrsim$1.5~M$_\odot$), and one outlier within the STAREVOL set.

From the above comparisons, there is no strong evidence that one set of evolutionary tracks is better than the other one. Therefore we provide the two masses and their uncertainties in the catalogue available at the CDS.

 In this procedure to determine masses we need metallicities as input for SPInS. We have used \feh\ values from the literature which are not homogeneous and therefore we have evaluated their impact on the resulting masses. We made two tests similar to those made for \Fbol. One test is to adopt a large uncertainty of 0.15 on \feh\ for all the stars, much larger than the original ones. The other test is to add or subtract 0.15 dex to the literature values of \feh\ for seven stars selected to cover the parameter space. Enlarging the metallicity uncertainty to 0.15 dex affects mainly the clump giants. 
Based on the BASTI tracks, only 5 stars in our sample have their mass affected by more than 30\%, and only 8 stars if we consider the STAREVOL tracks. The most critical stars are not common from one set to the other. This reinforces the interest of considering the masses computed by the two sets of stellar evolution models. Such differences could be explained by different inputs in the computation of the evolutionary tracks (e.g. mass loss, atmosphere models, etc.) which change the position in the HRD. It should be noted that at least 90\% of the stars in our sample experience a mass variation less than 10\%, while three quarters of the sample remain below 5\%, whatever the set of stellar models taken into account.  Changing the value of \feh\ by $\pm$0.15 dex for seven stars leads to a similar conclusion: the dwarfs are not affected, whatever their metallicity, while changes occur among giants. However, due to the dependency of \logg\ on the logarithm of mass in Eq.~(\ref{e:logg}), in the worst cases where the mass is changed by 30\%, the impact on \logg\ is limited to 0.11~dex and up to 0.5~dex for the most critical cases.

\subsection{Assessment of \logg}

\begin{figure}[h]
\centering
 \includegraphics[width=0.48\textwidth]{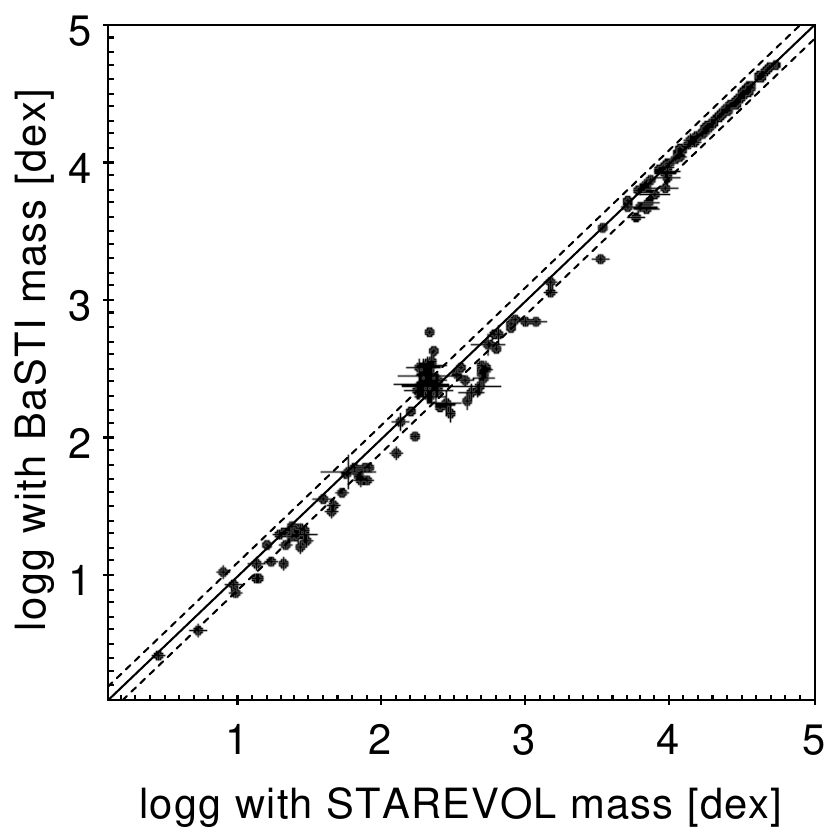}  
 \caption{Comparison of \logg\ determinations, using masses from SPInS with BaSTI or STAREVOL. The area between the two dashed lines indicates an agreement within 0.1 dex. }
\label{f:comp_logg_b_s}
\end{figure}

We computed the fundamental \logg\ of each star by applying Eq. (\ref{e:logg}) with the values of mass from SPInS, with both evolutionary tracks BaSTI and STAREVOL, and the radius deduced from \diam, with the propagation of their uncertainties. We consider here the 201 stars of the sample. The resulting uncertainties on \logg\ span from 0.004 to 0.13 (BaSTI) and 0.23 dex (STAREVOL), with a median value of 0.02 dex. 
Figure \ref{f:comp_logg_b_s} shows the comparison of \logg\ determinations, using the mass from SPInS with BaSTI or STAREVOL. The agreement is excellent for dwarfs with \logg$>$4. Below that value, \logg\ from STAREVOL is systematically larger by 0.06 dex than \logg\ from BaSTI, with an exception around \logg$_{\rm STAREVOL}$=2.3.

Following the comparisons made in the previous sections, for radii and masses, we used the seismic data to determine \logg\ in another and independent way, through the relation that gives \logg\ as a function of the maximum of the power spectrum of oscillation frequencies, $\nu_{\rm max}$, available for 42 stars, and the effective temperature:

\begin{equation}
   \label{e:numax}
   \log g \approx \log\nu_{\rm max} + 0.5\log T_{\rm eff} - \log\nu_{\rm max, \odot}, - 0.5\log T_{\rm eff, \odot} + \log g_\odot,
\end{equation}

\begin{figure}[h]
\centering
 \includegraphics[width=0.48\textwidth]{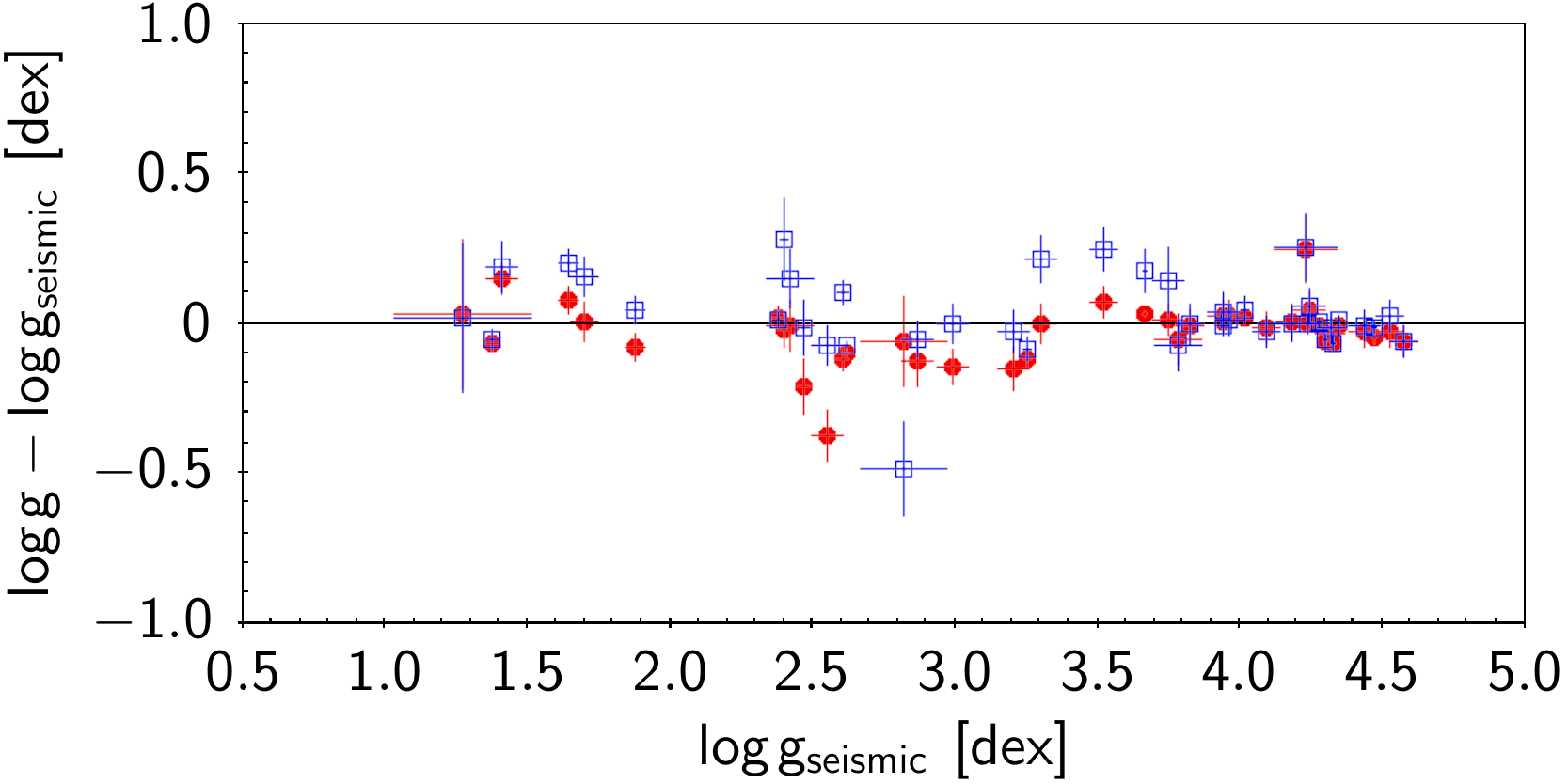} 
\caption{Comparison of our fundamental values of \logg\ to those determined from  $\nu_{\rm max}$ and our fundamental \teff. Red dots for SPInS masses using BaSTI models, blue open squares for STAREVOL models. }
 \label{f:comp_logg_seismic}
\end{figure}

The comparison of seismic and  fundamental \logg\ is shown in Fig. \ref{f:comp_logg_seismic}. For dwarfs, typically \logg$_{\rm seismic}>$3.8~dex, the agreement is very good except for one star, HIP92984. For this star we find \logg=4.48 dex with BaSTI and STAREVOL masses, while the seismic \logg\ is significantly lower, \logg=4.23. We have pointed out the ambiguity about the seismic parameters of that star in the previous section. If this star is excluded, the differences between seismic and fundamental \logg\ of dwarfs have a MAD of 0.02~dex. For giant stars, the dispersion is larger (MAD=0.07~dex), with \logg\ based on BaSTI lying slightly below the seismic values (median offset of $-0.06$~dex), while the \logg\ based on STAREVOL  tend to lie above (median offset of 0.01~dex). A few outliers, reaching nearly 0.5~dex, correspond to stars with one of the two masses giving a disagreement with the seismic \logg\ but not the other one. From that comparison, we cannot say that the agreement is better with BaSTI or with STAREVOL masses.

\begin{figure}[h]
\centering
 \includegraphics[width=0.48\textwidth]{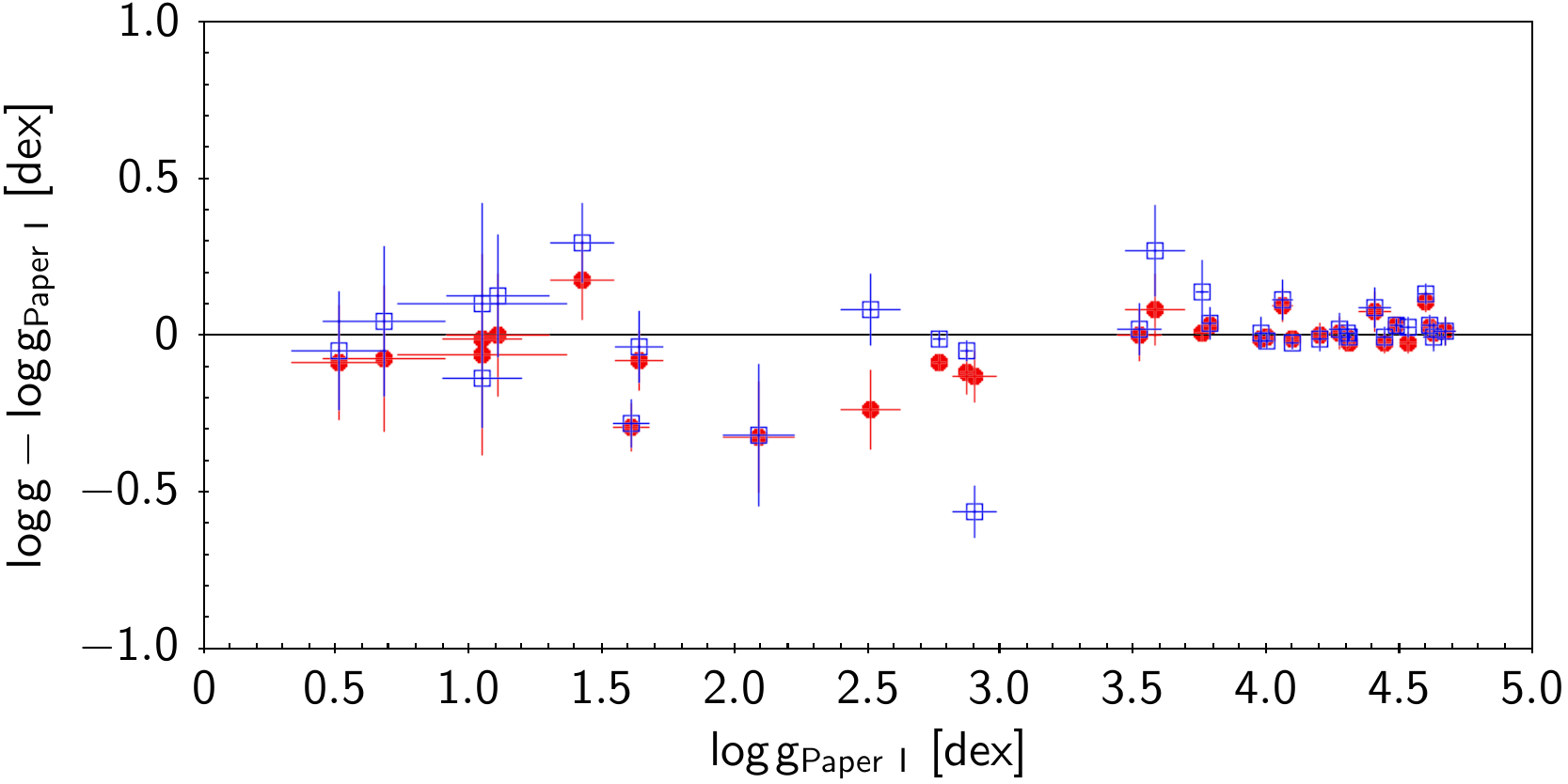} 
 \includegraphics[width=0.48\textwidth]{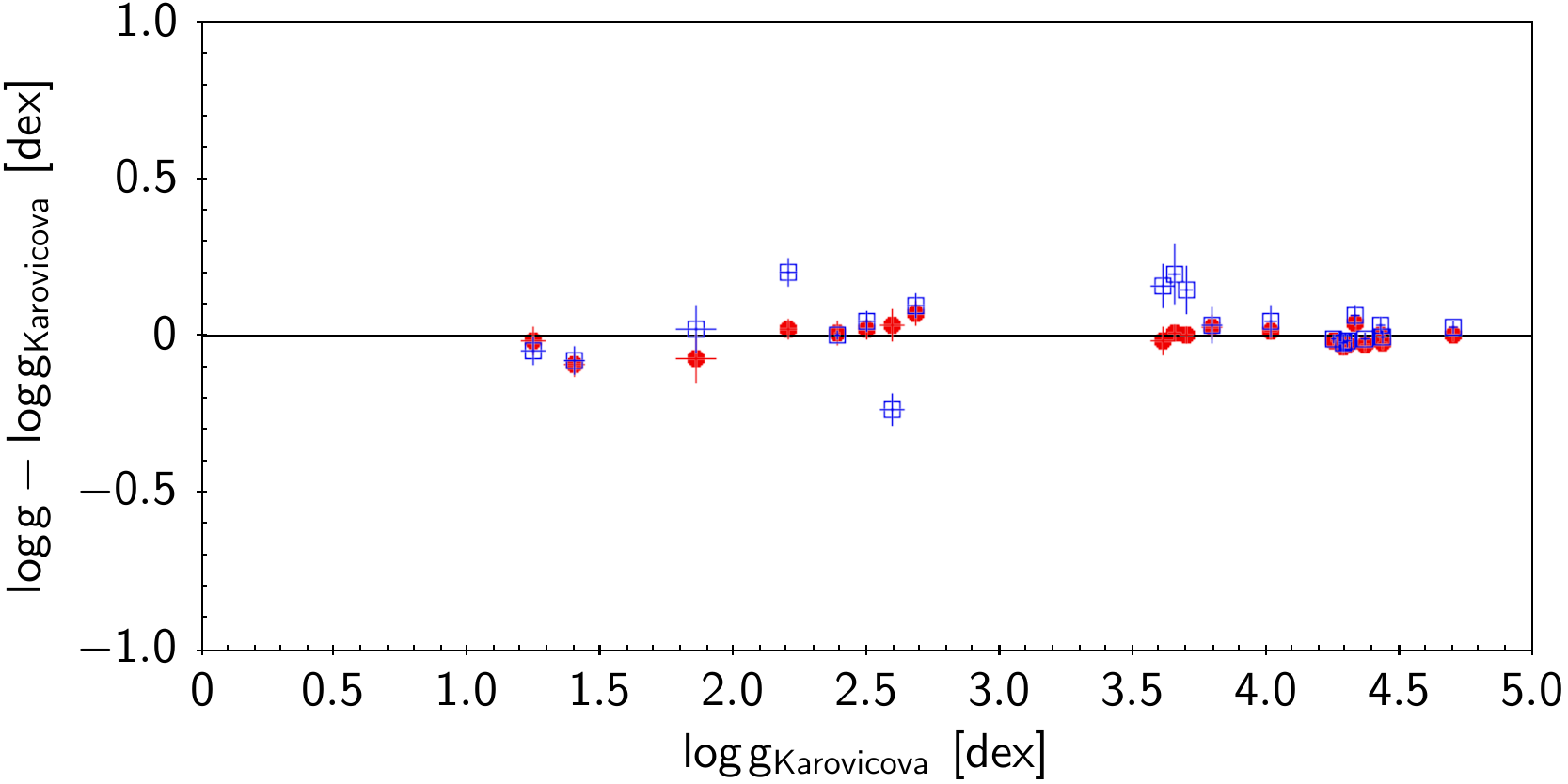}
\caption{Comparison of our fundamental values of \logg\ to those determined in Paper~I and by \cite{kar20,kar22a,kar22b}.  Red dots for SPInS masses using BaSTI models, blue open squares for STAREVOL models.}
 \label{f:comp_logg_others}
\end{figure}

We also compare our \logg\ determinations with those in Paper~I and in \cite{kar20,kar22a,kar22b} in Fig. \ref{f:comp_logg_others}. Offsets are negligible while dispersions (MAD) are 0.04 dex for Paper~I, whatever the tracks. The values of \logg\ generally agree well within the error bars except for one star, HIP37826 (Pollux), where the STAREVOL mass gives a discrepant \logg. The dispersions are 0.02 and 0.04 dex, using masses from BaSTI and STAREVOL respectively, for Karovicova et al.'s determinations based on masses obtained with Dartmouth stellar evolution tracks \citep{dot08}. 
There are two discrepant values with Karovicova et al.'s determinations: HIP98269 with the BaSTI mass, HIP70791 with the STAREVOL mass. We can draw similar conclusion as for the comparison to seismic \logg: a better agreement with one or the other set of evolutionary tracks is not obvious. Therefore, since there is no strong argument to adopt masses from BaSTI instead of STAREVOL and vice versa, we compute the average of the two values as the final \logg. 
This strategy allows us to mitigate some discrepancies among giants while it has no impact for most of the stars which have consistent masses whatever the used tracks.

\begin{figure}[h]
\centering
 \includegraphics[width=0.48\textwidth]{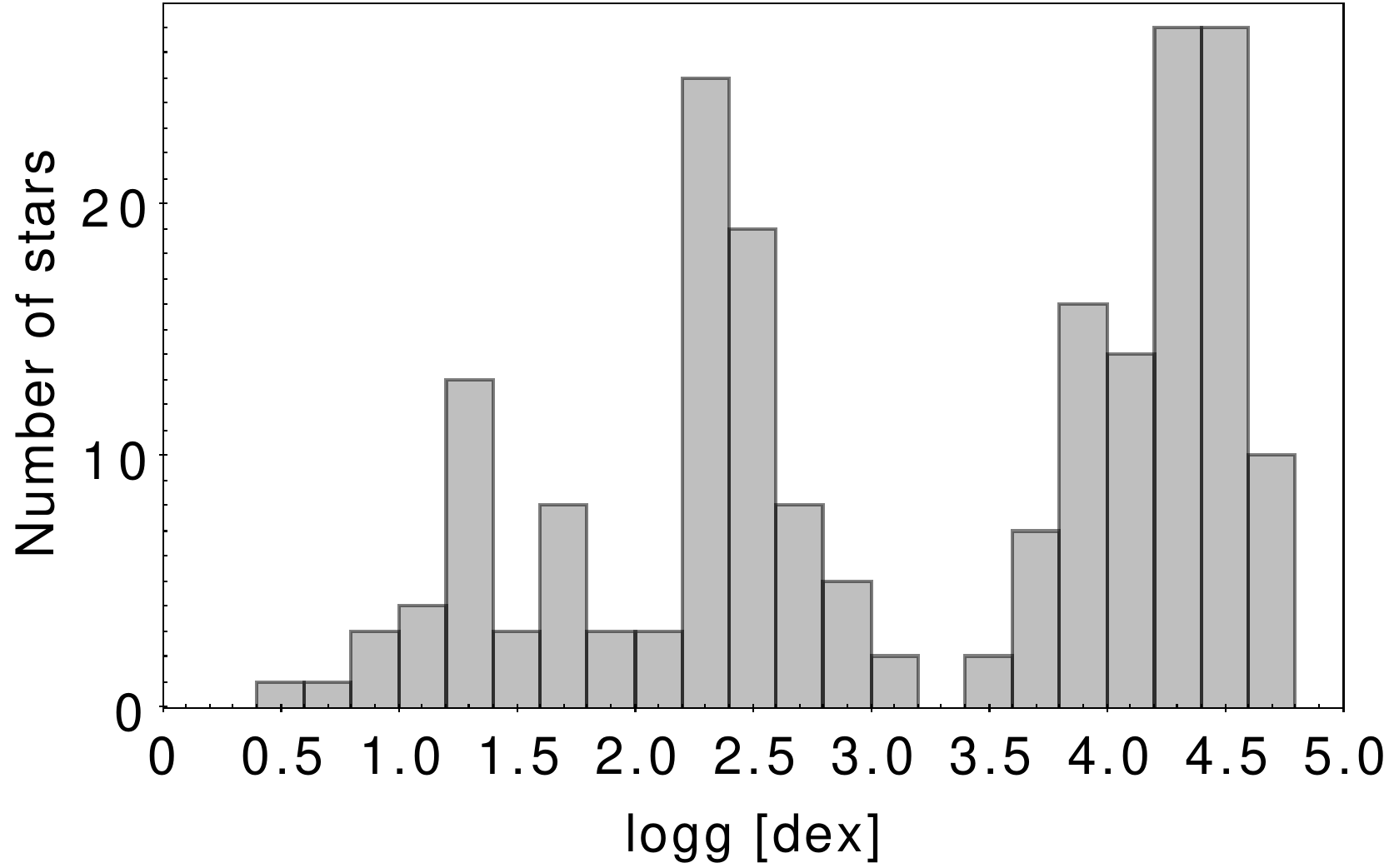} 
  \includegraphics[width=0.48\textwidth]{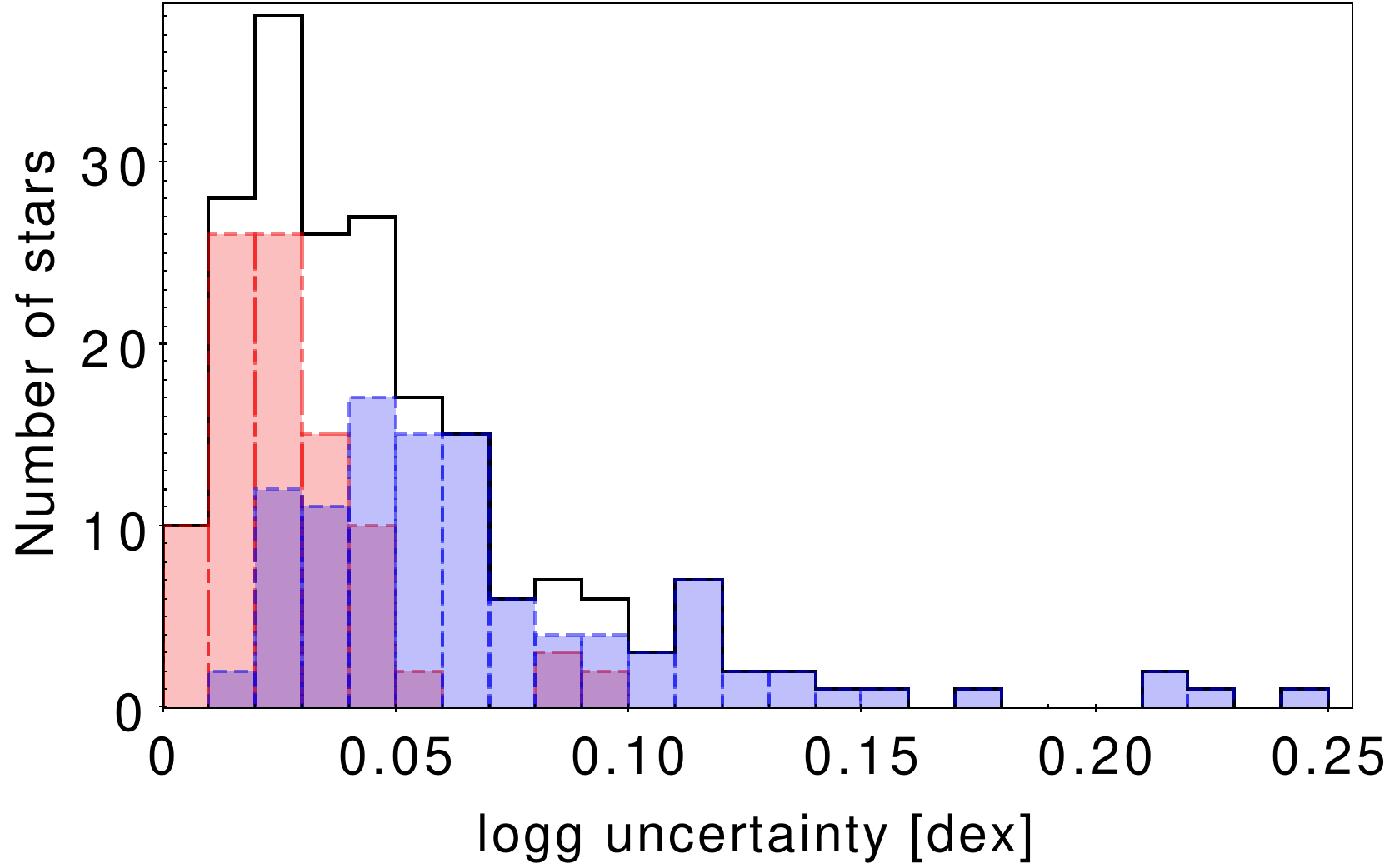} 
 \caption{Histogram of \logg\ (top panel) and uncertainty (bottom panel). The bottom panel shows the uncertainties for dwarfs (red) and giants (blue). }
 \label{f:logg_histo}
\end{figure}

The final \logg\ distribution and uncertainties are shown in Fig. \ref{f:logg_histo}. The bottom panel shows separately the uncertainties for dwarfs (\logg$>$3.8) and giants (\logg$\leq$3.8), highlighting the lower precision obtained for giants. The median uncertainty is 0.02~dex for dwarfs and 0.06~dex for giants. While 90\% of the dwarfs have an uncertainty lower than 0.05~dex, 90\% of the giants have an uncertainty higher than 0.03~dex. 

\section{The new set of Gaia FGK benchmark stars}
\label{s:sample}

The fundamental \teff\ and \logg\ determined for the 192 GBS V3 stars with a direct value of \diam\ are given in Table \ref{t:fund_teff_logg} of Appendix~\ref{s:appendix} while the nine other stars from V1 and V2 with an indirect \diam\ are provided in Table \ref{t:indirect}. The metallicity from the literature is provided for convenience, but will be redetermined homogeneously in the coming Paper VIII. The full catalogue with all the other parameters determined in this work is available in VizieR.

\begin{table}
\begin{center}
\caption{ \teff\ and \logg\ determined in this work for stars from V1 and V2 with an indirect value of \diam. \feh\ from the literature is given for indication.}
\label{t:indirect}
\begin{tabular}{r r c c r}
\hline\hline
HIP      &  HD    &   \teff  & \logg & \feh   \\
  &      &  (K)  &  (dex) & (dex)   \\
\hline
14086  &     18907      &     5143$\pm$ 56&         3.53$\pm$0.03      &         -0.63           \\
17147  &      22879     &     5962$\pm$ 86&         4.28$\pm$0.04      &         -0.84       \\
48152  &    84937       &     6484$\pm$106&         4.16$\pm$0.05      &         -2.12          \\
50382  &   298986        &    6343$\pm$ 43&         4.29$\pm$0.02      &         -1.33          \\
57360  &    102200       &    6205$\pm$ 45&         4.27$\pm$0.02      &         -1.23          \\
59490  &   106038        &    6172$\pm$ 42&         4.28$\pm$0.02      &         -1.33          \\
60172  &   107328        &    4576$\pm$ 87&         1.77$\pm$0.22      &         -0.38         \\
86796  &   160691        &    5974$\pm$ 60&         4.30$\pm$0.03      &         +0.29            \\
104659 &    201891     &      6040$\pm$ 44&         4.34$\pm$0.02      &         -1.02          \\
\hline
\end{tabular}
\end{center}
\end{table}

\begin{figure*}[h]
\centering
 \includegraphics[width=0.48\textwidth]{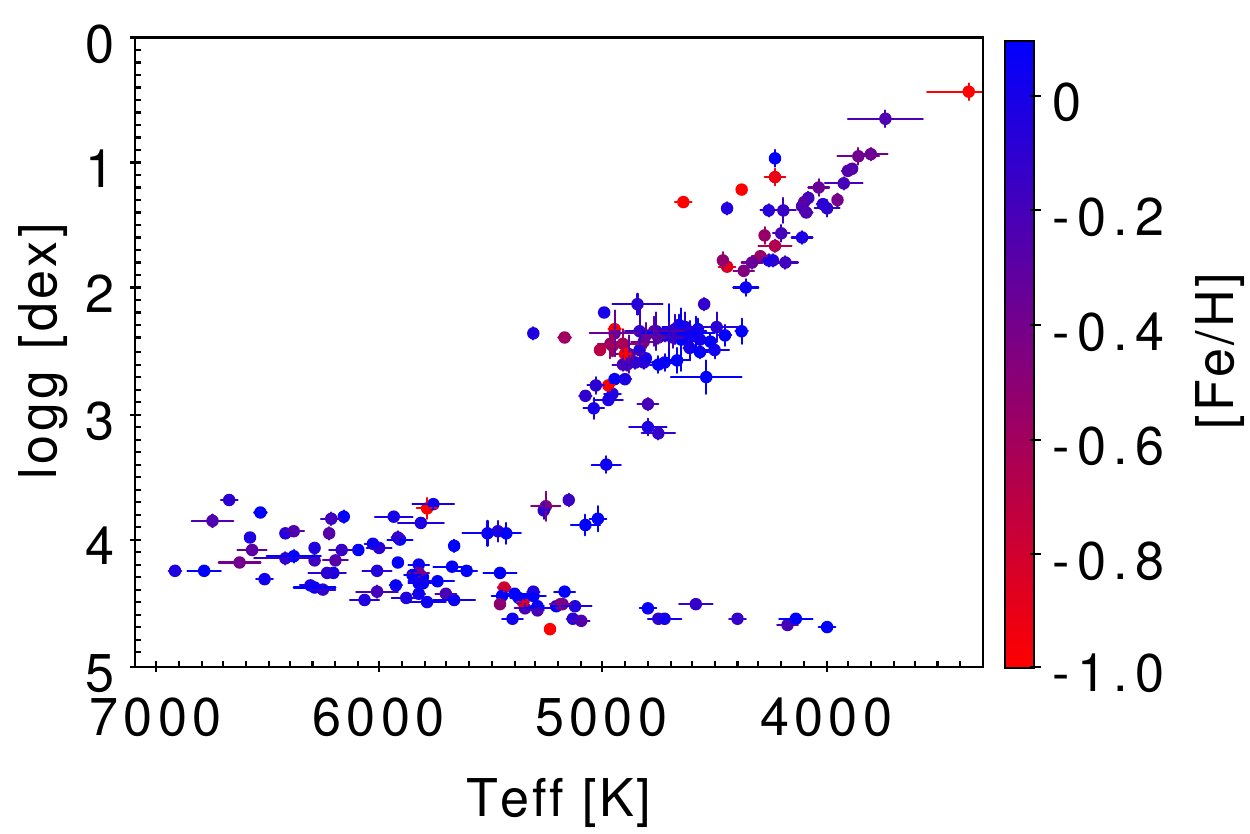} 
 \includegraphics[width=0.48\textwidth]{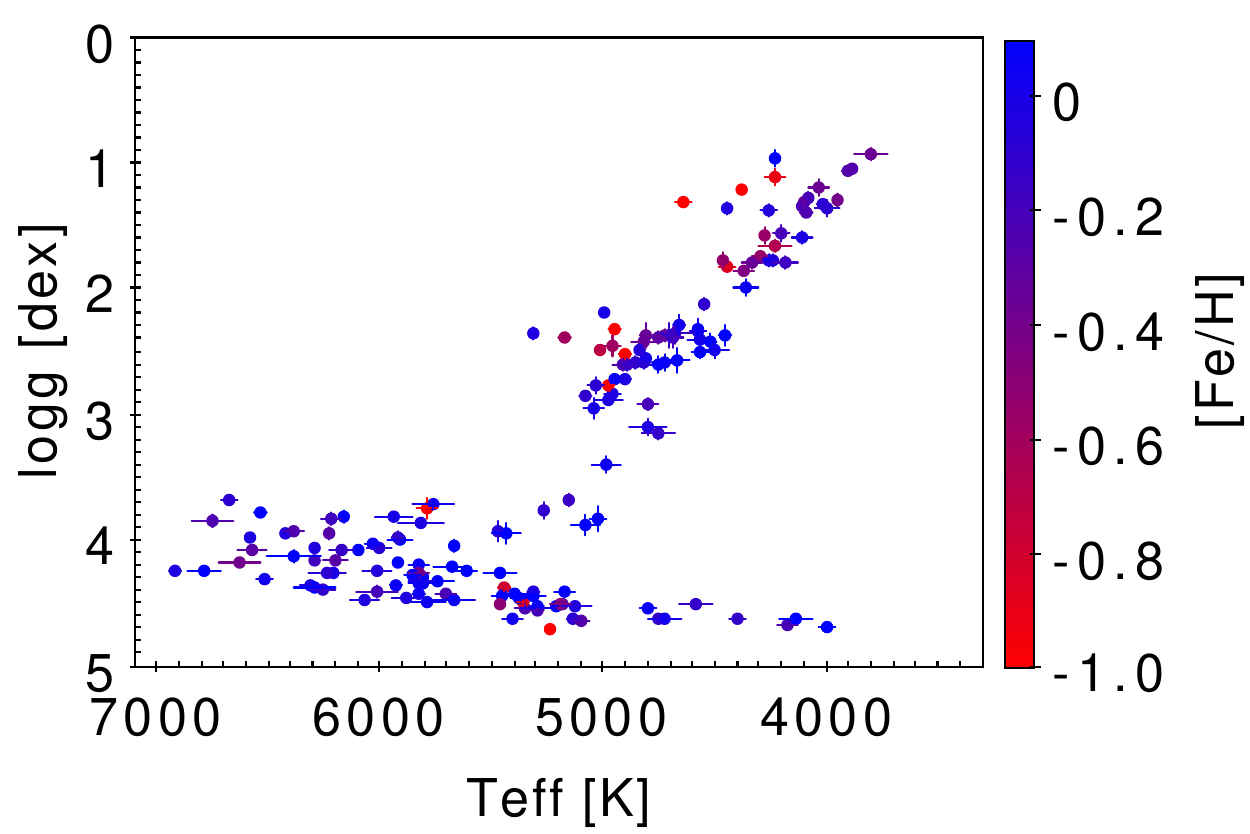} 
\caption{Kiel diagram with fundamental \teff\ and \logg. The colour scale is related to metallicities from the literature. The left panel shows the full sample of 192 stars while the right panel shows the stars with uncertainties on \teff\ and \logg\ better than 2\% and 0.1 dex respectively.}
 \label{f:kiel}
\end{figure*}

The Kiel diagram with fundamental \teff\ and \logg\ is shown in Fig. \ref{f:kiel} for the full sample of 192 stars and for a selection of the best stars, with an uncertainty on \teff\ and \logg\ better than 2\% and 0.1~dex, respectively. This selection of  165 stars mainly rejects giants with large uncertainties, as discussed in Sect. \ref{s:logg}, but still preserves a good distribution across the Kiel diagram. 

\begin{figure}[h]
\centering
\includegraphics[width=0.48\textwidth]{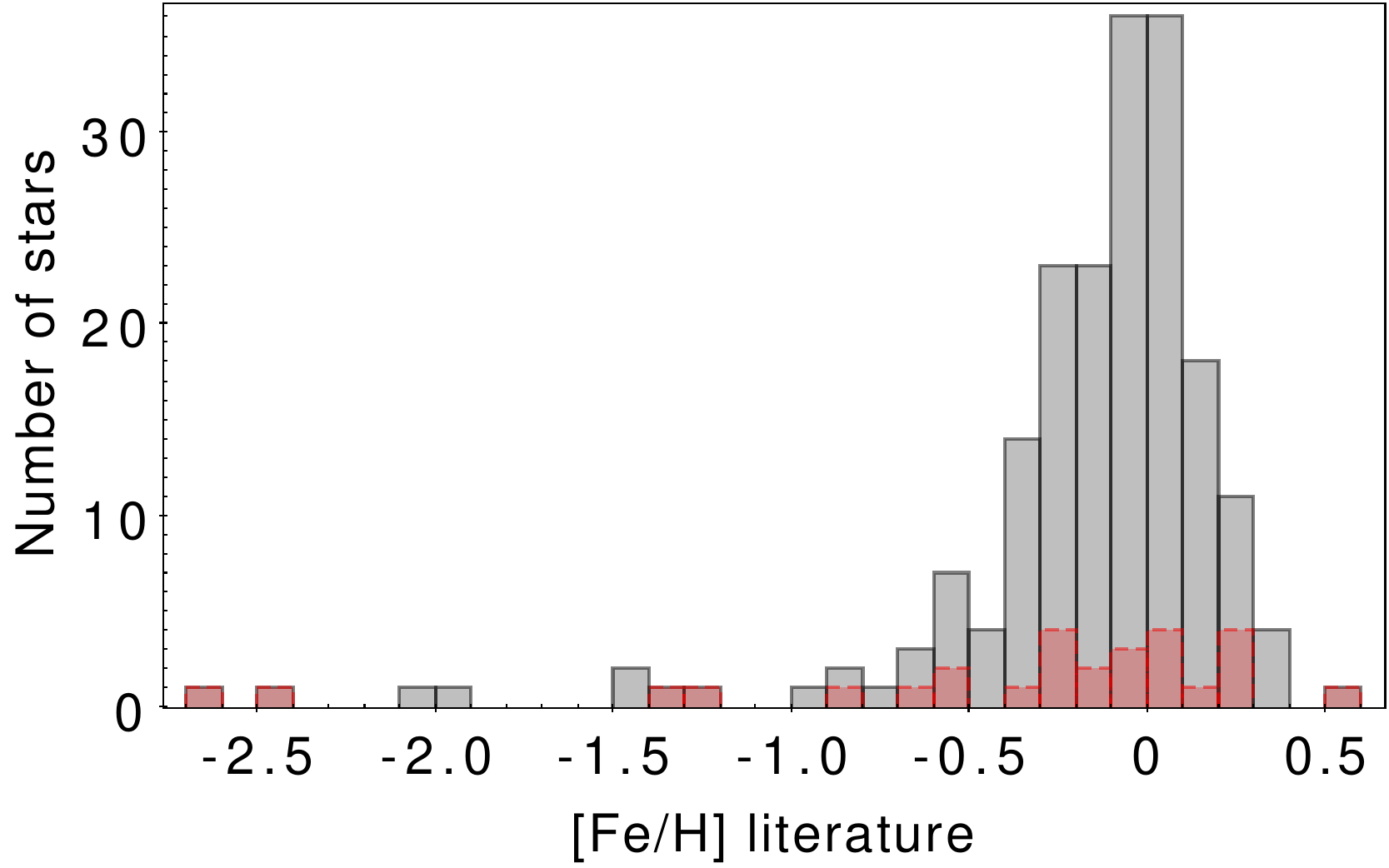} 
 \caption{Histogram of \feh\ for the 192 fundamental GBS V3 (grey) compared to the V1 version (red). }
 \label{f:feh_histo}
\end{figure}

The metallicity histogram of the GBS V3 is shown in Fig.~\ref{f:feh_histo}, compared to that of V1 (only considering stars with a direct \diam), highlighting a number of new metal-poor stars. This is however more evident in the interval $-1.0<$\feh$<-0.5$ than below \feh=$-1.0$. There were four stars in GBS V1 with $-1.0<$\feh$<-0.5$, a number increased to 14 in GBS V3. Four giant stars with \feh$<-1.0$ were added, which doubles the V1 number of stars with \feh$<-1.0$. The fundamental \teff\ and \logg\ for these eight GBS V3 stars are presented in Table \ref{t:mp} together with the values from Paper~I for the four stars in common. 

\begin{table*}[htbp]
\centering
\small{
\caption{Fundamental  \teff\ and \logg\ for the eight GBS V3 with \feh$<-1.0$ (from the literature), and comparison to the Paper~I values, when available. 
 }
\begin{tabular}{l | l | c | c | c | c | c   }
\hline\hline
HIP  &  other name  & \teff\ V3  &  \teff\ V1 & \logg\ V3 & \logg\ V1  & \feh\ \\
\hline
  HIP2413   & HD2665      & 4951$\pm$25  &   & 2.318$\pm$0.029 &   & -1.97 \\ 
  HIP5445   & HD6755      & 4977$\pm$27  &   & 2.767$\pm$0.027 &   & -1.43 \\ 
  HIP8837   & $\psi$ Phe  & 3362$\pm$183 & 3472$\pm$92  & 0.438$\pm$0.058 & 0.51$\pm$0.18 & -1.24 \\ 
  HIP57939  & HD103095    & 5235$\pm$18  & 4827$\pm$55  & 4.717$\pm$0.014 & 4.6$\pm$0.03 & -1.33 \\ 
  HIP68594  & HD122563    & 4642$\pm$35  & 4587$\pm$60  & 1.312$\pm$0.029 & 1.61$\pm$0.07 & -2.67 \\ 
  HIP76976  & HD140283    & 5788$\pm$45  & 5522$\pm$105 & 3.750$\pm$0.089 & 3.58$\pm$0.11 & -2.48 \\ 
  HIP92167  & HD175305    & 4902$\pm$30 &   & 2.533$\pm$0.017 &   & -1.45 \\ 
  HIP115949 & HD221170    & 4380$\pm$18  &   & 1.216$\pm$0.023 &   & -2.10 \\ 
\hline
\end{tabular}
\label{t:mp}
}
\end{table*}

One of the main purposes of the GBS is to calibrate or validate atmospheric parameters from spectroscopy. We therefore checked spectroscopic \teff\ and \logg\ available in different sources, using the subset of 165 most reliable GBS. 

\begin{figure}[h]
\centering
 \includegraphics[width=0.48\textwidth]{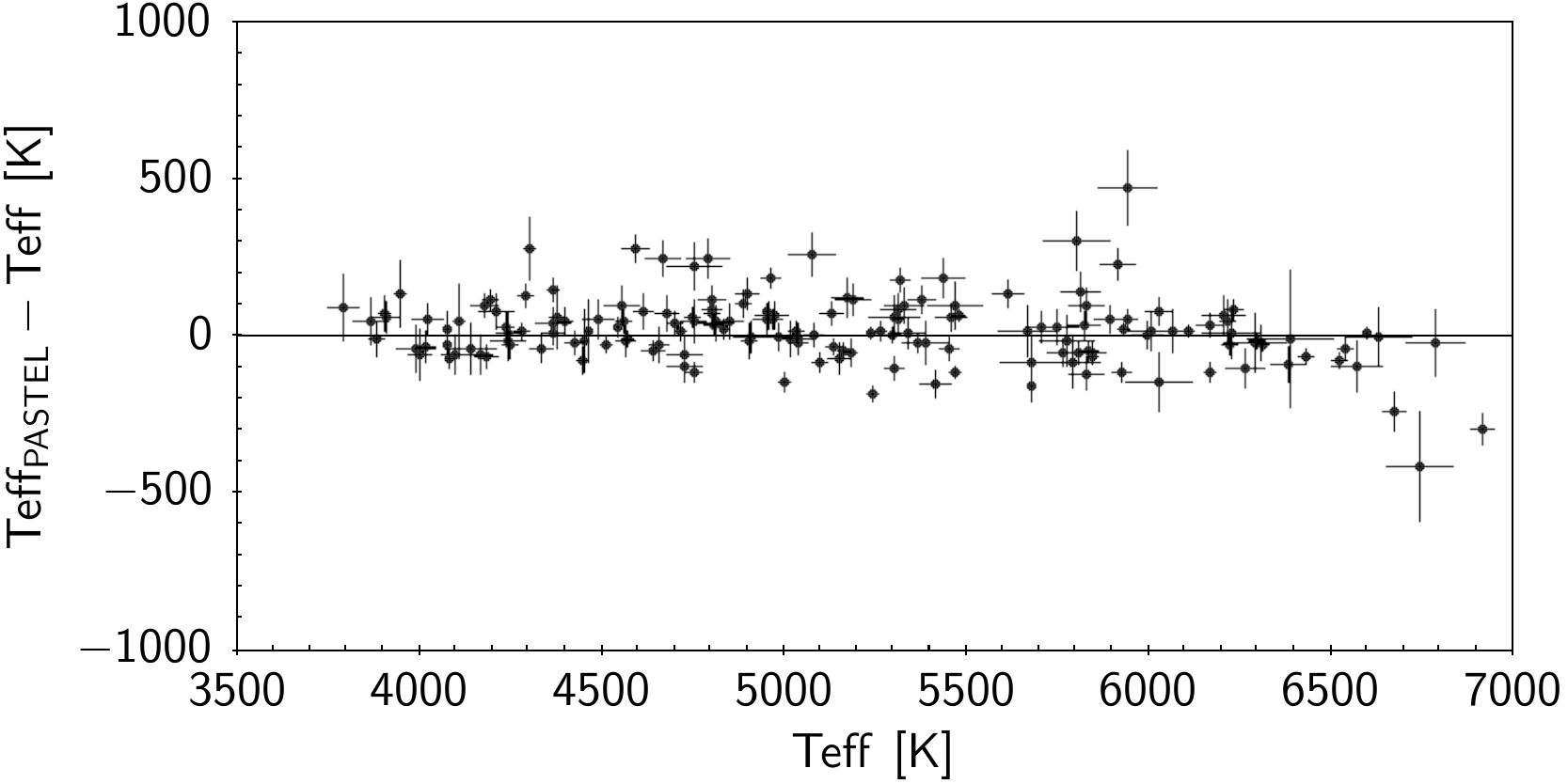} 
 \includegraphics[width=0.48\textwidth]{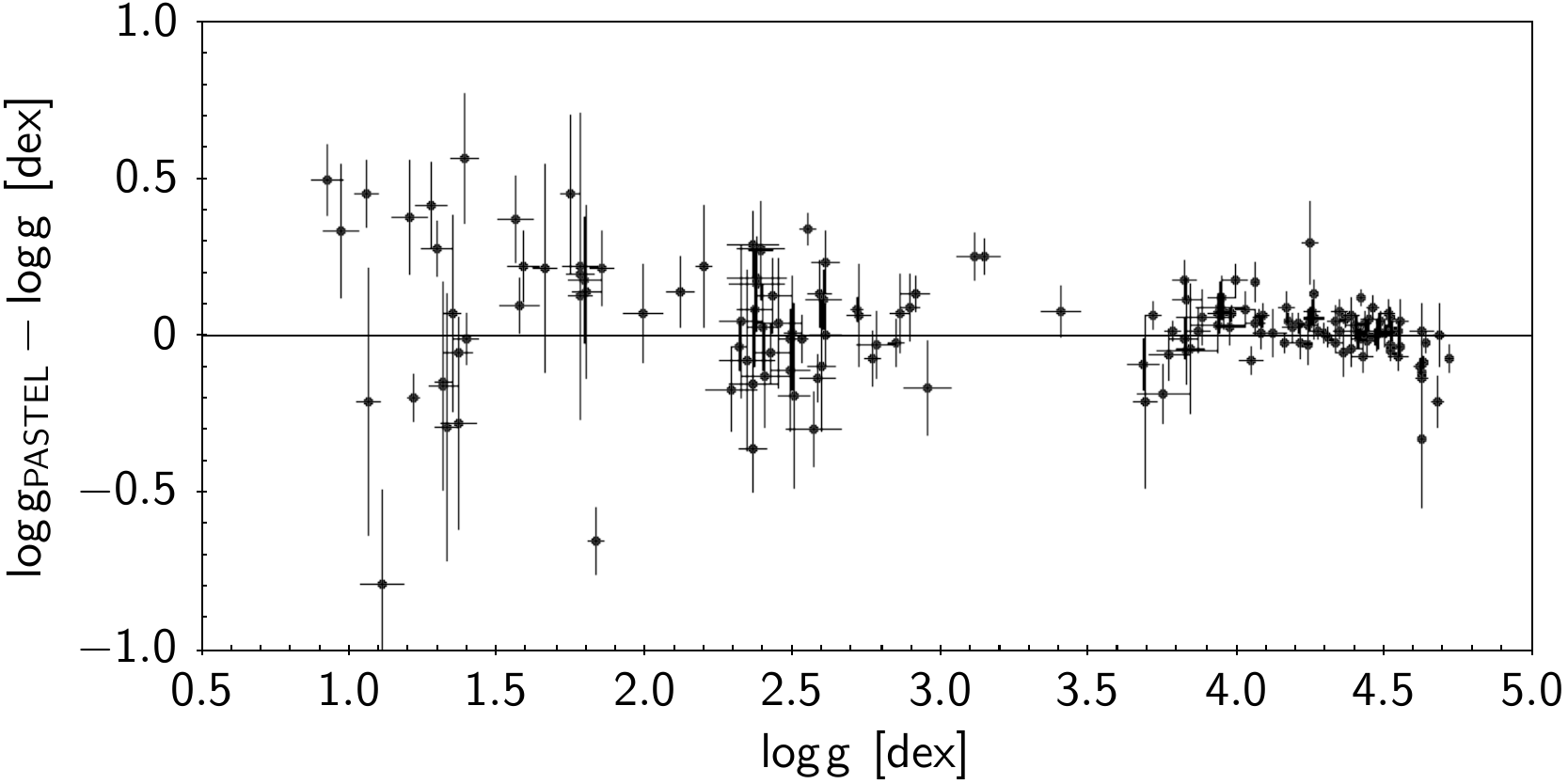} 
\caption{Comparison of our fundamental values of \teff\ (top panel) and \logg\ (bottom panel) to spectroscopic ones available in the PASTEL catalogue. }
 \label{f:comp_AP_lit}
\end{figure}

Fig. \ref{f:comp_AP_lit} compares our fundamental determinations with those available in the PASTEL catalogue, based on high-resolution, high signal-to-noise spectroscopy. Overall, the agreement on \teff\ is good with a dispersion of MAD=54 K and a slight offset of 12 K (median), the spectroscopic \teff\ being larger. Two extreme outliers have  differences larger than 400 K. For HIP86614 we suspect an uncertain angular diameter given its noisy squared visibility curve in \cite{boy12b} while for HIP108535 the only spectroscopic \teff\ in PASTEL is dubious. For that star we note a good agreement with the determination by \cite{pru11} based on a medium resolution spectrum. 

Concerning \logg\ we can see three regimes of precision, corresponding to dwarfs, clump giants, and cooler giants, with an increasing dispersion.  The dispersion among dwarfs is 0.04 dex (MAD). It rises to 0.1 dex among clump giants (2.0$<$\logg$<$3.5) with no offset, while for red giants there is a tendency of spectroscopic \logg\ to be larger than the fundamental ones by 0.16 dex (median offset) with a significant dispersion of 0.2 dex (MAD). The GBS can therefore be used to better understand and correct the spectroscopic gravities of evolved stars.

Focusing on the best studied stars we selected in the PASTEL catalogue the stars which are included in at least 15 spectroscopic studies at high resolution and high signal to noise ratio since 1990. The resulting 16 stars are all dwarfs or subgiants, with some of them also in common with Paper~I. In general there is a good agreement, within our uncertainties and the standard deviation from the literature values. Three stars, HIP14954, HIP57939 and HIP8159, exhibit a significant difference in \teff, larger than 150~K. 

HIP14954 (94 Cet) has been very much studied, with 41 spectroscopic determinations of \teff, likely because of its exoplanet discovered in 2000 \citep{que01}. The literature values range from 5916~K to 6424~K with a mean of 6176~K and a standard deviation of 84~K. Our determination is lower, \teff=5912$\pm$59~K, but still in agreement with the coolest spectroscopic determinations. Our fundamental value is in a very good agreement with that of \cite{boy13}, \teff=5916$\pm$98~K, independent from ours since we use the angular diameter from \cite{lig16}. It would be important to better understand why spectroscopy gives a higher \teff\ for that star because it has implications on the parameters of its exoplanet.

HIP8159 (109 Psc) also hosts an exoplanet and has several recent \teff\ from high resolution spectroscopy ranging between 5560~K and 5711~K. The fundamental determinations, from \cite{boy13} and from us (\teff=5438$\pm$61~K) based on the same \diam, are cooler than the mean spectroscopic value by $\sim$200~K. This discrepancy requires further investigation.

HIP57939 (HD103095) has 57 spectroscopic determinations of \teff\ after 1990, ranging from 4500~K to 5250~K with a mean of 
5057~K and a standard deviation of 18~K. Our fundamental determination \teff=5235$\pm$18~K is in agreement with the hottest spectroscopic determinations, e.g. by \cite{luc05}.

\begin{figure}[h]
\centering
 \includegraphics[width=0.48\textwidth]{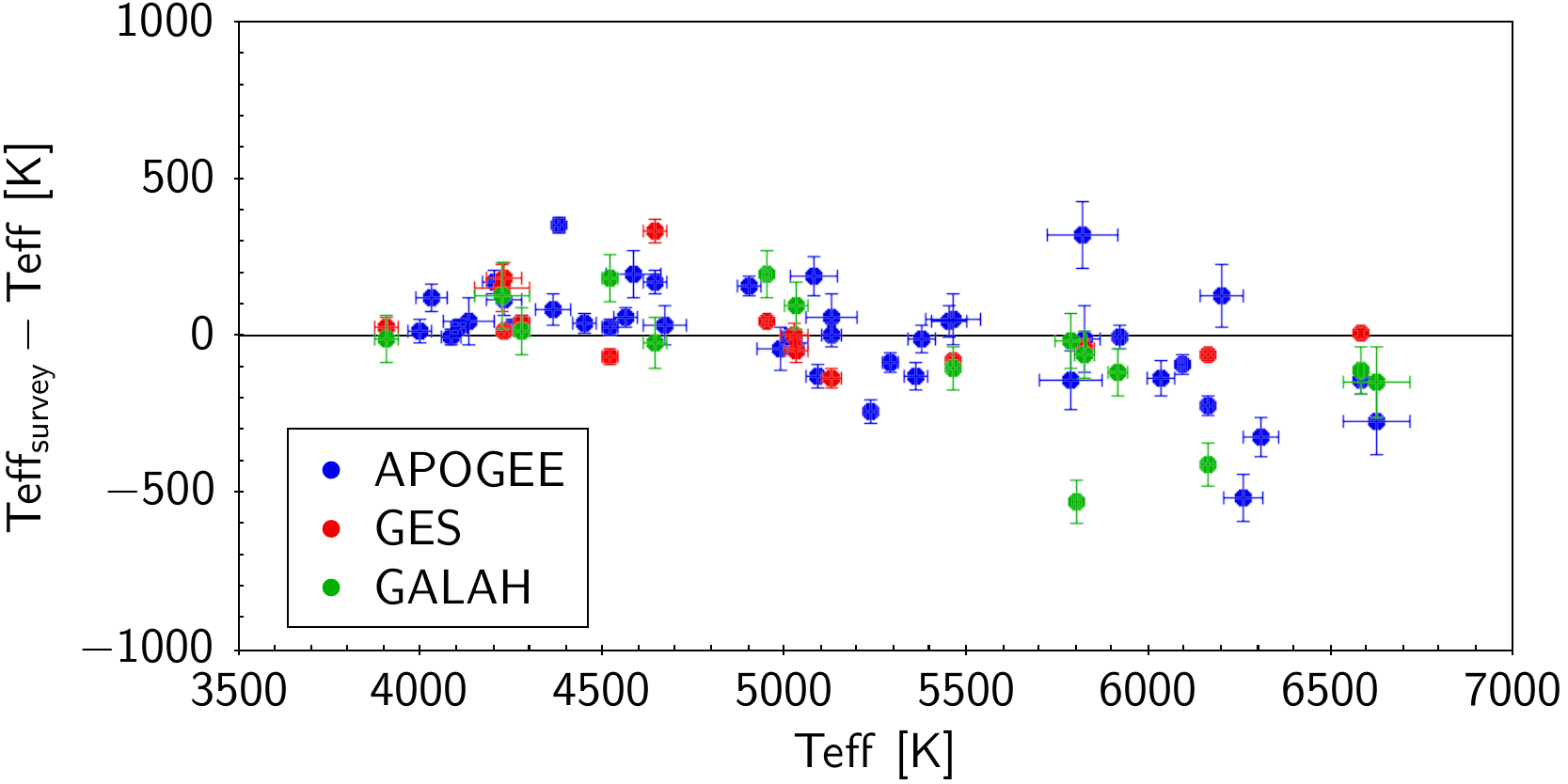} 
 \includegraphics[width=0.48\textwidth]{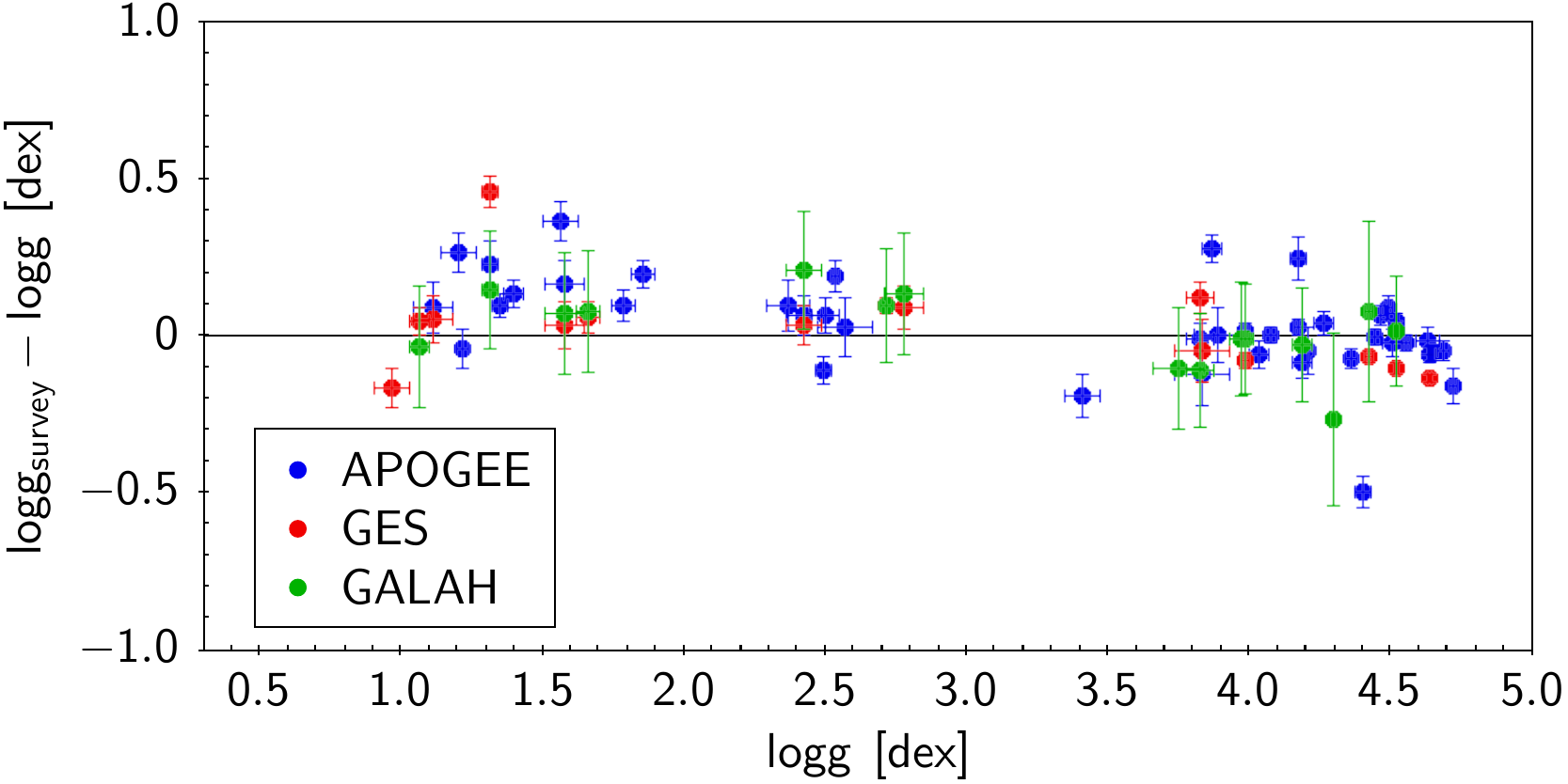} 
\caption{Comparison of \teff\ and \logg\ from this work with the spectroscopic ones in surveys: APOGEE DR17 (blue), Gaia-ESO (red), GALAH DR3 (green).}
 \label{f:comp_surveys}
\end{figure}

We also checked atmospheric parameters massively determined by large spectroscopic surveys against our fundamental determinations of the best GBS. We considered APOGEE DR17 \citep{maj17,apo}, the Gaia-ESO survey \citep{ran22,gil22} and GALAH DR3 \citep{bud21}  in the comparisons shown in Fig. \ref{f:comp_surveys}. Table \ref{t:comp_surveys} gives the median offsets and corresponding MAD for dwarfs and giants separately. Although GALAH and Gaia-ESO have less stars in common than APOGEE, we see the same trends in the three surveys. Their \teff\ and \logg\ for dwarfs are smaller on average than the fundamental ones, and vice-versa for the giants. These trends are worth to be investigated and better understood.

\begin{table}
\caption[]{Median difference (MED) and median absolute deviation (MAD) between our fundamental determinations of \teff\ and \logg\ and the spectroscopic ones from surveys (survey results minus our results), for N stars in common.}
\label{t:comp_surveys}
\begin{tabular}{l | r | r r | r r }
\hline
\noalign{\smallskip}
Sample & N & \multicolumn{2}{c|}{$\Delta$\teff} & \multicolumn{2}{c|}{$\Delta$\logg}  \\ 
       &   &   MED & MAD & MED & MAD \\ 
\noalign{\smallskip}
\hline
\noalign{\smallskip}
APOGEE dwarfs  & 26 & -21 & 111 & -0.02 & 0.04 \\
APOGEE giants  & 17 & 38 & 42 & 0.09 & 0.07 \\
\noalign{\smallskip}
Gaia-ESO dwarfs& 6 & -52 & 38 & -0.08 & 0.03 \\
Gaia-ESO giants& 9 & 39 & 92 & 0.05 & 0.02 \\
\noalign{\smallskip}
GALAH dwarfs   & 7 & -119 & 32 & -0.01 & 0.02 \\
GALAH  giants  & 8 & 53 & 73 & 0.08 & 0.05 \\
\noalign{\smallskip}
\hline
\end{tabular}
\end{table}

\begin{figure}[h]
\centering
 \includegraphics[width=0.48\textwidth]{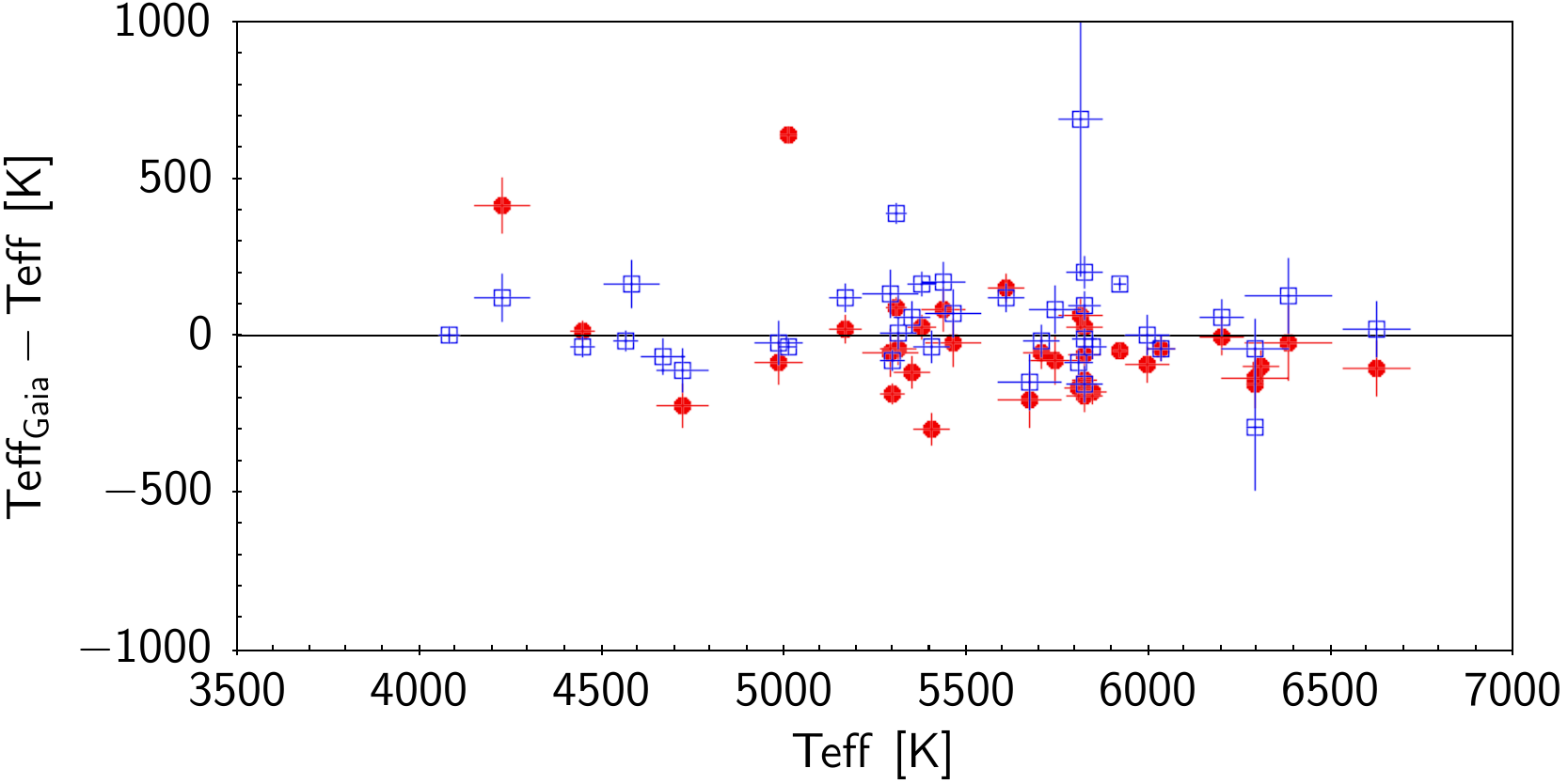} 
 \includegraphics[width=0.48\textwidth]{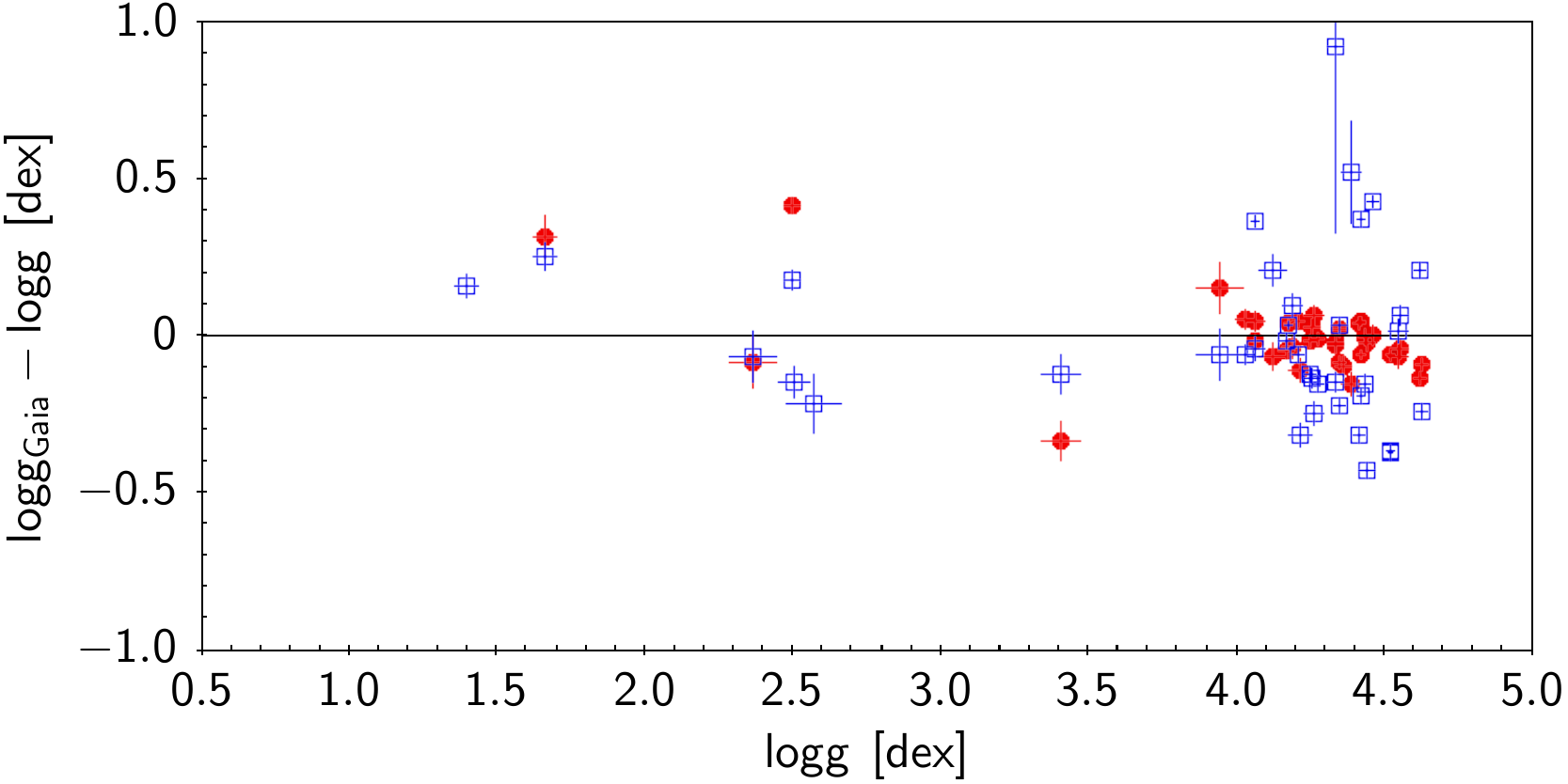} 
\caption{Comparison of \teff\ and \logg\ from this work with the photometric (red dots) and spectroscopic (blue squares) ones from the Gaia DR3 Golden Sample of Astrophysical Parameters \citep{gold}. }
 \label{f:comp_gold}
\end{figure}

Finally we also assessed the photometric and spectroscopic \teff\ and \logg\ of the Gaia DR3 Golden Sample of Astrophysical Parameters \citep{gold} with the best GBS, as shown in Fig.~\ref{f:comp_gold}. Photometric \teff\ are lower than fundamental ones by 58~K, while the offset of the spectroscopic \teff\ is negligible  (6~K). For \logg\ there is an excellent agreement of the photometric values with median offset of $-0.03$~dex and a dispersion (MAD) of 0.06~dex. Spectroscopic \logg, corrected as suggested by \cite{GSP-Spec}, are found smaller than the fundamental values by 0.06~dex, with a dispersion of 0.15~dex. These comparisons are based on 35 and 38 stars in common for the photometric and spectroscopic parameters respectively, mainly dwarfs.

\section{Conclusion}
\label{s:conclusion}
Large spectroscopic surveys usually calibrate or validate their determinations of atmospheric parameters using reference stars. Ideally they should adopt a common \teff\ and \logg\ scale in order to minimize systematic differences in abundances provided by different instruments and pipelines. The GBS are intended to provide such an anchor to the fundamental  \teff\ and \logg. GBS are also suited to help understand any shortcomings in the stellar models.

We have presented determinations of fundamental \teff\ and \logg, based on the Stefan-Boltzmann law and Newton's law of gravitation, for the third version of the GBS. Compared to the previous V1 and V2 versions, a significant improvement is the larger number of stars, 192 instead of $\sim$40, resulting from our systematic search of high quality angular diameters based on interferometric measurements. More accurate \logg\ were obtained thanks to the higher precision parallaxes which mostly come from Gaia DR3, while the improved \teff\ are in part a result of the homogenous photometric data from Gaia DR3, which feed into the \Fbol\ determination. \Fbol\ are now more precise and homogeneous owing to the methodology of SED fitting applied to a combination of photometric and spectrophotometric data including measurements made on BP/RP spectra from Gaia DR3. Better \Fbol\ also result from the adopted extinction values deduced from state-fo-the-art 3D maps of the solar neighbourhood. The most difficult part comes from the determination of masses from evolutionary tracks. We have shown that using two different grids can lead to differences of up to more than 50\% in masses, giving systematic offsets of about 0.06 dex in \logg\ among giants.   

At each stage of the compilation and determination of the parameters, we evaluated the uncertainties that we aimed to keep at the 1-2\% level. Our results were assessed by comparing them to other determinations of similar accuracy available in the literature. In general the comparison with literature data is satisfactory, with differences not exceeding 4\%. We can explain most of the outliers. We also determined seismic radii, masses and surface gravities for comparisons, using scaling relations and seismic parameters available for $\sim$40 stars. They show a good agreement for dwarfs but a trend in masses outside the 1-1.5M$_\odot$ range. From the different comparisons we are confident that our uncertainties in \teff\ are reliable. We reach the expected 1-2\% level in \teff. For \logg\ only dwarfs have such a level of accuracy. Uncertainties for giants are larger and reflect the difficulty to obtain reliable masses for them from evolutionary tracks.

The \teff\ and \logg\ presented here will be updated. Two steps of our determination process, \Fbol\ and masses, depend on \feh\ which we took from the literature. This is the subject of the coming paper VIII to determine abundances of the GBS-V3 from a large collection of high quality spectra. Some iterations will be needed to adjust \teff, \logg\ and \feh\ in a self-consistent way. In the meantime, we have evaluated the impact of using non-homogeneous metallicities through tests in which we modified the values of \feh\ and uncertainties by 0.15 dex in input of the SED fitting and of SPInS. We found that it has a negligible impact on \teff, and also on \logg\ for most of the stars, although a few giants have their mass affected by more than 30\% inducing a change of \logg\ by 0.11 to 0.5 dex.

In order to use the GBS V3 for calibration or validation of atmospheric parameters, we recommend the users to select the 165 stars with uncertainties on \teff\ and \logg\ lower than 2\% and 0.1~dex, respectively. We have used this subsample to assess \teff\ and \logg\ obtained by high-resolution and high-signal to noise spectroscopy (PASTEL catalogue), by medium-resolution spectroscopy (APOGEE, GALAH and Gaia-ESO surveys), and by Gaia photometry and spectroscopy. This has revealed some issues that need to be investigated to improve the future releases.

Due to the lack of metal-poor stars in the solar neighbourhood the GBS V3 do not yet cover the metallicity range in a uniform way. 
We still lack angular diameters for stars with \feh$<-1.0$ which are important targets in galactic archeology and stellar physics.
Interferometric measurements are still limited to stars brighter than V$\sim$8, and larger than \diam$\simeq$0.2~mas. There are however metal-poor candidates which are bright and large enough to fulfil these criteria. They could be observed with powerful interferometers, such as the new SPICA instrument on the CHARA array \citep{mou22} expected to provide an estimation of the stellar radius of such stars to 1\% precision. It would also be useful to remeasure partly or totally the GBS with \diam$<$1.2~mas, which show a large dispersion of the current measurements, exceeding the quoted uncertainties.

\begin{acknowledgements}
 CS, NL, and LC acknowledge financial support from "Programme National de Physique Stellaire" (PNPS) and from the "Programme National Cosmology et Galaxies (PNCG)" of CNRS/INSU, France. CS, PJ and LC acknowledge financial support from the French-Chilean program of cooperation ECOS C18U02 (ECOS-ANID 180049). We acknowledge support from FONDECYT Regular grant 1231057, 1200703, and Millenium Nucleus ERIS NCN2021\_017, Centros ANID Iniciativa Milenio.
 LC acknowledges the grant RYC2021-033762-I funded by MCIN/AEI/10.13039/501100011033 and by the European Union NextGenerationEU/PRTR.
U.H. acknowledges support from the Swedish National Space Agency (SNSA/Rymdstyrelsen).
DDBS acknowledges financial support from Becas-ANID scholar- ship 21220843. 

This work has made use of data from the European Space Agency (ESA) mission
{\it Gaia} (\url{https://www.cosmos.esa.int/gaia}), processed by the {\it Gaia}
Data Processing and Analysis Consortium (DPAC,
\url{https://www.cosmos.esa.int/web/gaia/dpac/consortium}). Funding for the DPAC
has been provided by national institutions, in particular the institutions
participating in the {\it Gaia} Multilateral Agreement. The preparation of this work has made extensive use of Topcat \citep{tay05}, of the Simbad and VizieR databases at CDS, Strasbourg, France, and of NASA's Astrophysics Data System Bibliographic Services. This publication makes use of VOSA, developed under the Spanish Virtual Observatory (https://svo.cab.inta-csic.es) project funded by MCIN/AEI/10.13039/501100011033/ through grant PID2020-112949GB-I00.
VOSA has been partially updated by using funding from the European Union's Horizon 2020 Research and Innovation Programme, under Grant Agreement no. 776403 (EXOPLANETS-A). The Stellar Parameters INferred Systematically (SPInS) code is a spin-off of the 
Asteroseismic Inference 
on a Massive Scale project,
one of the deliverables of the SpaceINN network, funded by the European Community's Seventh Framework Programme (FP7/2007-2013) under grant agreement no. 312844. SPInS was initially created for the 5th International Young Astronomer School Scientific Exploitation of Gaia Data held in Paris, (26 February - 2 March 2018), as a simple tool to estimate stellar ages, as well as other stellar properties.
\end{acknowledgements}

\bibliographystyle{aa}
\bibliography{main}


\begin{appendix} 

\section{Table with \teff\ and \logg\ of the GBS V3}
\label{s:appendix}

\onecolumn
\begin{longtable}{llcccccc}
\caption{Fundamental \teff\ and \logg\ and their uncertainties determined in this work for the 192 GBS V3, with \feh\ from the literature and the $\theta_{\rm LD}$ adopted measurement.}
\label{t:fund_teff_logg}
\tabularnewline
\hline
\hline
HIP     &  HD & $\theta_{\rm LD}$    &  Reference for $\theta_{\rm LD}$  &    \teff   &   ru\_\teff &  \logg   &      \feh      \\
      &      &  (mas)    &        &   (K)    &  (\%)   &     (dex)    &  (dex)   \\
\hline
\endfirsthead
\caption{continued.}\\
\hline\hline
HIP     &  HD & $\theta_{\rm LD}$  &   Reference for $\theta_{\rm LD}$    &     \teff   &   ru\_\teff &  \logg   &   \feh      \\
      &      &  (mas)   &       &   (K)     &  (\%)  &  (dex)    &  (dex)   \\
\hline
\endhead
\hline
\endfoot

HIP101345	&	HD195564	&	0.712	$\pm$	0.03	&	2013ApJ...771...40B	&	5514	$\pm$	116	&	2.1	&	3.95	$\pm$	0.09	&	0.05	\\
HIP10234	&	HD13468	&	0.886	$\pm$	0.01	&	2018A\&A...616A..68G	&	4676	$\pm$	80	&	1.7	&	2.37	$\pm$	0.09	&	-0.13	\\
HIP102422	&	HD198149	&	2.882	$\pm$	0.088	&	2016ApJS..227....4H	&	4751	$\pm$	73	&	1.5	&	3.15	$\pm$	0.05	&	-0.13	\\
HIP103598	&	HD200205	&	2.032	$\pm$	0.043	&	2010ApJ...710.1365B	&	4032	$\pm$	45	&	1.1	&	1.20	$\pm$	0.06	&	-0.36	\\
HIP104214	&	HD201091	&	1.775	$\pm$	0.013	&	2008A\&A...488..667K	&	4398	$\pm$	34	&	0.8	&	4.63	$\pm$	0.01	&	-0.13	\\
HIP104217	&	HD201092	&	1.581	$\pm$	0.022	&	2008A\&A...488..667K	&	4174	$\pm$	47	&	1.1	&	4.68	$\pm$	0.02	&	-0.21	\\
HIP106039	&	HD204381	&	1.524	$\pm$	0.017	&	2018A\&A...616A..68G	&	5079	$\pm$	31	&	0.6	&	2.86	$\pm$	0.04	&	-0.11	\\
HIP108535	&	HD209369	&	0.621	$\pm$	0.017	&	2016A\&A...586A..94L	&	6754	$\pm$	93	&	1.4	&	3.84	$\pm$	0.05	&	-0.24	\\
HIP108870	&	HD209100	&	1.758	$\pm$	0.012	&	2020MNRAS.493.2377R	&	4754	$\pm$	35	&	0.7	&	4.62	$\pm$	0.01	&	-0.13	\\
HIP109176	&	HD210027	&	1.206	$\pm$	0.053	&	2009ApJ...694.1085V	&	6419	$\pm$	141	&	2.2	&	4.15	$\pm$	0.06	&	-0.11	\\
HIP109937	&	HD211388	&	3.371	$\pm$	0.049	&	2018AJ....155...30B	&	4258	$\pm$	41	&	1.0	&	1.39	$\pm$	0.05	&	-0.05	\\
HIP11095	&	HD15248	&	0.949	$\pm$	0.019	&	2018A\&A...616A..68G	&	4721	$\pm$	52	&	1.1	&	2.59	$\pm$	0.06	&	0.06	\\
HIP111944	&	HD214868	&	2.731	$\pm$	0.02	&	2010ApJ...710.1365B	&	4203	$\pm$	34	&	0.8	&	1.56	$\pm$	0.06	&	-0.20	\\
HIP112440	&	HD215665	&	2.26	$\pm$	0.1	&	1999AJ....118.3032N	&	4848	$\pm$	109	&	2.2	&	2.12	$\pm$	0.08	&	-0.07	\\
HIP112447	&	HD215648	&	1.091	$\pm$	0.008	&	2012ApJ...746..101B	&	6223	$\pm$	23	&	0.4	&	3.95	$\pm$	0.04	&	-0.27	\\
HIP112731	&	HD216174	&	1.5980	$\pm$	0.0120	&	2016AJ....152...66B	&	4297	$\pm$	36	&	0.8	&	1.75	$\pm$	0.03	&	-0.55	\\
HIP112748	&	HD216131	&	2.496	$\pm$	0.04	&	2003AJ....126.2502M	&	4961	$\pm$	40	&	0.8	&	2.85	$\pm$	0.03	&	-0.03	\\
HIP113357	&	HD217014	&	0.685	$\pm$	0.011	&	2013ApJ...771...40B	&	5746	$\pm$	72	&	1.3	&	4.33	$\pm$	0.02	&	0.18	\\
HIP114622	&	HD219134	&	1.106	$\pm$	0.007	&	2012ApJ...757..112B	&	4800	$\pm$	34	&	0.7	&	4.55	$\pm$	0.01	&	0.06	\\
HIP114855	&	HD219449	&	2.22	$\pm$	0.031	&	2018AJ....155...30B	&	4631	$\pm$	69	&	1.5	&	2.32	$\pm$	0.11	&	-0.03	\\
HIP114971	&	HD219615	&	2.3400	$\pm$	0.0400	&	2015A\&A...573A.138B	&	4970	$\pm$	43	&	0.9	&	2.44	$\pm$	0.12	&	-0.53	\\
HIP115227	&	HD220009	&	2.045	$\pm$	0.034	&	2015A\&A...582A..49H	&	4227	$\pm$	77	&	1.8	&	1.66	$\pm$	0.04	&	-0.66	\\
HIP115620	&	HD220572	&	1.092	$\pm$	0.013	&	2018A\&A...616A..68G	&	4756	$\pm$	41	&	0.9	&	2.61	$\pm$	0.06	&	0.08	\\
HIP115949	&	HD221170	&	0.596	$\pm$	0.005	&	2020A\&A...640A..25K	&	4380	$\pm$	19	&	0.4	&	1.22	$\pm$	0.02	&	-2.10	\\
HIP116771	&	HD222368	&	1.082	$\pm$	0.009	&	2012ApJ...746..101B	&	6169	$\pm$	56	&	0.9	&	4.09	$\pm$	0.02	&	-0.14	\\
HIP12114	&	HD16160	&	1.03	$\pm$	0.007	&	2012ApJ...757..112B	&	4583	$\pm$	76	&	1.6	&	4.52	$\pm$	0.01	&	-0.13	\\
HIP12486	&	HD16815	&	2.248	$\pm$	0.014	&	2018A\&A...616A..68G	&	4720	$\pm$	23	&	0.5	&	2.37	$\pm$	0.06	&	-0.36	\\
HIP12530	&	HD16765A	&	0.497	$\pm$	0.007	&	2013ApJ...771...40B	&	6310	$\pm$	51	&	0.8	&	4.36	$\pm$	0.02	&	-0.02	\\
HIP12777	&	HD16895	&	1.103	$\pm$	0.008	&	2012ApJ...746..101B	&	6206	$\pm$	23	&	0.4	&	4.26	$\pm$	0.01	&	0.01	\\
HIP13288	&	HD17824	&	1.391	$\pm$	0.015	&	2018A\&A...616A..68G	&	4980	$\pm$	64	&	1.3	&	2.89	$\pm$	0.03	&	-0.02	\\
HIP13328	&	HD17709	&	4.056	$\pm$	0.041	&	2003AJ....126.2502M	&	3799	$\pm$	72	&	1.9	&	0.93	$\pm$	0.06	&	-0.36	\\
HIP14060	&	HD18784	&	1.036	$\pm$	0.014	&	2018A\&A...616A..68G	&	4652	$\pm$	80	&	1.7	&	2.30	$\pm$	0.12	&	0.00	\\
HIP14135	&	HD18884	&	12.2	$\pm$	0.04	&	2006A\&A...460..855W	&	3738	$\pm$	170	&	4.5	&	0.66	$\pm$	0.07	&	-0.24	\\
HIP14632	&	HD19373	&	1.246	$\pm$	0.007	&	2012ApJ...746..101B	&	5921	$\pm$	17	&	0.3	&	4.17	$\pm$	0.01	&	0.09	\\
HIP14838	&	HD19787	&	1.87	$\pm$	0.12	&	1999AJ....118.3032N	&	4703	$\pm$	161	&	3.4	&	2.36	$\pm$	0.23	&	0.11	\\
HIP14954	&	HD19994	&	0.761	$\pm$	0.01	&	2016A\&A...586A..94L	&	5912	$\pm$	59	&	1.0	&	4.00	$\pm$	0.04	&	0.20	\\
HIP15457	&	HD20630	&	0.936	$\pm$	0.024	&	2012ApJ...746..101B	&	5786	$\pm$	87	&	1.5	&	4.50	$\pm$	0.04	&	0.04	\\
HIP15776	&	HD21019	&	0.606	$\pm$	0.015	&	2013ApJ...771...40B	&	5259	$\pm$	66	&	1.3	&	3.73	$\pm$	0.12	&	-0.45	\\
HIP16537	&	HD22049	&	2.087	$\pm$	0.011	&	2020MNRAS.493.2377R	&	5130	$\pm$	30	&	0.6	&	4.63	$\pm$	0.01	&	-0.08	\\
HIP16852	&	HD22484	&	1.081	$\pm$	0.014	&	2012ApJ...746..101B	&	6000	$\pm$	59	&	1.0	&	4.06	$\pm$	0.03	&	-0.08	\\
HIP1686	&	HD1671	&	0.6	$\pm$	0.006	&	2016A\&A...586A..94L	&	6674	$\pm$	34	&	0.5	&	3.69	$\pm$	0.04	&	-0.09	\\
HIP17086	&	HD22798	&	0.792	$\pm$	0.021	&	2020A\&A...639A..67N	&	4652	$\pm$	88	&	1.9	&	2.40	$\pm$	0.25	&	0.24	\\
HIP171	&	HD224930	&	0.716	$\pm$	0.007	&	2020A\&A...640A..25K	&	5445	$\pm$	30	&	0.6	&	4.39	$\pm$	0.03	&	-0.79	\\
HIP17378	&	HD23249	&	2.343	$\pm$	0.009	&	2020MNRAS.493.2377R	&	5026	$\pm$	38	&	0.8	&	3.83	$\pm$	0.10	&	0.09	\\
HIP17595	&	HD23526	&	0.915	$\pm$	0.021	&	2018A\&A...616A..68G	&	4763	$\pm$	81	&	1.7	&	2.33	$\pm$	0.15	&	-0.15	\\
HIP17738	&	HD23940	&	1.093	$\pm$	0.021	&	2018A\&A...616A..68G	&	4815	$\pm$	52	&	1.1	&	2.43	$\pm$	0.06	&	-0.34	\\
HIP18859	&	HD25457	&	0.582	$\pm$	0.016	&	2018ApJ...858...71S	&	6295	$\pm$	91	&	1.5	&	4.39	$\pm$	0.04	&	0.10	\\
HIP19849	&			&	1.504	$\pm$	0.006	&	2012ApJ...757..112B	&	5181	$\pm$	21	&	0.4	&	4.51	$\pm$	0.01	&	-0.29	\\
HIP2021	&	HD2151	&	2.257	$\pm$	0.019	&	2007MNRAS.380L..80N	&	5917	$\pm$	25	&	0.4	&	3.97	$\pm$	0.04	&	-0.12	\\
HIP21421	&	HD29139	&	20.58	$\pm$	0.03	&	2005A\&A...433..305R	&	3921	$\pm$	80	&	2.0	&	1.17	$\pm$	0.05	&	-0.20	\\
HIP22449	&	HD30652	&	1.526	$\pm$	0.004	&	2012ApJ...746..101B	&	6518	$\pm$	35	&	0.5	&	4.31	$\pm$	0.01	&	0.03	\\
HIP22453	&	HD30504	&	2.803	$\pm$	0.013	&	2009MNRAS.394.1925V	&	4097	$\pm$	20	&	0.5	&	1.32	$\pm$	0.05	&	-0.34	\\
HIP22479	&	HD30814	&	1.31	$\pm$	0.01	&	2018A\&A...616A..68G	&	4901	$\pm$	27	&	0.6	&	2.73	$\pm$	0.04	&	-0.02	\\
HIP2413	&	HD2665	&	0.395	$\pm$	0.004	&	2020A\&A...640A..25K	&	4951	$\pm$	25	&	0.5	&	2.32	$\pm$	0.03	&	-1.97	\\
HIP24813	&	HD34411	&	0.981	$\pm$	0.015	&	2012ApJ...746..101B	&	5823	$\pm$	45	&	0.8	&	4.20	$\pm$	0.02	&	0.06	\\
HIP25993	&	HD36848	&	1.386	$\pm$	0.098	&	2012A\&A...539A..58C	&	4537	$\pm$	163	&	3.6	&	2.71	$\pm$	0.13	&	0.19	\\
HIP26019	&	HD36874	&	1.118	$\pm$	0.011	&	2018A\&A...616A..68G	&	4616	$\pm$	32	&	0.7	&	2.47	$\pm$	0.10	&	0.00	\\
HIP27435	&	HD38858	&	0.572	$\pm$	0.009	&	2013ApJ...771...40B	&	5705	$\pm$	47	&	0.8	&	4.43	$\pm$	0.02	&	-0.22	\\
HIP27530	&	HD39523	&	1.939	$\pm$	0.016	&	2018A\&A...616A..68G	&	4583	$\pm$	80	&	1.7	&	2.36	$\pm$	0.13	&	0.15	\\
HIP27621	&	HD39640	&	1.251	$\pm$	0.017	&	2018A\&A...616A..68G	&	4851	$\pm$	41	&	0.8	&	2.60	$\pm$	0.05	&	-0.11	\\
HIP27913	&	HD39587	&	1.051	$\pm$	0.009	&	2012ApJ...746..101B	&	5883	$\pm$	63	&	1.1	&	4.47	$\pm$	0.02	&	-0.03	\\
HIP28011	&	HD39910	&	1.09	$\pm$	0.008	&	2018A\&A...616A..68G	&	4565	$\pm$	33	&	0.7	&	2.50	$\pm$	0.05	&	0.26	\\
HIP28139	&	HD40020	&	1.012	$\pm$	0.023	&	2018A\&A...616A..68G	&	4671	$\pm$	59	&	1.3	&	2.57	$\pm$	0.09	&	0.13	\\
HIP29575	&	HD43023	&	0.842	$\pm$	0.014	&	2020A\&A...639A..67N	&	5043	$\pm$	43	&	0.9	&	2.96	$\pm$	0.08	&	0.00	\\
HIP3031	&	HD3546	&	1.77	$\pm$	0.08	&	1999AJ....118.3032N	&	4909	$\pm$	111	&	2.3	&	2.44	$\pm$	0.11	&	-0.62	\\
HIP30565	&	HD46116	&	1.145	$\pm$	0.031	&	2018A\&A...616A..68G	&	4880	$\pm$	68	&	1.4	&	2.53	$\pm$	0.15	&	-0.32	\\
HIP3093	&	HD3651	&	0.722	$\pm$	0.007	&	2016A\&A...586A..94L	&	5297	$\pm$	31	&	0.6	&	4.52	$\pm$	0.02	&	0.14	\\
HIP3137	&	HD3750	&	1.003	$\pm$	0.02	&	2018A\&A...616A..68G	&	4610	$\pm$	51	&	1.1	&	2.40	$\pm$	0.13	&	0.03	\\
HIP32362	&	HD48737	&	1.401	$\pm$	0.009	&	2012ApJ...746..101B	&	6537	$\pm$	25	&	0.4	&	3.79	$\pm$	0.03	&	0.14	\\
HIP32851	&	HD49933A	&	0.445	$\pm$	0.012	&	2011A\&A...534L...3B	&	6628	$\pm$	89	&	1.3	&	4.19	$\pm$	0.04	&	-0.39	\\
HIP3456	&	HD4211	&	1.1	$\pm$	0.011	&	2018A\&A...616A..68G	&	4572	$\pm$	35	&	0.8	&	2.34	$\pm$	0.10	&	0.01	\\
HIP36444	&	HD60060	&	0.948	$\pm$	0.01	&	2018A\&A...616A..68G	&	4814	$\pm$	32	&	0.7	&	2.59	$\pm$	0.04	&	-0.08	\\
HIP36732	&	HD60341	&	1.19	$\pm$	0.022	&	2018A\&A...616A..68G	&	4563	$\pm$	53	&	1.2	&	2.40	$\pm$	0.08	&	0.01	\\
HIP37279	&	HD61421	&	5.406	$\pm$	0.006	&	2021AJ....162..198B	&	6582	$\pm$	5	&	0.1	&	3.98	$\pm$	0.02	&	-0.02	\\
HIP3765	&	HD4628	&	0.868	$\pm$	0.004	&	2012ApJ...757..112B	&	5093	$\pm$	33	&	0.6	&	4.64	$\pm$	0.01	&	-0.26	\\
HIP37664	&	HD62713	&	1.446	$\pm$	0.012	&	2018A\&A...616A..68G	&	4661	$\pm$	32	&	0.7	&	2.29	$\pm$	0.08	&	0.02	\\
HIP37826	&	HD62509	&	8.018	$\pm$	0.043	&	2016ApJS..227....4H	&	4810	$\pm$	14	&	0.3	&	2.55	$\pm$	0.03	&	0.02	\\
HIP3850	&	HD4747	&	0.39	$\pm$	0.007	&	2019ApJ...873...83W	&	5351	$\pm$	48	&	0.9	&	4.55	$\pm$	0.03	&	-0.23	\\
HIP40526	&	HD69267	&	5.03	$\pm$	0.03	&	1999AJ....118.3032N	&	4080	$\pm$	15	&	0.4	&	1.28	$\pm$	0.05	&	-0.20	\\
HIP40693	&	HD69830	&	0.674	$\pm$	0.014	&	2015ApJ...800..115T	&	5317	$\pm$	58	&	1.1	&	4.45	$\pm$	0.03	&	-0.03	\\
HIP40843	&	HD69897	&	0.706	$\pm$	0.013	&	2013ApJ...771...40B	&	6203	$\pm$	57	&	0.9	&	4.17	$\pm$	0.02	&	-0.28	\\
HIP4151	&	HD5015	&	0.865	$\pm$	0.01	&	2012ApJ...746..101B	&	6033	$\pm$	37	&	0.6	&	4.03	$\pm$	0.03	&	0.05	\\
HIP4257	&	HD5268	&	0.767	$\pm$	0.035	&	2020A\&A...639A..67N	&	4944	$\pm$	116	&	2.3	&	2.36	$\pm$	0.18	&	-0.35	\\
HIP43587	&	HD75732	&	0.724	$\pm$	0.011	&	2016A\&A...586A..94L	&	5169	$\pm$	44	&	0.8	&	4.41	$\pm$	0.02	&	0.32	\\
HIP43813	&	HD76294	&	3.196	$\pm$	0.017	&	2018AJ....155...30B	&	4836	$\pm$	14	&	0.3	&	2.49	$\pm$	0.04	&	-0.05	\\
HIP45343	&	HD79210	&	0.871	$\pm$	0.014	&	2012ApJ...757..112B	&	3997	$\pm$	34	&	0.9	&	4.69	$\pm$	0.02	&	0.17	\\
HIP45860	&	HD80493	&	7.538	$\pm$	0.075	&	2003AJ....126.2502M	&	3881	$\pm$	20	&	0.5	&	1.06	$\pm$	0.04	&	-0.26	\\
HIP4587	&	HD5722	&	0.995	$\pm$	0.019	&	2018A\&A...616A..68G	&	4914	$\pm$	50	&	1.0	&	2.61	$\pm$	0.05	&	-0.23	\\
HIP46853	&	HD82328	&	1.632	$\pm$	0.005	&	2012ApJ...746..101B	&	6217	$\pm$	44	&	0.7	&	3.83	$\pm$	0.05	&	-0.17	\\
HIP47080	&	HD82885	&	0.821	$\pm$	0.012	&	2012ApJ...746..101B	&	5452	$\pm$	46	&	0.8	&	4.44	$\pm$	0.02	&	0.34	\\
HIP47431	&	HD83618	&	3.462	$\pm$	0.033	&	2018AJ....155...30B	&	4238	$\pm$	22	&	0.5	&	1.78	$\pm$	0.04	&	-0.06	\\
HIP47908	&	HD84441	&	2.643	$\pm$	0.015	&	2009MNRAS.394.1925V	&	5314	$\pm$	17	&	0.3	&	2.36	$\pm$	0.05	&	-0.03	\\
HIP48455	&	HD85503	&	2.887	$\pm$	0.016	&	2018AJ....155...30B	&	4519	$\pm$	23	&	0.5	&	2.43	$\pm$	0.06	&	0.27	\\
HIP49081	&	HD86728	&	0.771	$\pm$	0.012	&	2012ApJ...746..101B	&	5610	$\pm$	46	&	0.8	&	4.25	$\pm$	0.02	&	0.20	\\
HIP49637	&	HD87837	&	3.33	$\pm$	0.04	&	1999AJ....118.3032N	&	4106	$\pm$	47	&	1.2	&	1.59	$\pm$	0.06	&	-0.02	\\
HIP49908	&	HD88230	&	1.268	$\pm$	0.04	&	2001ApJ...551L..81L	&	4132	$\pm$	72	&	1.7	&	4.63	$\pm$	0.04	&	0.21	\\
HIP50564	&	HD89449	&	0.731	$\pm$	0.026	&	2013MNRAS.434.1321M	&	6385	$\pm$	120	&	1.9	&	4.12	$\pm$	0.05	&	0.10	\\
HIP50887	&	HD90043	&	0.659	$\pm$	0.009	&	2018MNRAS.477.4403W	&	4801	$\pm$	83	&	1.7	&	3.11	$\pm$	0.06	&	-0.03	\\
HIP51459	&	HD90839	&	0.794	$\pm$	0.014	&	2012ApJ...746..101B	&	6259	$\pm$	56	&	0.9	&	4.40	$\pm$	0.02	&	-0.12	\\
HIP5336	&	HD6582	&	0.973	$\pm$	0.009	&	2008ApJ...683..424B	&	5358	$\pm$	31	&	0.6	&	4.49	$\pm$	0.01	&	-0.83	\\
HIP544	&	HD166	&	0.624	$\pm$	0.009	&	2013ApJ...771...40B	&	5378	$\pm$	40	&	0.7	&	4.46	$\pm$	0.02	&	0.11	\\
HIP5445	&	HD6755	&	0.369	$\pm$	0.004	&	2020A\&A...640A..25K	&	4977	$\pm$	27	&	0.5	&	2.77	$\pm$	0.03	&	-1.43	\\
HIP54539	&	HD96833	&	4.131	$\pm$	0.007	&	2018AJ....155...30B	&	4543	$\pm$	6	&	0.1	&	2.12	$\pm$	0.05	&	-0.11	\\
HIP5458	&	HD6833	&	0.852	$\pm$	0.008	&	2020A\&A...640A..25K	&	4447	$\pm$	42	&	0.9	&	1.83	$\pm$	0.02	&	-0.86	\\
HIP55219	&	HD98262	&	4.561	$\pm$	0.016	&	2018AJ....155...30B	&	4187	$\pm$	61	&	1.4	&	1.38	$\pm$	0.10	&	-0.14	\\
HIP56127	&	HD99998	&	3.21	$\pm$	0.02	&	1999AJ....118.3032N	&	3852	$\pm$	90	&	2.3	&	0.95	$\pm$	0.06	&	-0.39	\\
HIP56343	&	HD100407	&	2.386	$\pm$	0.021	&	2005A\&A...436..253T	&	5034	$\pm$	34	&	0.7	&	2.78	$\pm$	0.07	&	-0.09	\\
HIP56997	&	HD101501	&	0.91	$\pm$	0.009	&	2012ApJ...746..101B	&	5310	$\pm$	28	&	0.5	&	4.42	$\pm$	0.01	&	-0.05	\\
HIP57477	&	HD102328	&	1.606	$\pm$	0.006	&	2010ApJ...710.1365B	&	4450	$\pm$	31	&	0.7	&	2.37	$\pm$	0.08	&	0.19	\\
HIP57757	&	HD102870	&	1.431	$\pm$	0.006	&	2012ApJ...746..101B	&	6093	$\pm$	13	&	0.2	&	4.08	$\pm$	0.02	&	0.13	\\
HIP57939	&	HD103095	&	0.593	$\pm$	0.004	&	2020A\&A...640A..25K	&	5235	$\pm$	18	&	0.3	&	4.72	$\pm$	0.01	&	-1.33	\\
HIP61317	&	HD109358	&	1.133	$\pm$	0.034	&	2018AJ....155...30B	&	6013	$\pm$	91	&	1.5	&	4.41	$\pm$	0.04	&	-0.20	\\
HIP63584	&	HD113337	&	0.386	$\pm$	0.009	&	2019A\&A...627A..44B	&	6783	$\pm$	79	&	1.2	&	4.24	$\pm$	0.03	&	0.17	\\
HIP63608	&	HD113226	&	3.318	$\pm$	0.013	&	2018AJ....155...30B	&	4950	$\pm$	10	&	0.2	&	2.72	$\pm$	0.02	&	0.06	\\
HIP64394	&	HD114710	&	1.127	$\pm$	0.011	&	2012ApJ...746..101B	&	5930	$\pm$	30	&	0.5	&	4.37	$\pm$	0.01	&	0.06	\\
HIP65721	&	HD117176	&	0.998	$\pm$	0.004	&	2015ApJ...806...60K	&	5473	$\pm$	22	&	0.4	&	3.94	$\pm$	0.08	&	-0.06	\\
HIP6592	&	HD8651	&	1.228	$\pm$	0.013	&	2018A\&A...616A..68G	&	4685	$\pm$	43	&	0.9	&	2.39	$\pm$	0.08	&	-0.20	\\
HIP671	&	HD360	&	0.906	$\pm$	0.015	&	2018A\&A...616A..68G	&	4679	$\pm$	79	&	1.7	&	2.32	$\pm$	0.11	&	-0.09	\\
HIP67459	&	HD120477	&	4.72	$\pm$	0.04	&	1999AJ....118.3032N	&	3950	$\pm$	20	&	0.5	&	1.30	$\pm$	0.05	&	-0.40	\\
HIP67927	&	HD121370	&	2.134	$\pm$	0.012	&	2014ApJ...781...90B	&	6161	$\pm$	18	&	0.3	&	3.82	$\pm$	0.05	&	0.25	\\
HIP68594	&	HD122563	&	0.925	$\pm$	0.011	&	2020A\&A...640A..25K	&	4642	$\pm$	35	&	0.8	&	1.32	$\pm$	0.03	&	-2.67	\\
HIP69673	&	HD124897	&	21.05	$\pm$	0.21	&	2008A\&A...485..561L	&	4277	$\pm$	23	&	0.5	&	1.58	$\pm$	0.07	&	-0.55	\\
HIP70497	&	HD126660	&	1.109	$\pm$	0.007	&	2012ApJ...746..101B	&	6292	$\pm$	20	&	0.3	&	4.06	$\pm$	0.02	&	-0.03	\\
HIP70791	&	HD127243	&	0.971	$\pm$	0.007	&	2020A\&A...640A..25K	&	5015	$\pm$	23	&	0.5	&	2.50	$\pm$	0.02	&	-0.70	\\
HIP7083	&	HD9362	&	2.301	$\pm$	0.021	&	2018A\&A...616A..68G	&	4750	$\pm$	28	&	0.6	&	2.40	$\pm$	0.05	&	-0.31	\\
HIP71053	&	HD127665	&	3.901	$\pm$	0.008	&	2018AJ....155...30B	&	4181	$\pm$	56	&	1.3	&	1.80	$\pm$	0.05	&	-0.17	\\
HIP71681	&	HD128621	&	5.999	$\pm$	0.025	&	2017A\&A...597A.137K	&	5207	$\pm$	12	&	0.2	&	4.53	$\pm$	0.01	&	0.24	\\
HIP71683	&	HD128620	&	8.502	$\pm$	0.038	&	2017A\&A...597A.137K	&	5804	$\pm$	13	&	0.2	&	4.29	$\pm$	0.01	&	0.20	\\
HIP72567	&	HD130948	&	0.569	$\pm$	0.011	&	2013ApJ...771...40B	&	5812	$\pm$	57	&	1.0	&	4.33	$\pm$	0.03	&	-0.01	\\
HIP7294	&	HD9408	&	1.64	$\pm$	0.09	&	1999AJ....118.3032N	&	4774	$\pm$	131	&	2.7	&	2.35	$\pm$	0.11	&	-0.30	\\
HIP73184	&	HD131977	&	1.098	$\pm$	0.014	&	2020MNRAS.493.2377R	&	4724	$\pm$	71	&	1.5	&	4.63	$\pm$	0.02	&	0.02	\\
HIP73568	&	HD133124	&	3.055	$\pm$	0.077	&	2018AJ....155...30B	&	3994	$\pm$	56	&	1.4	&	1.37	$\pm$	0.06	&	-0.10	\\
HIP74666	&	HD135722	&	2.764	$\pm$	0.03	&	2003AJ....126.2502M	&	4810	$\pm$	30	&	0.6	&	2.38	$\pm$	0.09	&	-0.35	\\
HIP74793	&	HD136726	&	2.149	$\pm$	0.023	&	2018AJ....155...30B	&	4253	$\pm$	25	&	0.6	&	1.78	$\pm$	0.04	&	-0.02	\\
HIP74975	&	HD136202	&	0.785	$\pm$	0.023	&	2013ApJ...771...40B	&	5820	$\pm$	99	&	1.7	&	3.87	$\pm$	0.04	&	-0.02	\\
HIP7607	&	HD9927	&	3.649	$\pm$	0.007	&	2018AJ....155...30B	&	4356	$\pm$	56	&	1.3	&	1.99	$\pm$	0.07	&	0.05	\\
HIP7643	&	HD10142	&	0.964	$\pm$	0.006	&	2018A\&A...616A..68G	&	4705	$\pm$	28	&	0.6	&	2.38	$\pm$	0.10	&	-0.13	\\
HIP76976	&	HD140283	&	0.327	$\pm$	0.005	&	2018MNRAS.475L..81K	&	5788	$\pm$	45	&	0.8	&	3.75	$\pm$	0.09	&	-2.48	\\
HIP77052	&	HD140538	&	0.597	$\pm$	0.015	&	2013ApJ...771...40B	&	5667	$\pm$	91	&	1.6	&	4.47	$\pm$	0.04	&	0.05	\\
HIP78072	&	HD142860	&	1.217	$\pm$	0.005	&	2012ApJ...746..101B	&	6296	$\pm$	16	&	0.2	&	4.16	$\pm$	0.01	&	-0.18	\\
HIP79672	&	HD146233	&	0.676	$\pm$	0.006	&	2011A\&A...526L...4B	&	5824	$\pm$	30	&	0.5	&	4.42	$\pm$	0.01	&	0.03	\\
HIP7981	&	HD10476	&	1.0	$\pm$	0.004	&	2013ApJ...771...40B	&	5129	$\pm$	71	&	1.4	&	4.52	$\pm$	0.01	&	-0.04	\\
HIP80843	&	HD148897	&	1.917	$\pm$	0.045	&	2021ApJ...922..163V	&	4227	$\pm$	51	&	1.2	&	1.11	$\pm$	0.07	&	-0.92	\\
HIP8102	&	HD10700	&	2.005	$\pm$	0.011	&	2020MNRAS.493.2377R	&	5463	$\pm$	16	&	0.3	&	4.52	$\pm$	0.01	&	-0.51	\\
HIP81300	&	HD149661	&	0.724	$\pm$	0.011	&	2012ApJ...757..112B	&	5408	$\pm$	47	&	0.9	&	4.62	$\pm$	0.02	&	0.03	\\
HIP8159	&	HD10697	&	0.547	$\pm$	0.012	&	2013ApJ...771...40B	&	5438	$\pm$	61	&	1.1	&	3.94	$\pm$	0.08	&	0.14	\\
HIP81693	&	HD150680	&	2.367	$\pm$	0.051	&	2003AJ....126.2502M	&	5760	$\pm$	96	&	1.7	&	3.72	$\pm$	0.03	&	0.03	\\
HIP83000	&	HD153210	&	3.608	$\pm$	0.041	&	2016ApJS..227....4H	&	4499	$\pm$	68	&	1.5	&	2.49	$\pm$	0.07	&	0.02	\\
HIP8362	&	HD10780	&	0.763	$\pm$	0.018	&	2012ApJ...746..101B	&	5293	$\pm$	77	&	1.5	&	4.55	$\pm$	0.03	&	0.03	\\
HIP8404	&	HD11037	&	0.89	$\pm$	0.019	&	2020A\&A...639A..67N	&	4834	$\pm$	59	&	1.2	&	2.33	$\pm$	0.21	&	-0.13	\\
HIP84862	&	HD157214	&	0.725	$\pm$	0.012	&	2013ApJ...771...40B	&	5825	$\pm$	49	&	0.8	&	4.26	$\pm$	0.03	&	-0.39	\\
HIP84950	&	HD157681	&	1.908	$\pm$	0.013	&	2016AJ....152...66B	&	4108	$\pm$	23	&	0.6	&	1.35	$\pm$	0.03	&	-0.23	\\
HIP85235	&	HD158633	&	0.573	$\pm$	0.01	&	2013ApJ...771...40B	&	5188	$\pm$	51	&	1.0	&	4.51	$\pm$	0.02	&	-0.46	\\
HIP85258	&	HD157244	&	5.997	$\pm$	0.037	&	2015A\&A...582A..49H	&	4232	$\pm$	17	&	0.4	&	0.97	$\pm$	0.06	&	0.50	\\
HIP86614	&	HD162003	&	0.949	$\pm$	0.025	&	2012ApJ...746..101B	&	5936	$\pm$	89	&	1.5	&	3.82	$\pm$	0.04	&	0.01	\\
HIP86742	&	HD161096	&	4.498	$\pm$	0.032	&	2016ApJS..227....4H	&	4577	$\pm$	18	&	0.4	&	2.33	$\pm$	0.07	&	0.06	\\
HIP86974	&	HD161797	&	1.88	$\pm$	0.008	&	2018AJ....155...30B	&	5665	$\pm$	16	&	0.3	&	4.05	$\pm$	0.04	&	0.25	\\
HIP87808	&	HD163770	&	3.15	$\pm$	0.003	&	2009MNRAS.394.1925V	&	4448	$\pm$	23	&	0.5	&	1.37	$\pm$	0.04	&	-0.03	\\
HIP87833	&	HD164058	&	9.86	$\pm$	0.128	&	2003AJ....126.2502M	&	4018	$\pm$	32	&	0.8	&	1.33	$\pm$	0.04	&	-0.08	\\
HIP88348	&	HD164922	&	0.4120	$\pm$	0.0100	&	2016ApJ...830...46F	&	5392	$\pm$	66	&	1.2	&	4.43	$\pm$	0.03	&	0.18	\\
HIP8837	&	HD11695	&	8.13	$\pm$	0.2	&	2004A\&A...413..711W	&	3362	$\pm$	183	&	5.4	&	0.44	$\pm$	0.06	&	-1.24	\\
HIP89047	&	HD167042	&	0.831	$\pm$	0.0068	&	2020A\&A...640A...2S	&	4987	$\pm$	66	&	1.3	&	3.41	$\pm$	0.06	&	0.04	\\
HIP8928	&	HD11977	&	1.528	$\pm$	0.013	&	2018A\&A...616A..68G	&	4890	$\pm$	30	&	0.6	&	2.61	$\pm$	0.05	&	-0.15	\\
HIP89348	&	HD168151	&	0.664	$\pm$	0.014	&	2016A\&A...586A..94L	&	6569	$\pm$	69	&	1.1	&	4.09	$\pm$	0.03	&	-0.29	\\
HIP89962	&	HD168723	&	3.062	$\pm$	0.048	&	2016ApJS..227....4H	&	4801	$\pm$	47	&	1.0	&	2.92	$\pm$	0.04	&	-0.20	\\
HIP90344	&	HD170693	&	2.041	$\pm$	0.043	&	2010ApJ...710.1365B	&	4367	$\pm$	46	&	1.1	&	1.86	$\pm$	0.04	&	-0.45	\\
HIP9094	&	HD11964	&	0.607	$\pm$	0.015	&	2013ApJ...771...40B	&	5082	$\pm$	63	&	1.2	&	3.88	$\pm$	0.08	&	0.11	\\
HIP91949	&	HD173701	&	0.332	$\pm$	0.006	&	2012ApJ...760...32H	&	5315	$\pm$	48	&	0.9	&	4.44	$\pm$	0.03	&	0.30	\\
HIP92043	&	HD173667	&	1.0	$\pm$	0.006	&	2012ApJ...746..101B	&	6426	$\pm$	20	&	0.3	&	3.95	$\pm$	0.03	&	-0.04	\\
HIP92167	&	HD175305	&	0.484	$\pm$	0.006	&	2020A\&A...640A..25K	&	4902	$\pm$	30	&	0.6	&	2.53	$\pm$	0.02	&	-1.45	\\
HIP92512	&	HD175306	&	2.189	$\pm$	0.007	&	2015ApJ...809..159R	&	4464	$\pm$	23	&	0.5	&	1.78	$\pm$	0.06	&	-0.53	\\
HIP92984	&	HD175726	&	0.346	$\pm$	0.007	&	2012ApJ...760...32H	&	6070	$\pm$	62	&	1.0	&	4.48	$\pm$	0.03	&	-0.04	\\
HIP93427	&	HD177153	&	0.2890	$\pm$	0.0060	&	2012ApJ...760...32H	&	6016	$\pm$	63	&	1.0	&	4.24	$\pm$	0.03	&	-0.05	\\
HIP93429	&	HD176678	&	2.463	$\pm$	0.012	&	2018AJ....155...30B	&	4491	$\pm$	105	&	2.3	&	2.31	$\pm$	0.11	&	-0.08	\\
HIP9440	&	HD12438	&	1.091	$\pm$	0.016	&	2018A\&A...616A..68G	&	4954	$\pm$	42	&	0.8	&	2.45	$\pm$	0.08	&	-0.58	\\
HIP94755	&	HD181096	&	0.443	$\pm$	0.007	&	2019MNRAS.489..928S	&	6390	$\pm$	51	&	0.8	&	3.94	$\pm$	0.03	&	-0.25	\\
HIP95362	&	HD182736	&	0.4360	$\pm$	0.0030	&	2012ApJ...760...32H	&	5268	$\pm$	18	&	0.3	&	3.77	$\pm$	0.07	&	-0.11	\\
HIP95447	&	HD182572	&	0.845	$\pm$	0.025	&	2012ApJ...746..101B	&	5673	$\pm$	84	&	1.5	&	4.22	$\pm$	0.04	&	0.38	\\
HIP96014	&	HD184293	&	1.548	$\pm$	0.022	&	2016AJ....152...66B	&	4336	$\pm$	42	&	1.0	&	1.79	$\pm$	0.05	&	-0.35	\\
HIP96100	&	HD185144	&	1.254	$\pm$	0.011	&	2012ApJ...746..101B	&	5289	$\pm$	24	&	0.5	&	4.56	$\pm$	0.02	&	-0.21	\\
HIP96441	&	HD185395	&	0.749	$\pm$	0.007	&	2016A\&A...586A..94L	&	6914	$\pm$	33	&	0.5	&	4.24	$\pm$	0.01	&	-0.03	\\
HIP96837	&	HD185958	&	1.764	$\pm$	0.012	&	2009MNRAS.394.1925V	&	4990	$\pm$	20	&	0.4	&	2.20	$\pm$	0.02	&	0.00	\\
HIP96895	&	HD186408	&	0.539	$\pm$	0.007	&	2013MNRAS.433.1262W	&	5849	$\pm$	39	&	0.7	&	4.28	$\pm$	0.02	&	0.08	\\
HIP96901	&	HD186427	&	0.490	$\pm$	0.006	&	2013MNRAS.433.1262W	&	5807	$\pm$	36	&	0.6	&	4.35	$\pm$	0.02	&	0.07	\\
HIP97527	&	HD187637	&	0.231	$\pm$	0.006	&	2012ApJ...760...32H	&	6236	$\pm$	81	&	1.3	&	4.26	$\pm$	0.03	&	-0.09	\\
HIP98036	&	HD188512	&	2.079	$\pm$	0.011	&	2020MNRAS.493.2377R	&	5155	$\pm$	15	&	0.3	&	3.68	$\pm$	0.06	&	-0.15	\\
HIP98269	&	HD189349	&	0.417	$\pm$	0.005	&	2022A\&A...658A..48K	&	5175	$\pm$	31	&	0.6	&	2.39	$\pm$	0.04	&	-0.59	\\
HIP98337	&	HD189319	&	6.089	$\pm$	0.011	&	2021AJ....162..198B	&	3904	$\pm$	30	&	0.8	&	1.06	$\pm$	0.04	&	-0.26	\\
HIP98624	&	HD188887	&	1.595	$\pm$	0.011	&	2018A\&A...616A..68G	&	4381	$\pm$	28	&	0.6	&	2.35	$\pm$	0.10	&	0.11	\\
HIP98767	&	HD190360	&	0.698	$\pm$	0.019	&	2016A\&A...586A..94L	&	5463	$\pm$	75	&	1.4	&	4.26	$\pm$	0.03	&	0.22	\\
HIP98819	&	HD190406	&	0.584	$\pm$	0.01	&	2012ApJ...751...97C	&	5825	$\pm$	50	&	0.9	&	4.35	$\pm$	0.02	&	0.05	\\
HIP99663	&	HD192781	&	1.859	$\pm$	0.002	&	2010ApJ...710.1365B	&	4084	$\pm$	25	&	0.6	&	1.40	$\pm$	0.04	&	-0.24	\\

\end{longtable}
\end{appendix}

\end{document}